\documentclass[modern,trackchanges]{aastex62}

\usepackage{bm} 
\usepackage{amsmath}
\usepackage{multirow}
\usepackage{rotating}

\usepackage{scrextend,rotating}
\makeatletter
\newenvironment{rotatepage}
        {\clearpage%
            \if@twoside%
                \ifthispageodd{\pagebreak[4]\global\pdfpageattr\expandafter{\the\pdfpageattr/Rotate 90}}{%
                \pagebreak[4]\global\pdfpageattr\expandafter{\the\pdfpageattr/Rotate 270}}%
            \else%
                \pagebreak[4]\global\pdfpageattr\expandafter{\the\pdfpageattr/Rotate 90}%
            \fi%
        }%
        {\clearpage\pagebreak[4]\global\pdfpageattr\expandafter{\the\pdfpageattr/Rotate 0}}%
\makeatother

\newcommand*\diff{\mathop{}\!\mathrm{d}}
\newcommand{\Secref}[1]{\hyperref[#1]{Section~\ref*{#1}}}
\newcommand{\Appref}[1]{\hyperref[#1]{Appendix~\ref*{#1}}}

\graphicspath{{./}{figures/}}
\submitjournal{AJ}

\shorttitle{Deep exploration of HR 8799 bcd.}
\shortauthors{Ruffio et al.}

\begin{document}

\title{Deep exploration of the planets HR 8799 b, c, and d with moderate resolution spectroscopy}

\correspondingauthor{Jean-Baptiste Ruffio}
\email{jruffio@caltech.edu}

\author[0000-0003-2233-4821]{Jean-Baptiste Ruffio}
\affiliation{Department of Astronomy, California Institute of Technology, Pasadena, CA 91125, USA}
\affiliation{Kavli Institute for Particle Astrophysics and Cosmology, Stanford University, Stanford, CA 94305, USA}

\author[0000-0002-9936-6285]{Quinn M. Konopacky}
\affiliation{Center for Astrophysics and Space Science, University of California, San Diego; La Jolla, CA 92093, USA}

\author[0000-0002-7129-3002]{Travis Barman}
\affiliation{Lunar and Planetary Laboratory, University of Arizona, Tucson, AZ 85721, USA}

\author[0000-0003-1212-7538]{Bruce Macintosh}
\affiliation{Kavli Institute for Particle Astrophysics and Cosmology, Stanford University, Stanford, CA 94305, USA}

\author[0000-0002-9803-8255]{Kielan K. Wilcomb}
\affiliation{Center for Astrophysics and Space Science, University of California, San Diego; La Jolla, CA 92093, USA}

\author[0000-0002-4918-0247]{Robert J. De Rosa}
\affiliation{European Southern Observatory, Alonso de C\'{o}rdova 3107, Vitacura, Santiago, Chile}

\author[0000-0003-0774-6502]{Jason J. Wang}
\altaffiliation{51 Pegasi b Fellow}
\affiliation{Department of Astronomy, California Institute of Technology, Pasadena, CA 91125, USA}

\author[0000-0002-1483-8811]{Ian Czekala}
\altaffiliation{NASA Hubble Fellowship Program Sagan Fellow}
\affiliation{Department of Astronomy and Astrophysics, 525 Davey Laboratory, The Pennsylvania State University, University Park, PA 16802, USA}
\affiliation{Center for Exoplanets and Habitable Worlds, 525 Davey Laboratory, The Pennsylvania State University, University Park, PA 16802, USA}
\affiliation{Center for Astrostatistics, 525 Davey Laboratory, The Pennsylvania State University, University Park, PA 16802, USA}
\affiliation{Institute for Computational \& Data Sciences, The Pennsylvania State University, University Park, PA 16802, USA}
\affiliation{Department of Astronomy, 501 Campbell Hall, University of California, Berkeley, CA 94720-3411, USA}

\author[0000-0002-4164-4182]{Christian Marois}
\affiliation{National Research Council of Canada Herzberg, 5071 West Saanich Rd, Victoria, BC, Canada V9E 2E7}
\affiliation{University of Victoria, 3800 Finnerty Rd, Victoria, BC, Canada V8P 5C2}

\begin{abstract}
The four directly imaged planets orbiting the star HR 8799 are an ideal laboratory to probe atmospheric physics and formation models. We present more than a decade's worth of Keck/OSIRIS observations of these planets, which represent the most detailed look at their atmospheres to-date by its resolution and signal to noise ratio. 
We present the first direct detection of HR 8799 d, the second-closest known planet to the star, at moderate spectral resolution with Keck/OSIRIS (K-band; $R\approx4,000$). Additionally, we uniformly analyze new and archival OSIRIS data (H and K band) of HR 8799 b, c, and d. 
First, we show detections of water (H\textsubscript{2}O) and carbon monoxide (CO) in the three planets and discuss the ambiguous case of methane (CH\textsubscript{4}) in the atmosphere of HR 8799b.
Then, we report radial velocity (RV) measurements for each of the three planets. The RV measurement of HR 8799 d is consistent with predictions made assuming coplanarity and orbital stability of the HR 8799 planetary system. 
Finally, we perform a uniform atmospheric analysis on the OSIRIS data, published photometric points, and low resolution spectra. We do not infer any significant deviation from to the stellar value of the carbon to oxygen ratio (C/O) of the three planets, which therefore does not yet yield definitive information about the location or method of formation. However, constraining the C/O ratio for all the HR 8799 planets is a milestone for any multiplanet system, and particularly important for large, widely separated gas giants with uncertain formation processes.
\end{abstract}


\section{Introduction}

There are two main pathways to form sub-stellar companions around young stars \citep{Lissauer1987,Boss1997,Boley2009}. On one hand, planets like those in our solar system are believed to form by core accretion, in which a core slowly accretes solid material before reaching a critical mass that triggers runaway accretion of the gas. On the other hand, sub-stellar companions can, for example, form by gravitational instability of the protoplanetary disk or turbulent fragmentation of the protostellar cloud. In the transitional mass range between planets and brown-dwarfs, the formation mechanisms of directly imaged planets ($>2\,\mathrm{M}_\mathrm{Jup}$; effective temperature $500-2000\,\mathrm{K}$; semi-major axis $>5\,\mathrm{au}$) are still outstanding \citep{Kratter2010,Nielsen2019}. 
The present-day composition of a planet's atmosphere may contain evidence that can be used to probe its formation history, including its original location, migration pathways, and the fraction of gas and solids it accreted \citep{Oberg2011}.
Indeed, following the core accretion paradigm, the composition of planetesimals in the protoplanetary disk are controlled by the different ice-lines, where volatiles freeze out of the gas phase.
Alternatively, objects forming through gravitational instability are expected to inherit the bulk composition of the original protostellar gas, giving the same composition as the star. 
For these reasons, abundance ratios such as the carbon-to-oxygen ratio (C/O) are considered to be important tracers of planet formation.
Evidently, other factors like the efficiency of the mixing within the planet’s atmosphere, planetesimal drift, migration, or core dredging, could also strongly influence the final composition of an atmosphere \citep{Mordasini2016,Madhusudhan2017}. 

The ability of high-contrast imaging instruments to spatially resolve the light of the planet from the star makes direct imaging an excellent way to characterize exoplanets. 
The system of four planets ($10-70\,\mathrm{au}$, $5-13\,\mathrm{M}_\mathrm{Jup}$) orbiting HR 8799 was the first multi-planet system to be directly imaged \citep{Marois2008science,Marois2010}. 
The host star is $\sim1.5M_\odot$ \citep{Gray1999,Baines2012}, $\sim 42\,\mathrm{Myr}$ old \citep{Zuckerman2011,Malo2013,Bell2015}, and is located at $41.3\pm0.2\,\mathrm{pc}$ \citep{Gaia2018}.
From orbital monitoring and modeling, the four planets are thought to be within a few degrees of coplanarity, and at or near 1:2:4:8 resonance \citep{Fabrycky2010,Soummer2011,Pueyo2015,Konopacky2016,Zurlo2016, Wertz2017, Wang2018, ONeil2019,Gravity2019,Ruffio2019RV,Gozdziewski2020}.
They are bracketed by both a warm and a cold debris disk similar to the asteroid and Kuiper Belts in the Solar system \citep{Su2009,Reidemeister2009,Hughes2011,Matthews2014,Contro2016,Booth2016,Read2018,Gozdziewski2018,Wilner2018,Geiler2019}.
HR 8799b, c, d, and e, which have been observed with a broad wavelength coverage ($1-5\,\mu\mathrm{m}$) and spectral resolution up to $R=15,000$ \citep{WangJi2018}, are among the best studied exoplanets and as such represent a unique opportunity for comparative planetology. 
From low resolution spectroscopy and photometry, the planets were found to significantly differ from field brown-dwarfs of similar effective temperature \citep{Bowler2010, Currie2011, Madhusudhan2011, Skemer2014, Bonnefoy2016, Zurlo2016}. They likely feature thick, patchy clouds, and disequilibrium chemistry \citep{Hinz2010, Galicher2011, Skemer2012, Ingraham2014,Currie2014, Greenbaum2018}.

At low spectral resolution ($R=\lambda/\Delta\lambda\approx30$), quantitative assessment of planetary molecular abundances is limited by degeneracies between abundances, clouds, and pressure or gravity. Higher resolution spectra can help break these degeneracies. Moderate ($R\sim4,000$) and high ($R>25,000$) resolution spectroscopy can resolve individual absorption lines, providing unambiguous detection of specific molecules and their abundance. Additionally, spectroscopy of sub-stellar companions at these resolutions has enabled radial velocity (RV) measurements \citep{Brogi2014,Snellen2014,Ruffio2019RV}, spin measurements \citep{Lissauer1991,Snellen2014, Schwarz2016, Bryan2018,Bryan2020}, and the creation of ``surface'' maps \citep[through Doppler mapping;][]{Crossfield2014Nature,Crossfield2014}.

Infrared moderate resolution spectroscopy with Keck/OSIRIS ($R\approx4000$; \citealt{Larkin2006}) has allowed the unambiguous detection of water (H\textsubscript{2}O) and carbon monoxide (CO) \citep{Konopacky2013,Barman2015}. 
While methane (CH\textsubscript{4}) is abundant in the atmosphere of cool field brown-dwarfs, its lack of strong detection in the atmosphere of the HR 8799 planets is evidence of disequilibrium chemistry \citet{Barman2011}. An initial detection of CH\textsubscript{4} in HR 8799b by \citet{Barman2015} was not confirmed by \citet{PetitditdelaRoche2018}. Further analyses and observations are therefore warranted to confirm the abundance of methane and the magnitude of the eddy diffusion coefficient.
Moderate resolution spectroscopy in K band is particularly useful estimate to the C/O of an exoplanet, which can then be compared to the C/O of the star. The C/O of the host star HR 8799 is close to solar with $C/O = 0.54^{+0.12}_{-0.09}$ \citep{WangJi2020}.
Using OSIRIS spectra and self-consistent atmospheric models, slightly super stellar C/O were found for HR 8799b and c, $[C/O]_c = 0.65^{+0.1}_{-0.05}$ \citet{Konopacky2013} and $[C/O]_b \in [0.58,0.7]$ \citet{Barman2015}, suggesting core accretion scenarios. 
However, there exists a significant scatter in C/O estimates between different studies. 
There exists two complementary paradigms for atmospheric characterization: pre-computed self-consistent forward models, and atmospheric retrievals which fit for the temperature-pressure profile but do not ensure overall chemical equilibrium. Due to the computational cost of generating models at high spectral resolution, forward model grids are useful for high-resolution data, but remain limited by the small number of varied parameters and a coarse sampling. Atmospheric retrievals of lower resolution spectra infer a C/O near unity for HR 8799b, but stellar C/O for the other three planets: $[C/O]_b = 0.97\pm0.01$ \citep{Lee2013}, $[C/O]_b = 0.92\pm0.01$ and $[C/O]_c = 0.55\pm0.1$ \citep{Lavie2017}, $[C/O]_e = 0.6^{+0.07}_{-0.08}$ \citep{Molliere2020}, and $[C/O]_c = 0.58^{0.06}_{-0.06}$ with weak priors \citep{WangJi2020}. \citet{WangJi2020} also shows that the C/O can vary significantly, from 0.39 to 0.58 in this case, depending on the priors used and the data included in the fit of HR 8799c.

Methods to measure stellar radial velocities for exoplanet Doppler detection are generally classified into two categories: cross correlation \citep[e.g.][]{Baranne1996} and forward modeling \citep[e.g][]{Gao2016,Cale2019}; although both techniques fundamentally rely on likelihood maximization. Exoplanet atmospheres have mostly been studied at high spectral resolution spectroscopy using the simpler cross correlation function technique.
In \citet{Ruffio2019RV}, we developed a forward model for high-contrast companions at high spectral resolution. Forward modeling relies on minimal pre-processing of the data, and allows the direct estimation of any atmospheric parameters (RV, C/O etc.) with their associated uncertainties. The joint estimation of the planet signal and the starlight avoids the over-subtraction of the planet signal when subtracting the starlight. Another significant advantage is the uncertainties corresponding to the starlight subtraction can be analytically marginalized out of the joint posterior probability distribution. We note that in the presence of flat priors, this method and the conventional cross correlation function are both maximum likelihood estimations of the planet signal.
Using the forward modeling approach, we were able to measure the RV of HR 8799b and c, and resolve the 3D orientation of the system \citep{Ruffio2019RV}. This new framework allowed a first preliminary detection of HR 8799d at moderate spectral resolution in \citet{Ruffio2019thesis}. 
The framework was improved in \citet{wilcomb2020} to correct for systematics and account for intra-night telluric variability. 

In this work, we present new deep HR 8799d observations and perform a uniform analysis of the three planets.
First, we describe the observations and data reduction in \Secref{sec:obs}, and redefine the statistical model in \Secref{sec:model}. The detection of H\textsubscript{2}O, CO, and CH\textsubscript{4} is discussed in \Secref{sec:ccf} along with the extracted spectra for the three planets. Then, we present the first RV measurement of HR 8799d and revise the RV of HR 8799b and c in \Secref{sec:RV}. In \Secref{sec:atmcharac}, we fit a grid of atmospheric models, including a parametrization of C/O, to the OSIRIS data as well as photometric data points and lower resolution spectra.
Finally, we discuss the results in \Secref{sec:discussion} and conclude in \Secref{sec:conclusion}.

\section{Observations}
\label{sec:obs}

OSIRIS is an infrared integral field spectrograph on the W. M. Keck I telescope with moderate spectral resolution ($R=\lambda/\diff\lambda\sim4000$; \citealt{Larkin2006,Mieda2014,Boehle2016,Lockhart2019}). 
The smallest platescale ($20\,\mathrm{mas/pix}$; $\sim 1.3\times0.3''$ field of view) was used for all but the 2018 observations ($35\,\mathrm{mas/pix}$). The spectral cubes were reduced using the OSIRIS data reduction pipeline \citep{Krabbe2004,Lyke2017}. 

Observing logs for the science and the telluric standards can be found in \Appref{app:apposirisobs}. The wavelength, transmission, and point spread function (PSF) calibration are described in detail in \cite{Ruffio2019RV}. The wavelength solution was calibrated using the $\mathrm{OH}^-$ emission lines from the sky observations.
Standard star observations are used to calibrate a super-sampled PSF and the transmission spectrum, which are subsequently used in the forward model of the data. The adaptive optics correction was sometimes turned off to avoid detector persistence. In this case, or when the PSF was truncated, the spectrum was only used to calibrate the transmission. In 2020, we changed to observing strategy and started only using the host star for calibration. Persistence issues were avoided by only integrating the star at the edge of the field of view and avoiding the center where the planet would be located. 

Using the method described below, the planets were detected in a number of exposures totalling $12.7$ hours out of $14.4$ hours integrated on HR 8799b in K-band ($1.97-2.38\mu\mathrm{m}$) and $0.8$ out of $6.8$ hours in H-band ($1.47-1.80\mu\mathrm{m}$), $7.7$ out of $8.3$ hours in K-band and $2.8$ out of $9.5$ hours in H-band for HR 8799c, and finally $9.2$ out of $20.0$ hours in K-band for HR 8799 d. We can only use exposures in which the planet is detected for any subsequent analysis because OSIRIS pointing and dithering precision does not allow a blind localization of the planet. While a subset of this data has already been published \citep{Barman2011,Konopacky2013}, the H-band data on HR 8799c and the K-band data for HR 8799d are presented here for the first time.

\section{Data modeling}
\label{sec:model}

We use the method developed in \citet{Ruffio2019RV} and improved in \citet{wilcomb2020}. The general concept is to define a linear model of the data that includes the planet signal, the starlight, and a set a principal components of the residuals to include possible systematics. The linear model can be easily optimized and analytically marginalized to quickly calculate the posterior probability density of any other non linear parameters such as RV, temperature, surface gravity, and C/O. For completeness, we redefine the model below, such that
\begin{equation}
     {\bm d} = {\bm M}{\bm \phi} + {\bm n},
\end{equation}
where ${\bm d}$ is the data, ${\bm M}$ is the linear model (which is itself a function of non-linear parameters, e.g. RV, C/O, etc.), ${\bm \phi}$ are the linear parameters, and ${\bm n}$ is the noise. Each term is described in the following sub-sections in more detail.

\subsection{The data}
The data vector ${\bm d}$ contains the pixels of a postage-stamp sized spectral cube centered at the location of interest with $5\times5$ spaxels in the spatial dimensions and $N_\lambda = 1665$ pixels in the wavelength direction in K-band. The spectrum of each spaxel is high-pass filtered using a periodicity cutoff of $83\,\mathrm{pix}$ (i.e., $21\,\mathrm{nm}$) in Fourier space \citep[See][]{Ruffio2019RV}. Bad pixels are simply removed from the data vector; they are identified by the OSIRIS data reduction pipeline and additional sigma clipping.

\subsection{The noise}
We assume uncorrelated centered Gaussian noise, which means that the random vector ${\bm n}$ has zero mean and it is described by a diagonal covariance matrix $\Sigma = s^2 \Sigma_0$. $\Sigma_0$ is a fixed diagonal matrix containing a quantity proportional to the error at each pixel, and $s$ a scaling factor that is fitted for. In the following, inferred parameters are marginalized over the scaling factor. The diagonal elements of $\Sigma_0$ are defined as the square root of the continuum of the spectrum at each location. The values are clipped at half their median to avoid vanishing errors that would bias the result.

\subsection{The model matrix}
The model of the signal includes a representation of the companion, the starlight at the location of the companion, and a set of principal components used to account for the imperfection of the telluric model or other systematics. The latter is a way to account for the time variability of the tellurics when using limited standard star observations.
It is defined by the $N_d \times N_{\bm \phi}$ matrix ${\bm M}$, where $N_d$ is the size of ${\bm d}$ and $N_{\bm \phi}$ is the number of linear parameters,
\begin{equation}
    {\bm M} = [{\bm c}_{\mathrm{0,planet}},{\bm c}_1,\dots,{\bm c}_{25},{\bm r}_{\mathrm{pc}1},\dots,{\bm r}_{\mathrm{pc}K}].
    \label{eq:modelwithPCA}
\end{equation}
The first column vector ${\bm c}_{\mathrm{0,planet}}$ represents the planet signal. It is defined as the instrument PSF scaled by an atmospheric model of the companion and multiplied by the transmission of the atmosphere and the instrument. 
The next 25 column vectors, $\{{\bm c}_{i}\}$ with $1\le i \le 25$, represent the starlight spectrum at the $5\times5$ spaxels centered at the location of the companion. The starlight spectra need to be scaled to match the intensity and chromaticity of the speckle field at each location. With $i$ being the spaxel index, the starlight is defined as
\begin{equation}
{\bm c}_i={\bm d}_{i,L} \frac{(\mathcal{T}\mathcal{S}_{\mathrm{star}})_H}{(\mathcal{T}\mathcal{S}_{\mathrm{star}})_L},
\end{equation}
where ${\bm d}_{i,L}$ is the low-pass filtered spectrum of the data for the spaxel $i$, $\mathcal{T}$ is the transmission of the atmosphere and the instrument, and $\mathcal{S}_{\mathrm{star}}$ is the model spectrum of the star from the Phoenix library \citep{Husser2013}. The subscript $H$ and $L$ respectively designate the high-pass and low-pass filtered version of the spectrum.

The last $K$ column vectors $\{{\bm r}_{\mathrm{pc}i}\}$ are defined from the principal components of the residuals of a preliminary fit following the recipe below \citep{wilcomb2020}.\newline
The pre-processing steps include:
\begin{enumerate}
\item At each location in the field of view, we fit and subtract the model with the starlight components only.
\item Then we extract a spectrum from the $5\times5\times N_\lambda$ data cube of residuals by fitting the PSF at each wavelength.
\item The spectrum is normalized to the local continuum by dividing it by the low-pass filtered data at the same location.
\item The principal components are then calculated from the resulting normalized spectra over the entire field of view, but excluding the region within 5 spaxels of the real companion.
\end{enumerate}
For any subsequent modeling of the data and at any location of interest:
\begin{enumerate}
\item The principal components are scaled back to the local continuum.
\item Then, the 1D principal component spectra are applied to the 3D PSF model resulting in $K$ column vectors $({\bm r}_{\mathrm{pc}1},\dots,{\bm r}_{\mathrm{pc}K})$ matching the dimensions of the data ${\bm d}$. 
\end{enumerate}
This shows the flexibility of the linear model that we defined previously. Indeed, the model can be adapted or improved at will without changing the general framework.
One advantage of this approach, compared to directly subtracting the projection of the data on a set of principal components \citep{Hoeijmakers2018,PetitditdelaRoche2018,WangJi2018}, is to avoid the over subtraction of the planet signal. For example, over-subtraction at moderate resolution could occur when telluric lines that appear in the principal components could subtract real molecular features from the planet, hence biasing subsequent abundance estimates. In the following, we use $K=10$ principal components as in \citet{wilcomb2020}.
The forward model remains sensitive to the shape of the continuum of the spectrum despite the fact that the spectra are high-pass filtered. Indeed, the shape of the continuum remains encoded in the depth of the different spectral lines in the atmospheric model and the tellurics.

\subsection{The parameters}
The parameters of interest are the radial velocity of the companion, and the atmospheric parameters: temperature ($T$), surface gravity ($\mathrm{log}(g)$), and the ratio of carbon abundance to oxygen abundance (C/O). These are non-linear parameters, grouped in the symbol ${\bm \psi}$, that define the companion component ${\bm c}_{\mathrm{0,planet}}$ in the matrix ${\bm M}$. Their estimation and the calculation of the posteriors are discussed in \Secref{sec:inference}. For simplicity, the atmospheric model is fixed when trying to detect the location of the planet and when estimating its radial velocity.

Then, there are 1+25+K linear parameters in ${\bm \phi}$, which correspond to the amplitude of the companion signal, the amplitude of the starlight at each of the $5\times5$ spaxels centered at the position of the companion, and the amplitude of the K principal components. The last parameter is the scaling factor, $s$, of the noise. The linear parameters, as well as the noise scaling factor, are all nuisance parameters that will be marginalized over when estimating the non-linear parameters.

\subsection{The Likelihood function}
The likelihood function is defined as,
\begin{equation}
    \mathcal{L}({\bm \psi},{\bm \phi},s^2) =\frac{1}{\sqrt{(2\pi)^{N_d}\vert{\bm \Sigma}_0\vert s^{2N_d}}} \exp\left\lbrace-\frac{1}{2s^2}({\bm d}-{\bm M}_{{\bm \psi}}{\bm \phi})^{\top}{\bm \Sigma}_0^{-1}({\bm d}-{\bm M}_{{\bm \psi}}{\bm \phi})\right\rbrace.
\end{equation}
We also define $\chi^2=\chi_0^2/s^2$, with
\begin{equation}
    \chi_0^2 = ({\bm d}-{\bm M}{\bm \phi})^{\top}{\bm \Sigma}_0^{-1}({\bm d}-{\bm M}{\bm \phi}).
\end{equation}

This likelihood will be used to
\begin{enumerate}
\item detect the planet in the raw spectral cubes, 
\item directly detect specific molecules in their atmospheres, 
\item and estimate the RV and the C/O ratio of the three planets.
\end{enumerate}
We briefly describe the methods in the following sections (See Appendix of \citealt{Ruffio2019RV} for complete mathematical derivations).

\subsection{Planet and molecule detection}
\label{sec:plmoldetec}

OSIRIS is an integral field spectrograph and the lack of precise pointing requires the detection of the companion in individual exposures before they can be combined.
We choose an approach comparable to the cross correlation function (CCF) for the detection of the planet itself as well as molecules in its atmosphere.

As discussed in \citet{Ruffio2019RV}, the commonly used CCF is a simplified maximum likelihood approach in which one estimates the signal to noise ratio (S/N) of the amplitude of the molecular spectral template in the data. We can produce a similar output using the data model defined previously by extracting the amplitude of the companion component, which is the first element of the ${\bm \phi}$ vector, and its associated error.
Using the model defined in this \Secref{sec:model}, the S/N of the companion is defined as,
\begin{equation}
\mathrm{S/N} = \frac{ \tilde{{\bm \phi}}[0] }{ \sqrt{ \mathrm{cov}(\tilde{{\bm \phi}})[0,0]}}.
\label{eq:generalizedSNR}
\end{equation}
The tilde signifies that the linear parameters ${\bm \phi}$ is set to its optimal value with
\begin{equation}
\tilde{{\bm \phi}} = ({\bm M}^{\top}{\bm \Sigma_0}^{-1}{\bm M})^{-1}{\bm M}^{\top}{\bm \Sigma_0}^{-1}{\bm d},
\label{eq:pseudoinv}
\end{equation}
although it is numerically faster and more stable to estimate $\tilde{{\bm \phi}}$ from a linear least square solver that does not require matrix inversion.
The covariance matrix is defined as
\begin{equation}
    \mathrm{cov}(\tilde{{\bm \phi}}) = \tilde{s}^2 ({\bm M}^{\top}{\bm \Sigma_0}^{-1}{\bm M})^{-1},
    \label{eq:cov}
\end{equation}
and
\begin{equation}
\tilde{s}^2 = \frac{1}{N_d} \chi^2_{0,{\bm \phi}=\tilde{{\bm \phi}}}.
\end{equation}

The S/N from \autoref{eq:generalizedSNR} is computed at each spaxel location in the field of view, and for a set of RV shifts of the companion spectrum resulting in a S/N cube. 
The S/N defined in \autoref{eq:generalizedSNR} might not always be accurate due to the simplifying assumptions in the modeling the noise.
As a mitigation strategy, we subtract the median of the S/N cube in the RV dimension and then in the spatial direction. In a similar fashion, we divide the S/N by the sample standard deviation computed at each location using the spectral direction. Then, individual S/N cubes for each exposures are registered and added together into a final cube for each planet. We finally ensure that the final S/N cube as a whole has a unit standard deviation after masking the locations within 5 pixels ($\sim0.1''$) of the companion. 
The standard deviation calculations are performed after masking the spaxels at the locations of the planet. 

For the detection of the planet, we use the best fit atmospheric models described in \citet{Barman2015,Konopacky2013} as spectral templates. The same template is used for HR 8799d as for HR 8799c, which is justified by the similar atmospheric parameters retrieved in \Secref{sec:atmcharacOSIRIS}.
The S/N maps at the expected RV of the star for each HR 8799d exposure are shown in \Appref{app:apposirisobs} for 10 principal components. In \Appref{app:noiseanalysis}, we show that the forward model is able to detect HR 8799d while it is $\sim 30$ times fainter than the starlight at the same location in only 10 minutes. The residuals are consistent with the photon noise limit and the planet S/N per bin is $\sim 2$.

Once the planet's location is known, molecules are detected using the molecular templates described in \citet{Konopacky2013,Barman2015}. The final S/N curves as a function of the RV shift of the molecular spectra are presented in \Secref{sec:ccf}.

\subsection{Inference}
\label{sec:inference}
In this work, we are defining two inference problems: first, the estimation of the RV of the companion assuming a fixed atmospheric model, and secondly, the estimation of the atmospheric parameters while marginalizing over the RV. The parameters of the atmospheric models are: effective temperature ($T_{\mathrm{eff}}$), surface gravity ($\mathrm{log}(g)$; with $g$ in $\mathrm{cm} \mathrm{s}^{-2}$), and carbon to oxygen ratio (C/O). In both cases, we also marginalize over the linear parameters ${\bm \phi}$ and the noise scaling parameter $s$; this can be done analytically \citep{Ruffio2019RV}.
Although it is technically possible, we do not marginalize over the location of the planet and assumed it fixed once the planet is detected.
If the parameters of interest are represented by a vector ${\bm \psi}$, where ${\bm \psi}=[\mathrm{RV}]$ or ${\bm \psi} = [T_{\mathrm{eff}}, \mathrm{log}(g), \mathrm{C/O}, \mathrm{RV}]$, its posterior can be written as
\begin{equation}
    \mathcal{P}({\bm \psi} \vert {\bm d})
    \propto  \frac{\mathcal{P}( {\bm \psi})}{\sqrt{\vert{\bm \Sigma}_0\vert \times \left\vert{\bm M}_{{\bm \psi}}^{\top}{\bm \Sigma}_0^{-1}{\bm M}_{{\bm \psi}}\right\vert}} \left(\frac{1}{\chi^2_{0,{\bm \phi}=\tilde{{\bm \phi}},{\bm \psi}}}\right)^{\frac{N_{\bm d}-N_{\bm \phi}+1}{2}},
    \label{eq:appRVpost}
\end{equation}
where $\mathcal{P}( {\bm \psi})$ is ${\bm \psi}$'s prior. The model and the number of parameters are such that it is computationally tractable to calculate the posterior on a discrete grid of the parameter space. $\mathcal{P}( {\bm \psi})$ is assumed to be flat over the finite extent of the grid. Uniform improper priors are also assumed for the linear parameters ${\bm \phi}$.
We have assumed the inverse squared prior for the scaling parameter $s$ which corresponds to Jeffrey's prior when jointly estimating the mean and standard deviation of a Gaussian distribution.
In the following, we express estimated parameters as the mode of the posterior distribution, and the error bars as the $68\%$ confidence interval that includes the highest values of the probability distribution function.

\section{Detection of molecules}
\label{sec:ccf}

\begin{figure}[bth]
\gridline{\fig{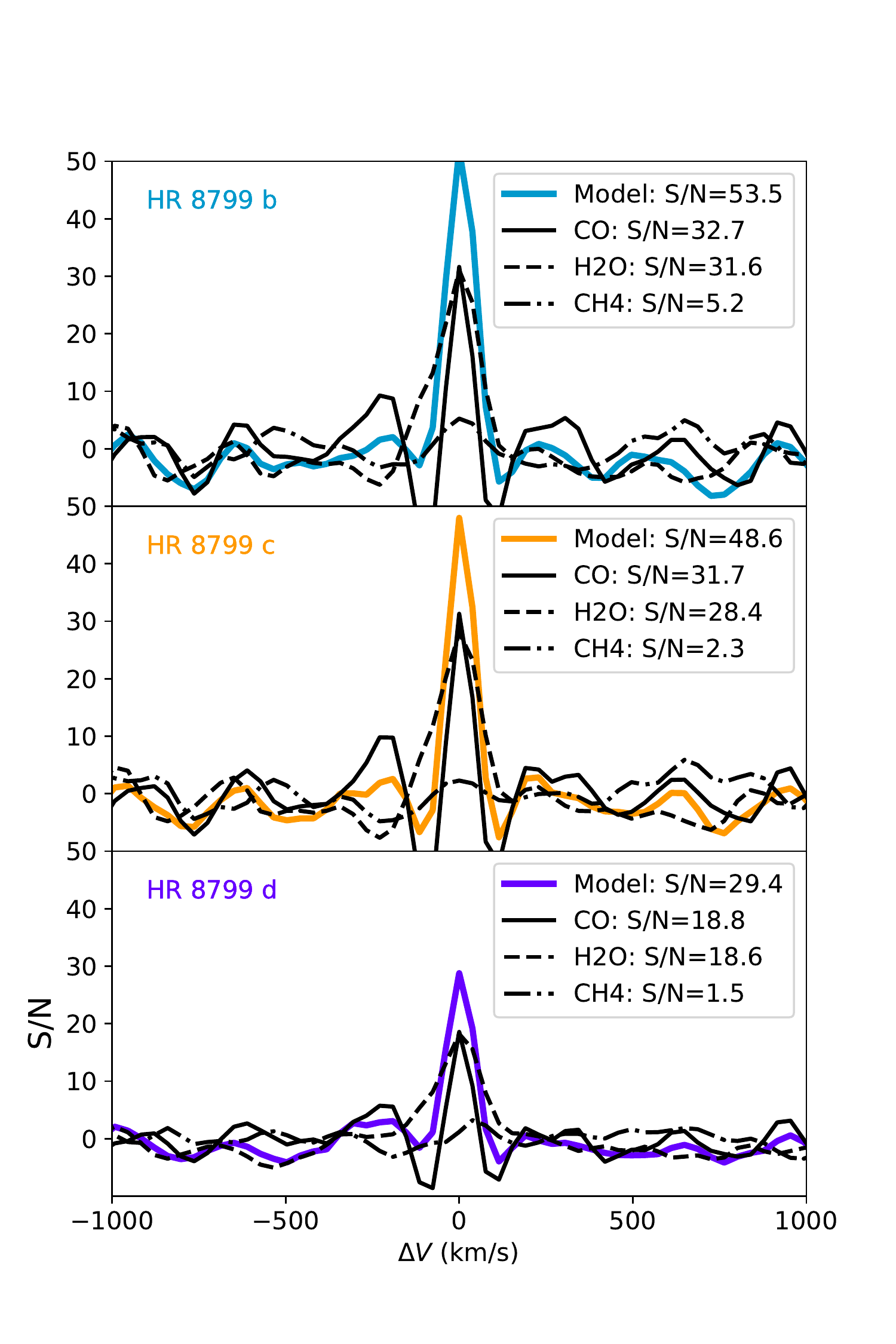}{0.49\textwidth}{(a) K band}
		\fig{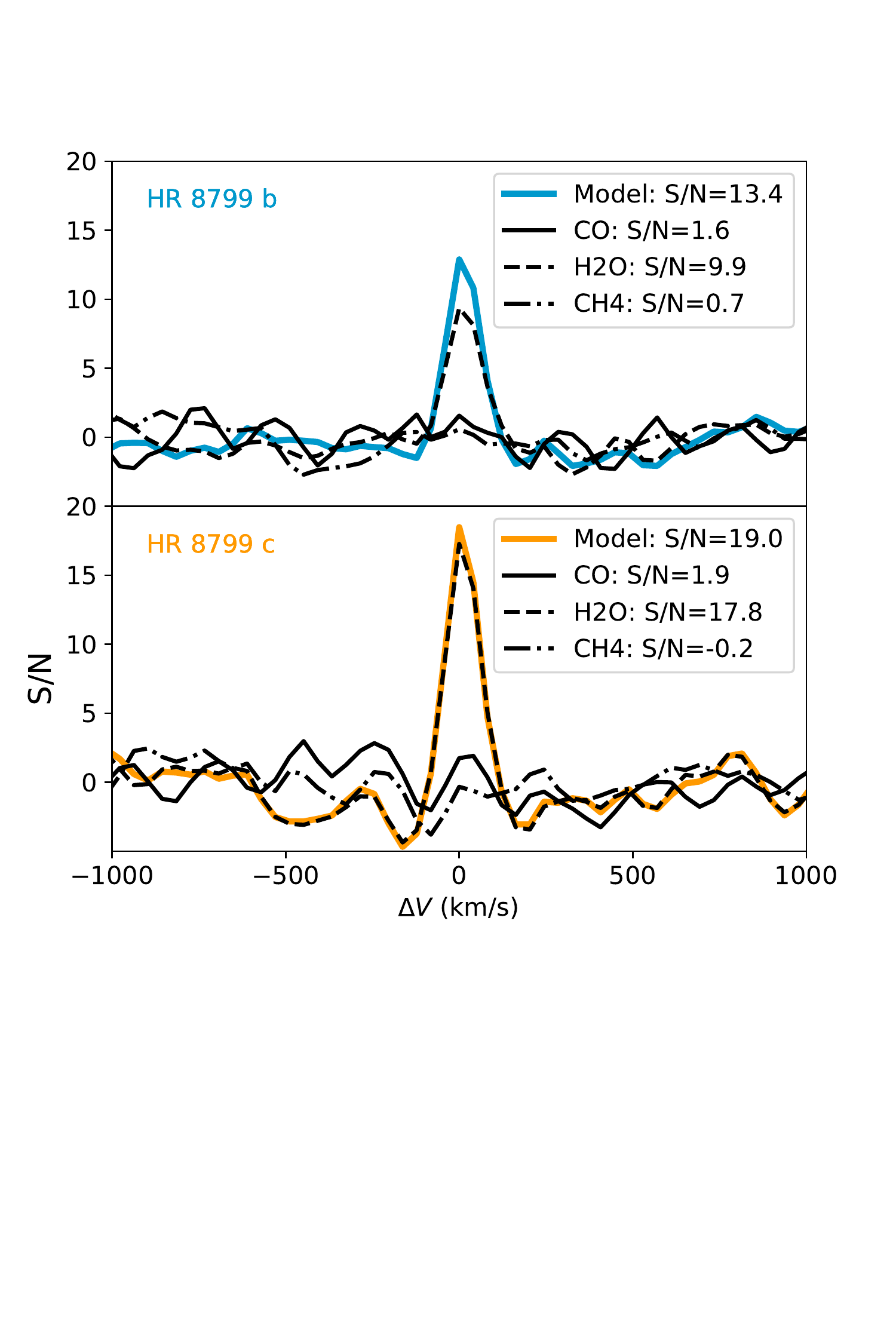}{0.49\textwidth}{(b) H band}
          }
\caption{Signal to noise ratio (S/N) of the detection of water (H\textsubscript{2}O), carbon monoxide (CO), and methane (CH\textsubscript{4}) in HR 8799b, c, and d as a function of the radial velocity (RV) shift of the model spectrum. Observations in (a) K band and (b) H band are shown separately. The curve labeled ``model'' includes a best fit atmospheric model from \citet{Barman2015} and \citet{Konopacky2013} for HR 8799b and c respectively. The HR 8799c model is used for HR 8799d as the two planets are thought to be sufficiently similar. The noise is empirically estimated from the data as described in \autoref{sec:plmoldetec} and masking the locations within 5 pixels ($\sim0.1''$) of the companion.}
\label{fig:CCFsummary}
\end{figure}

\begin{figure}
  \centering
  \includegraphics[width=1\linewidth]{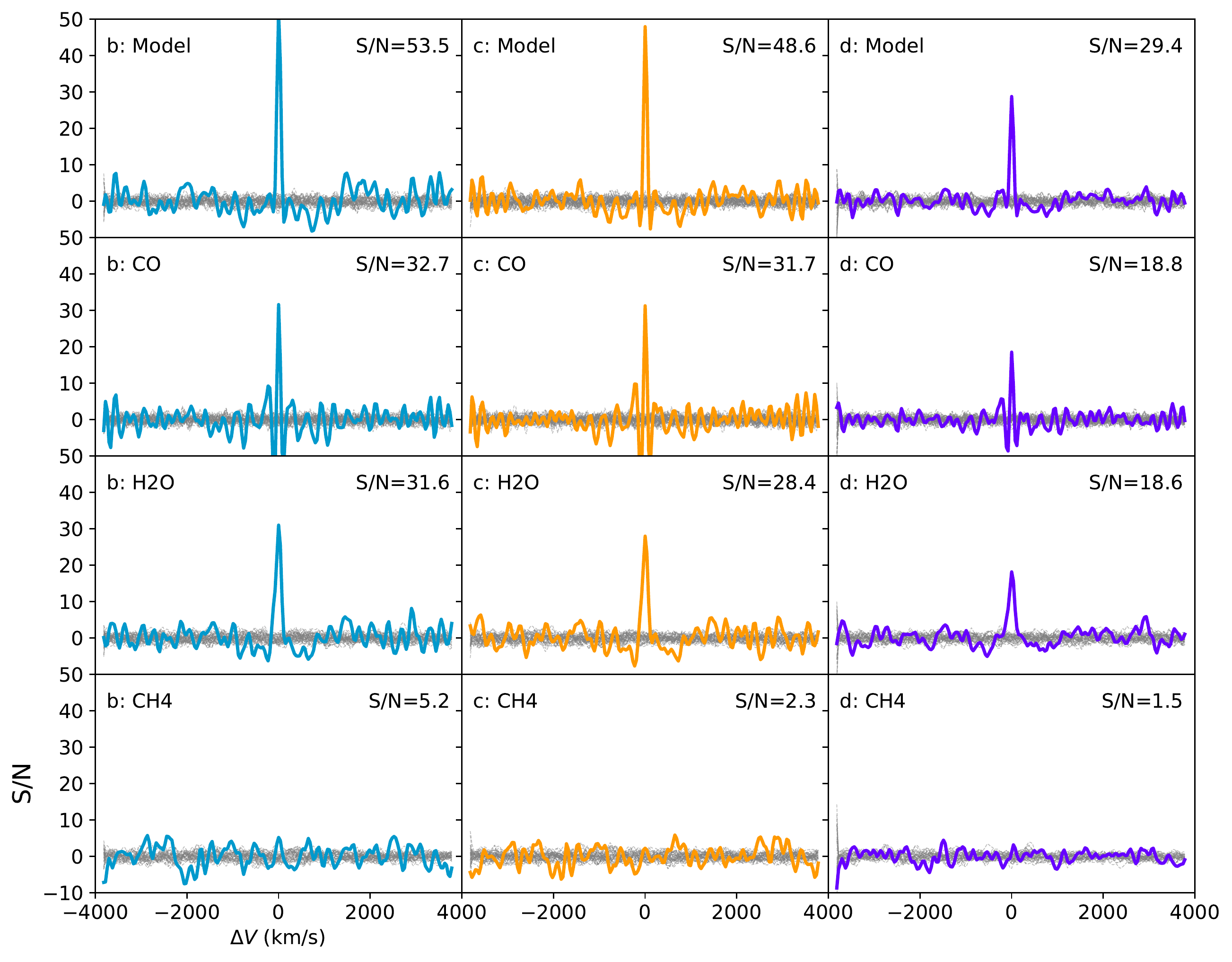}
  \caption{Similar to \autoref{fig:CCFsummary} for K band, but each curve is shown separately and compared to 50 samples of the noise (grey curves).}
  \label{fig:CCF_Kbb}
\end{figure}

\begin{figure}
  \centering
  \includegraphics[width=0.66\linewidth]{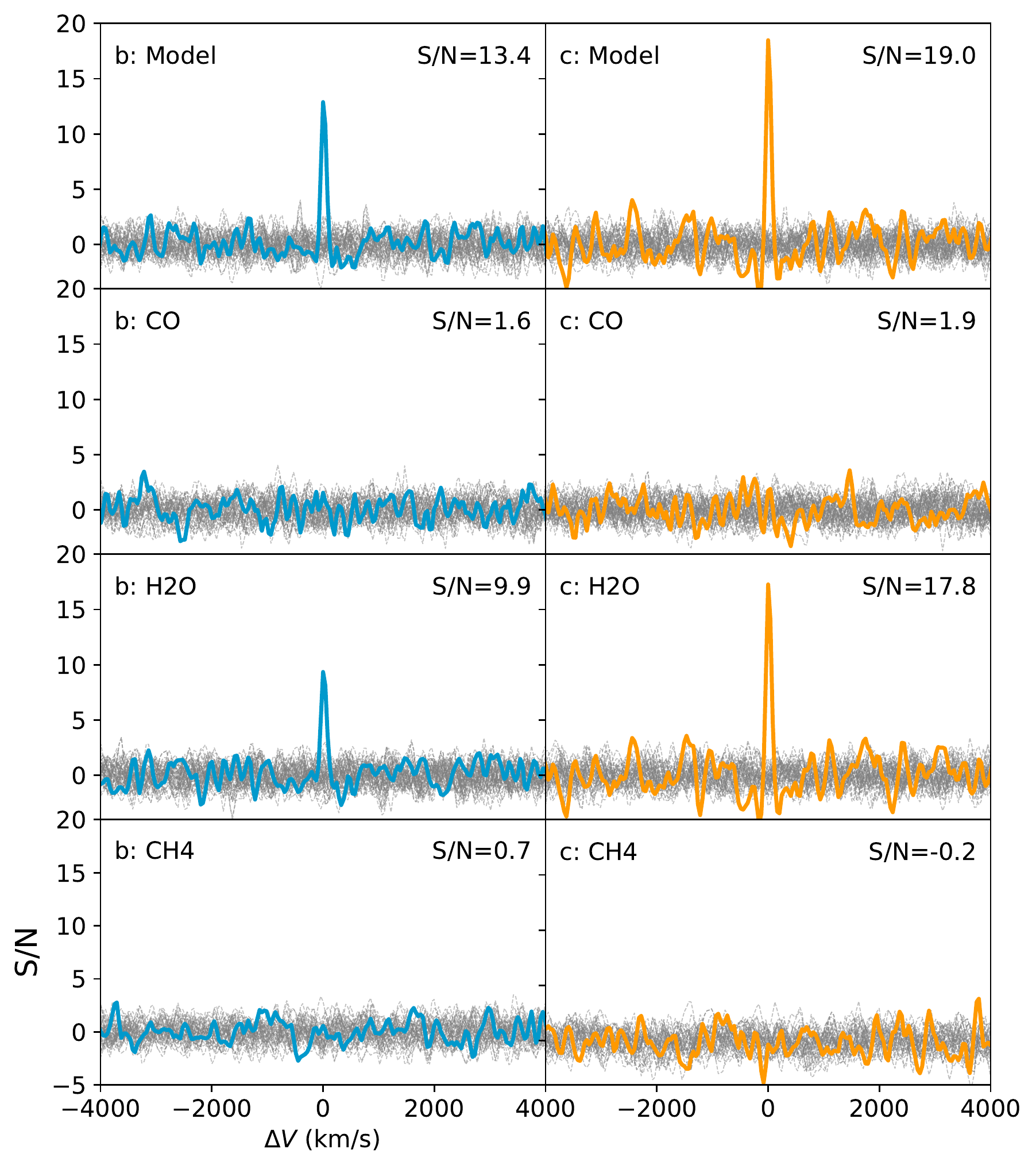}
  \caption{Similar to \autoref{fig:CCFsummary} for H band, but each curve is shown separately and compared to 50 samples of the noise (grey curves).}
  \label{fig:CCF_Hbb}
\end{figure}

\begin{figure}
  \centering
  \includegraphics[width=1\linewidth]{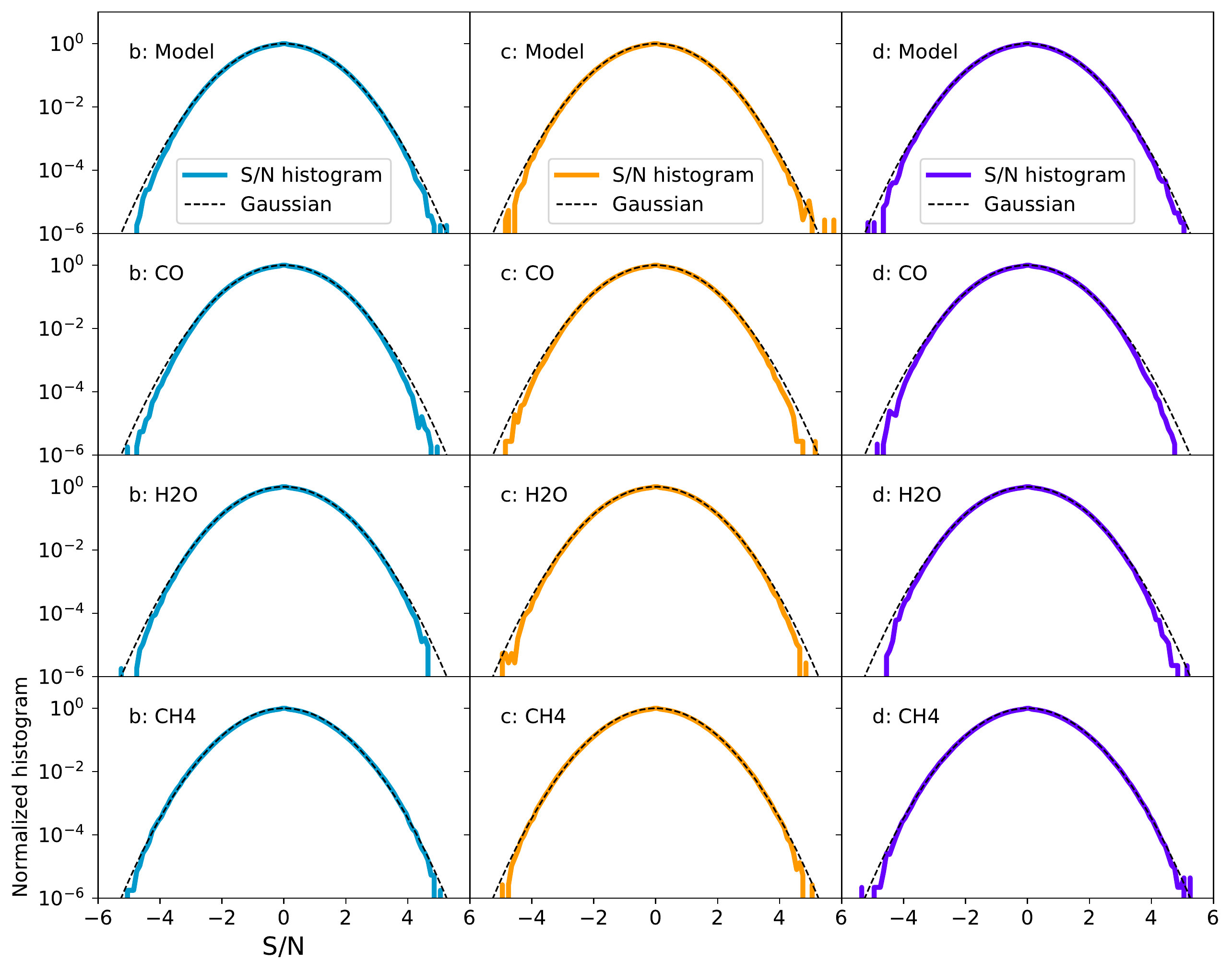}
  \caption{Histogram of the noise samples for each panel of \autoref{fig:CCF_Kbb}. The noise distribution is compared to a Gaussian normalized to a unit peak.}
  \label{fig:CCFhisto_Kbb}
\end{figure}

We use the maximum likelihood-based method described in \Secref{sec:plmoldetec} to compute the S/N of a spectral template as a function of its RV shift, which is conceptually comparable to a CCF. For each planet, we use four different templates: a full atmospheric model, which is both used for the detection of the planet and its RV estimation, and templates for H\textsubscript{2}O, CO, and CH\textsubscript{4}. Early results of this work were presented in \citet{Ruffio2019thesis}. The current analysis is improved in several ways: we have gathered significantly more data on HR 8799d, the logarithm of spectral energy distribution of the molecular models were mistakenly used in \citet{Ruffio2019thesis} resulting in slightly lower S/N, and principal components are now included in the model.
The S/N curves are summarized in \autoref{fig:CCFsummary}. The detections are compared to samples of the noise in \autoref{fig:CCF_Kbb} and \autoref{fig:CCF_Hbb} for H band and K band respectively. The histograms of the noise are shown in \autoref{fig:CCFhisto_Kbb}.
The detection S/Ns in H band and K band are given as a function of the number of principal components in \autoref{tab:CCFsnrKbb} and \autoref{tab:CCFsnrHbb} demonstrating the gains brought by the improved forward model. 

Here, we estimate the noise from the surrounding speckles so the S/N does not include sources of noise that are planet-specific such as additional photon noise or additional telluric residuals. This remains valid as long as the planet signal is negligible compared to starlight (i.e., speckles), which is likely true for HR 8799c and d, but could start to break at the large separation of HR 8799b.
Confusion, or crosstalk, between templates can also occur due to overlapping spectral lines and introduce systematics. The crosstalk between H\textsubscript{2}O and CO is minimal compared to the strength of their respective detection, and therefore can be neglected. In other words, the detections of H\textsubscript{2}O and CO are far too strong to be explained by a crosstalk with another molecule. An illustration of all the crosstalks between templates can be found in Figure D.11 in \citet{Ruffio2019thesis}.
The situation is different for methane, there is no clear detection for any of the three planets, and the amplitude of the signal (S/N$\sim5$) is consistent with the amplitude of the cross-talk between CH\textsubscript{4} and strong H\textsubscript{2}O and CO detections. 
In this case, the large S/N value for CH\textsubscript{4} and HR 8799b only signifies that the null-hypothesis is rejected (ie, there is indeed a planet at the considered location). In other words, the methane signal appears too weak to be confidently disentangled from the crosstalk and no detection can be claimed despite the apparent strength of the signal at the expected radial velocity of the planet.

CCF-like analyses are interesting because they provide an unambiguous and straightforward way to detect the molecules with the strongest signatures in the atmosphere of exoplanets. The estimation of molecular abundances or enrichment ratios will be addressed in \Secref{sec:atmcharac}.


\begin{deluxetable}{@{\extracolsep{4pt}}c|lrr|lrr|lrr} 
\tablewidth{0pc}
\tabletypesize{\scriptsize}
\tablecaption{K-band detection S/N of molecules as a function of the number of principal components used in the forward model. \label{tab:CCFsnrKbb}} 
\tablehead{ 
\colhead{}&
\multicolumn{3}{c}{HR 8799b}&
\multicolumn{3}{c}{HR 8799c}&
\multicolumn{3}{c}{HR 8799 d}\\
\cline{2-4} \cline{5-7} \cline{8-10}
  \colhead{\# PCs} & 
  \colhead{0} & \colhead{1} & \colhead{10} &
  \colhead{0} & \colhead{1} & \colhead{10} &
  \colhead{0} & \colhead{1} & \colhead{10} 
}
\startdata 
Model & 51.7 & 51.4 & 53.5 &
      42.5 & 43.4 & 48.6 &
      19.7 & 21.2 & 29.4  \\
CO    &  32 & 32 & 32.7 & 
       27.5 & 28 & 31.7 &
       12.2 & 13.2 & 18.8 \\
 $\mathrm{H}_2\mathrm{O}$ & 29 & 30 & 31.6 &
      23.7 & 24.5 & 28.4 &
      12.5 & 13.1 & 18.6  \\
 $\mathrm{CH}_4$   & 5.5 & 5.1 & 5.2 &
       1.7 & 2.1 & 2.3 &
       2.3 & 1.9 & 1.5
\enddata
\end{deluxetable}

\begin{deluxetable}{@{\extracolsep{4pt}}c|lrr|lrr} 
\tablewidth{0pc}
\tabletypesize{\scriptsize}
\tablecaption{Same as \autoref{tab:CCFsnrKbb} for H-band. \label{tab:CCFsnrHbb}} 
\tablehead{ 
\colhead{}&
\multicolumn{3}{c}{HR 8799b}&
\multicolumn{3}{c}{HR 8799c}\\
\cline{2-4} \cline{5-7}
  \colhead{\# PCs} & 
  \colhead{0} & \colhead{1} & \colhead{10} &
  \colhead{0} & \colhead{1} & \colhead{10} 
}
\startdata 
Model & 12.2 & 13.2 &13.4 &
      15.3 & 15.8 & 19  \\
CO    &  1.4 & 1.4 & 1.6 & 
       1.9 & 1.3 & 1.2  \\
 $\mathrm{H}_2\mathrm{O}$ & 9.0 & 9.6 & 9.9 &
      14.5 & 14.9 & 17.8   \\
 $\mathrm{CH}_4$   & 0.4 & 0.7 & 0.7 &
       -0.3 & -0.4 & -0.2
\enddata
\end{deluxetable}

\section{Radial velocity}
\label{sec:RV}

In this section, we present the first RV measurements of HR 8799d, and we revisit the values presented in \citet{Ruffio2019RV} for HR 8799b and c using an improved data model \citep{wilcomb2020}. For each exposure in which the planet is detected, we compute the RV posterior according to \Secref{sec:inference} with an RV sampling from $-80\,\mathrm{km}.\mathrm{s}^{-1}$ to $+80\,\mathrm{km}.\mathrm{s}^{-1}$ with a step size of $0.4\,\mathrm{km}.\mathrm{s}^{-1}$. This range of RVs corresponds to an absolute spectral shift of 2 spectral bins in either direction for OSIRIS. The coarse sampling of $0.4\,\mathrm{km}.\mathrm{s}^{-1}$ is satisfactory for single exposure uncertainties but it is later interpolated on a finer grid when combining exposures together. We use the same spectral templates as the one used for detecting the planets.

\begin{figure}
  \centering
  \includegraphics[width=1\linewidth]{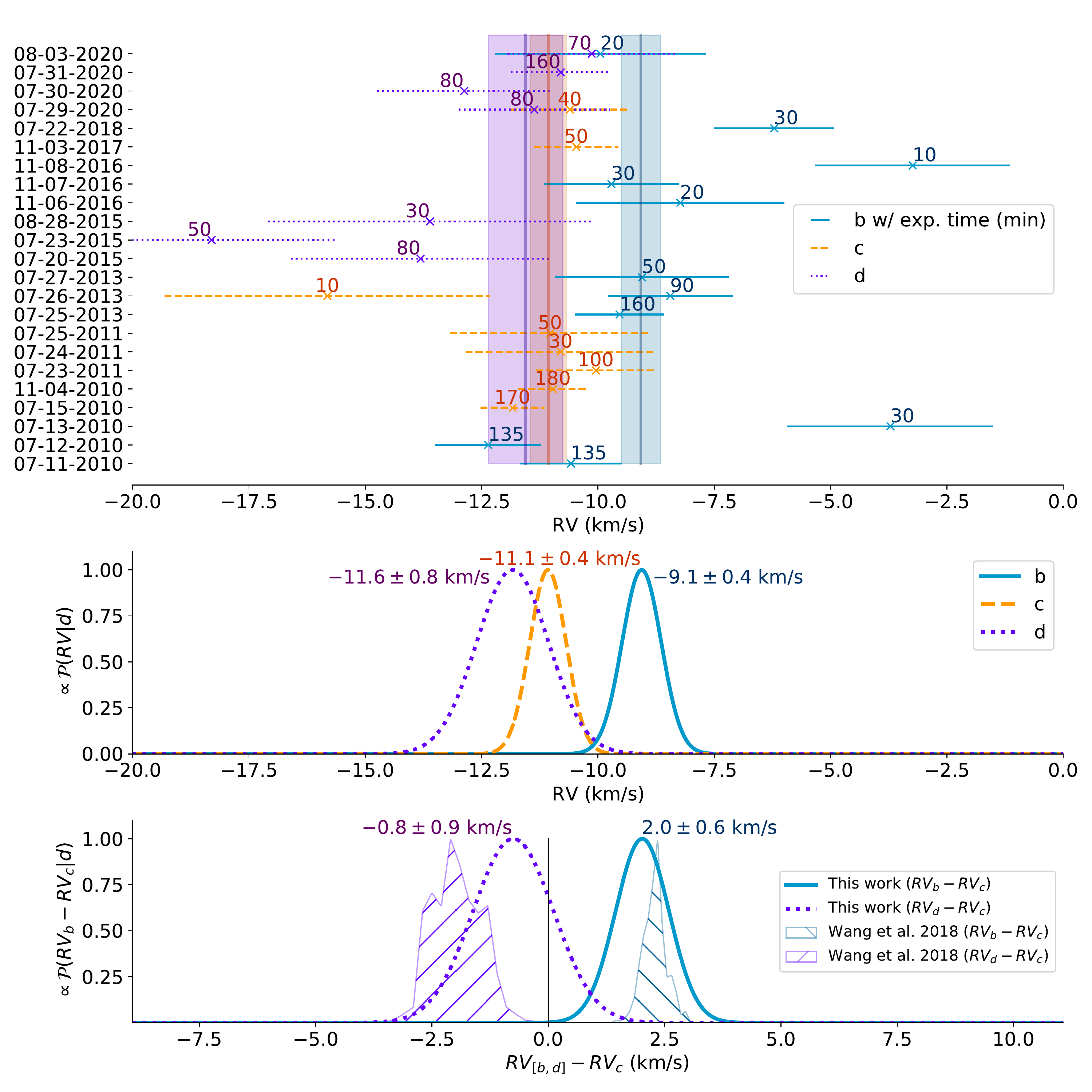}
  \caption{Radial velocity (RV) of the HR 8799b, c, and d using Keck/OSIRIS ($R\approx4,000$) after barycentric correction. The top panel features the nightly RV measurements. The total exposure time used for each epoch is indicated next to each data point in minutes. The vertical line and associated error bar represent the combined RV estimate for each planet assuming it is constant. The posteriors of the final RVs are shown in the middle panel. The bottom panel includes the differential RV posteriors for HR 8799b-c, and HR 8799 d-c respectively. The measurements are compared to the predictions (hatched posteriors) from astrometric measurements including assumptions of coplanarity and stability of system \citep{Wang2018}. Astrometry does not constrain the sign of the relative RV, but only the solutions with positive $RV_b - RV_c$ are shown here.}
  \label{fig:RV_bcd}
\end{figure}

\begin{figure}
  \centering
  \includegraphics[width=0.4\linewidth]{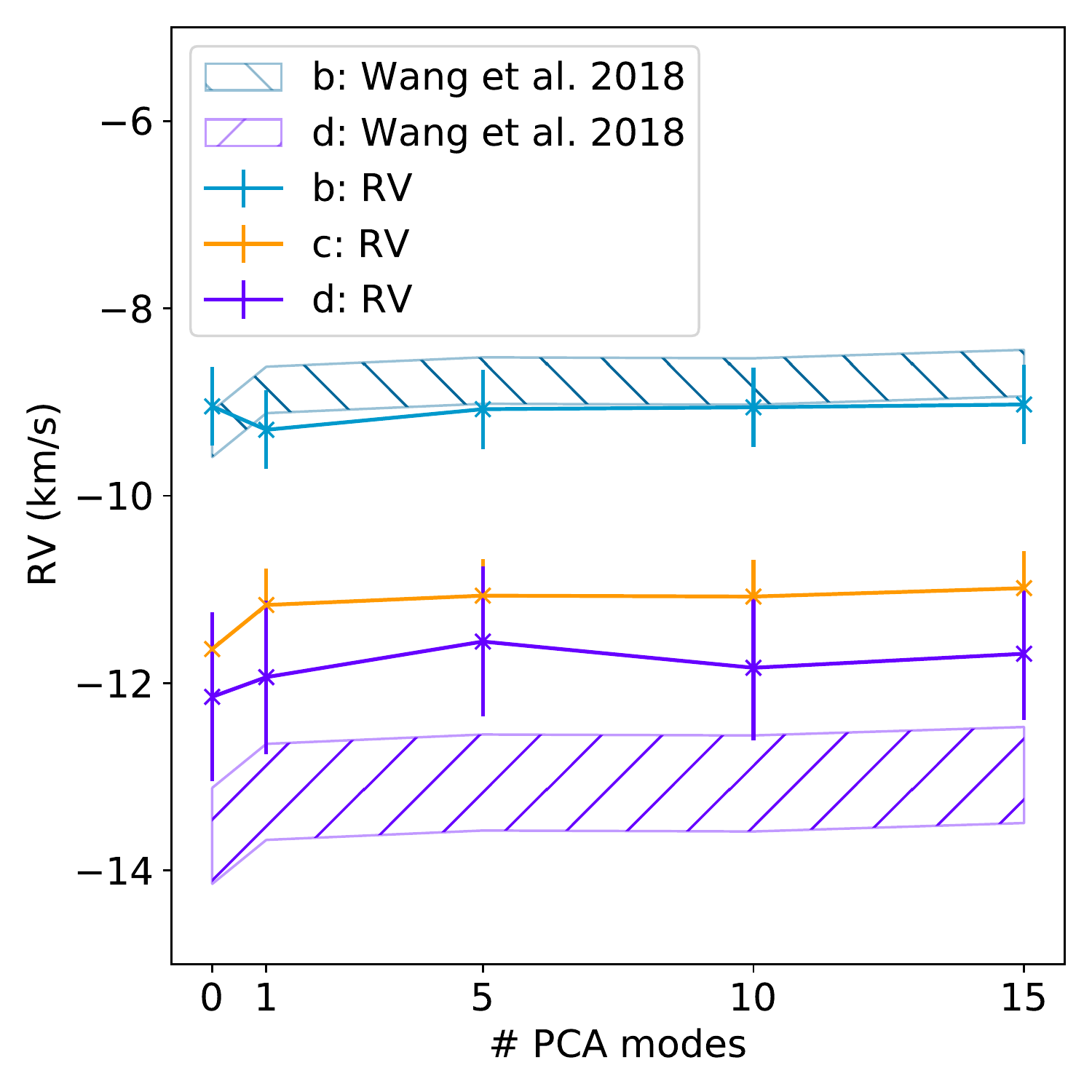}
  \caption{Combined radial velocities of HR 8799b, c, and d as a function of the number of principal components used in the forward model of the data. As a reference, we show the RV predictions for b and d (hatched contours) by adding the predicted differential RV ($RV_b - RV_c$ or $RV_d - RV_c$) calculated from visual orbits \citep{Wang2018} to the measured $RV_c$.}
  \label{fig:RV_bcd_PCA}
\end{figure}

\autoref{fig:RV_bcd} shows the nightly combined RVs of the three planets; they are also listed in \autoref{tab:RVepochs} with existing literature values. The updated RVs for HR 8799b and c are still consistent with \citet{Ruffio2019RV}. \autoref{fig:RV_bcd_PCA} shows the RV evolution as a function of the number of principal components used in the analysis.
The median planets' RV variations over ten years (2010-2020) are relatively small based on the visual orbits for HR 8799b and c: $\Delta RV_b = 0.1\,\mathrm{km}.\mathrm{s}^{-1}$ and $\Delta RV_c = 0.4\,\mathrm{km}.\mathrm{s}^{-1}$. The difference between 2015 and 2020 for HR 8799d is $\Delta RV_d = -0.4\,\mathrm{km}.\mathrm{s}^{-1}$. As a first analysis, we therefore compare the weighted mean RVs to the predictions from astrometry assuming coplanar and stable orbits \citep{Wang2018}. We achieve sub-kilometer per second precision on the RV of three planets. A full orbital analysis using individual nightly RVs is performed in \Secref{sec:orbits} assuming coplanar orbits.

\begin{table}
  \centering
\begin{tabular}{ |c|ccccc| } 
 \hline
  Planet & Date & MJD & RV (km/s) &  RV (km/s) & Time \\ 
   &  &  & (this work) &  \cite{Ruffio2019RV} & (min) \\ 
 \hline\hline
\multirow{11}{*}{b} 
& 2010-07-11 & $55388.55$ & $-10.6 \pm 1.1$ & $-10.2 \pm 1.2$ & 135 \\
& 2010-07-12 & $55389.54$ & $-12.4 \pm 1.1$ & $-10.1 \pm 1.1$ & 135 \\
& 2010-07-13 & $55390.62$ & $-3.7 \pm 2.2$ & $-4.0 \pm 2.4$ & 30 \\
& 2013-07-25 & $56498.55$ & $-9.5 \pm 1.0$ & $-8.9 \pm 1.0$ & 160 \\
& 2013-07-26 & $56499.50$ & $-8.4 \pm 1.3$ & $-9.4 \pm 1.4$ & 90 \\
& 2013-07-27 & $56500.51$ & $-9.1 \pm 1.9$ & $-6.5 \pm 2.3$ & 50 \\
& 2016-11-06 & $57698.32$ & $-8.2 \pm 2.2$ & $-10.8 \pm 2.2$ & 20 \\
& 2016-11-07 & $57699.32$ & $-9.7 \pm 1.5$ & $-11.8 \pm 1.5$ & 30 \\
& 2016-11-08 & $57700.34$ & $-3.2 \pm 2.1$ & & 10 \\
& 2018-07-22 & $58321.52$ & $-6.2 \pm 1.3$ & $-6.7 \pm 1.3$ & 30 \\
& 2020-08-03 & $59064.46$ & $-9.9 \pm 2.3$ & & 20 \\
 \hline
\multirow{8}{*}{c} 
& 2010-07-15 & $55392.55$ & $-11.8 \pm 0.7$ & $-11.8 \pm 0.7$ & 170 \\
& 2010-11-04 & $55504.31$ & $-11.0 \pm 0.7$ & $-11.5 \pm 0.8$ & 180 \\
& 2011-07-23 & $55765.54$ & $-10.0 \pm 1.3$ & $-10.8 \pm 1.3$ & 100 \\
& 2011-07-24 & $55766.56$ & $-10.8 \pm 2.1$ & & 30 \\
& 2011-07-25 & $55767.57$ & $-11.0 \pm 2.2$ & $-18.8 \pm 5.7$ & 50 \\
& 2013-07-26 & $56499.58$ & $-15.8 \pm 3.5$ & & 10 \\
& 2017-11-03 & $58060.25$ & $-10.5 \pm 0.9$ & & 50 \\
& 2020-07-29 & $59059.50$ & $-10.6 \pm 1.3$ & & 40 \\
 \hline
\multirow{7}{*}{d} 
& 2015-07-20 & $57223.51$ & $-13.8 \pm 2.8$ & & 80 \\
& 2015-07-23 & $57226.52$ & $-18.3 \pm 2.7$ & & 50 \\
& 2015-08-28 & $57262.54$ & $-13.6 \pm 3.5$ & & 30 \\
& 2020-07-29 & $59059.59$ & $-11.4 \pm 1.6$ & & 80 \\
& 2020-07-30 & $59060.56$ & $-12.9 \pm 1.9$ & & 80 \\
& 2020-07-31 & $59061.54$ & $-10.8 \pm 1.1$ & & 160 \\
& 2020-08-03 & $59064.55$ & $-10.1 \pm 1.8$ & & 70 \\
 \hline
   \multicolumn{2}{|c}{ } & &  &  \cite{WangJi2018} & \\ 
 \hline
c & 2016-2017 & & & $-8.9 \pm 2.5$ & 15.6 hr \\
HR 8799 & 2016-2017 & & & $-10.9 \pm 0.5$ & 15.6 hr \\
 \hline
\end{tabular}
\caption{Barycentric corrected radial velocities of HR 8799b, c, and d measured with Keck/OSIRIS. Litterature RVs for the planets and the host star are given for comparison.}
\label{tab:RVepochs}
\end{table}

\section{Orbital analysis}
\label{sec:orbits}

Following the same approach as in \citet{Ruffio2019RV}, we fit for the orbits of HR 8799b, c, and d assuming the three planets are coplanar. We use the same modified branch from the open source python package \texttt{orbitize!} \citep{blunt2019,Blunt2020} with the parallel tempered implementation of the affine-invariant ensemble sampler for Markov chain Monte Carlo \citep[ptemcee;][]{ForemanMackey2013,Vousden2016}. 
We compare the resulting orbital constraints when including both the relative astrometry and the planetary RV measurements to the astrometry only case. The data included in the fit are the RVs from \autoref{tab:RVepochs}, and the relative astrometry measurements from Keck/NIRC2 \citet{Konopacky2016}, and Gemini/GPI \citet{Wang2018}. The priors for each parameter are listed in \autoref{tab:RVresults}.
We use 16 temperatures, 512 walkers, and we only save the position of the walkers every 50 steps. For the RV case, we ran the chains for 400,000 steps ($\approx2\times10^{8}$ samples in total), which includes a 200,000 steps burn-in phase. The auto-correlation length is between 500 and 1500 steps depending on the parameters, which gives about 200 independent samples per walker and $\sim100,000$ in total. When not including the RV, the chains were 200,000 steps including a 100,000 steps burn-in phase.

\begin{figure}
  \centering
  \includegraphics[trim={0cm 0cm 0cm 0cm},clip,width=1\linewidth]{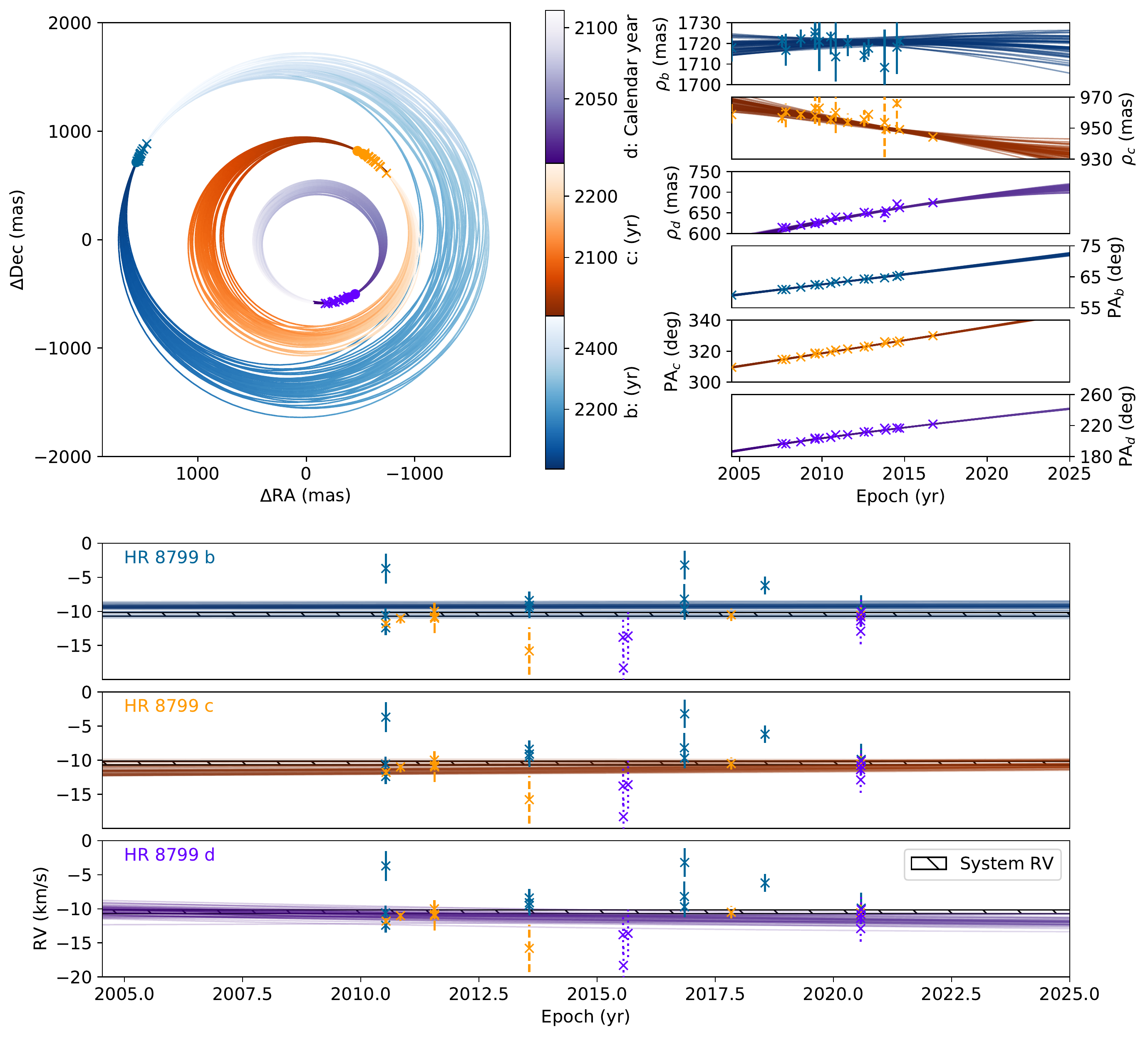}
  \caption{Orbits of HR 8799b (blue), c (orange), and d (pruple), randomly sampled from their posterior. The three planets are here assumed to be coplanar and the orbital fit include the radial velocity (RV) measurements of the planets. From top to bottom on the upper right, the panels show the separation of the planets (b to d), and their position angle. The error bars were converted from right ascension and declination to separation and position angle using a Monte Carlo approach. The bottom three panels are the RV measurement compared to the sample orbits (b to d). }
  \label{fig:orbits}
\end{figure}

\begin{figure}
  \centering
  \includegraphics[trim={0cm 0cm 0cm 0cm},clip,width=1\linewidth]{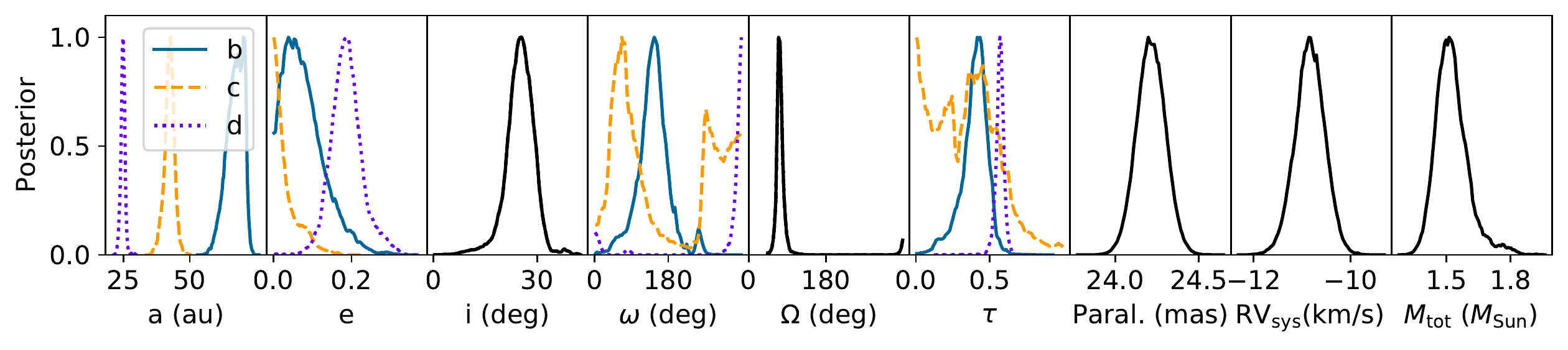}
  \caption{Marginalized posteriors of the orbital parameters from the joint coplanar fit of HR 8799b, c, and d using planetary RVs.}
  \label{fig:orbitspost}
\end{figure}

\begin{figure}[bth]
\gridline{\fig{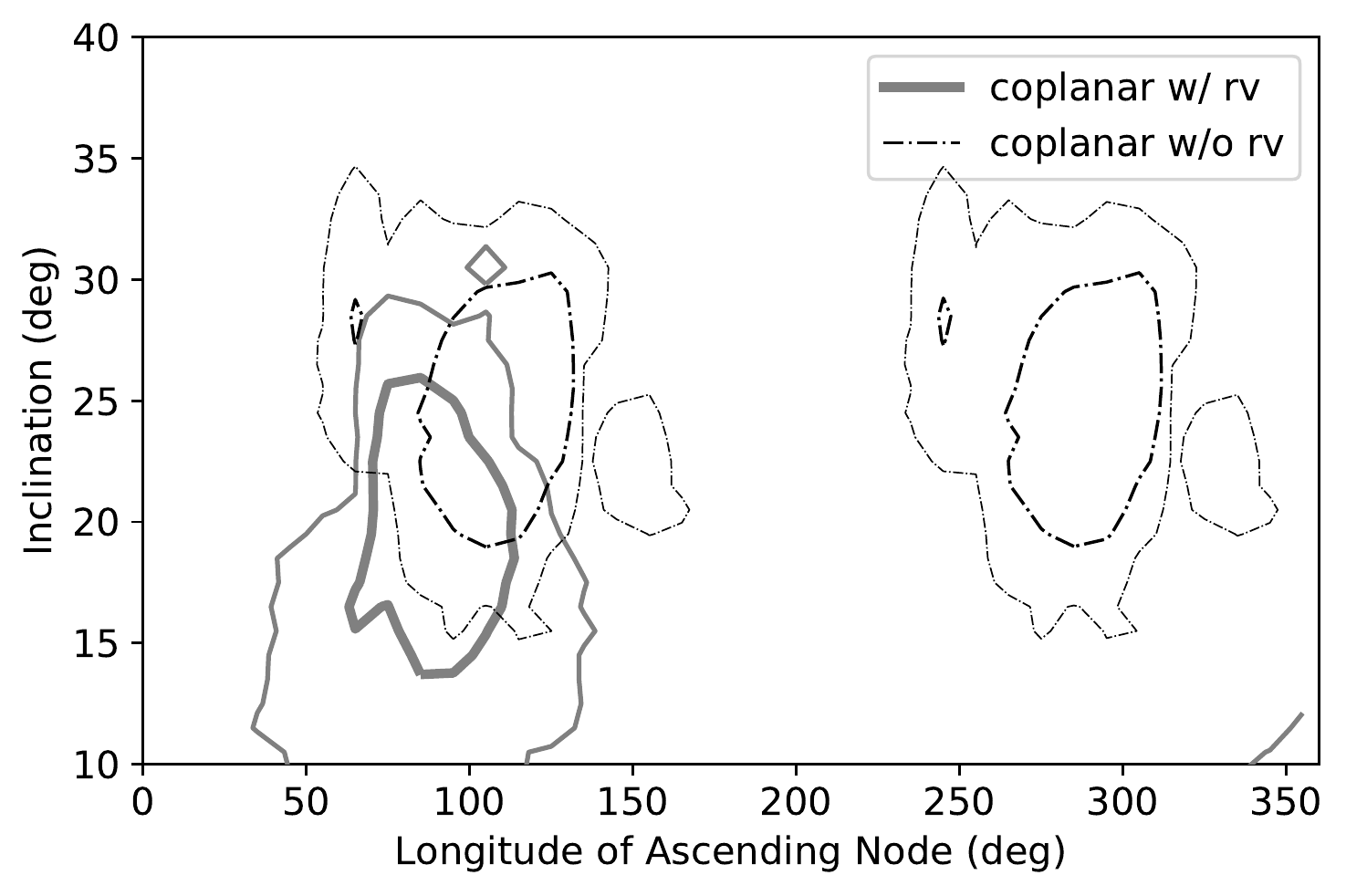}{0.49\textwidth}{(a) Orbital plane Orientation}
		\fig{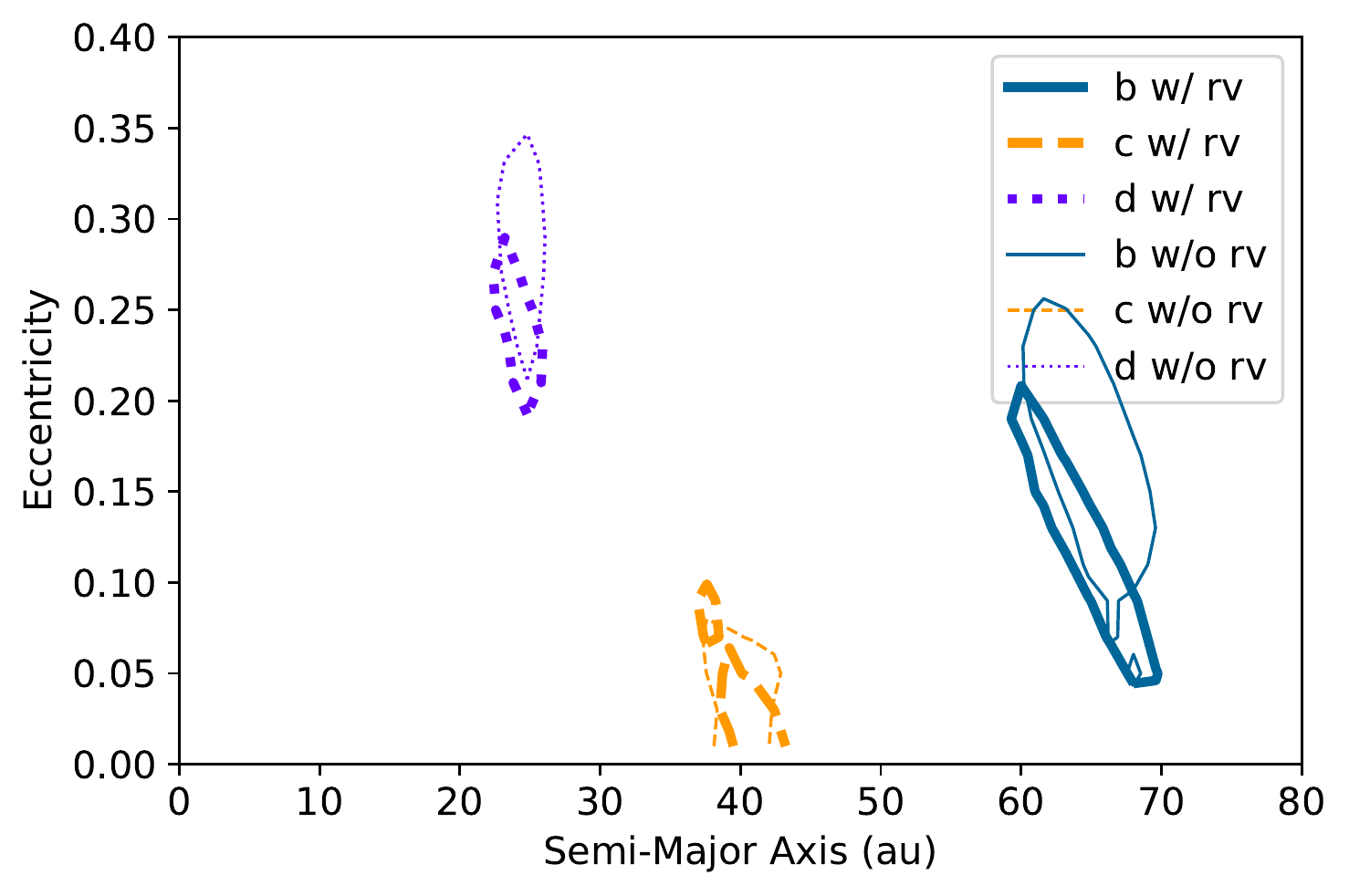}{0.49\textwidth}{(b) Shape of the orbits}
          }
\caption{Orbital parameters with or without the radial velocity (RV) measurements assuming b, c, and d are coplanar. The contours are the $68\%$ confidence intervals for (a) the orientation of the orbital plane and (b) the shape of the orbit for HR 8799b, c, and d. The dashed contours only include the astrometric data from direct imaging, while the solid contours also include the planetary RVs. The assumption of coplanarity forces HR 8799b, c, and d to share the same inclination and longitude of ascending node. For the relative astrometry only case, the posterior of the longitude of ascending node is extended from $\Omega\in[0,180]^{\circ}$ where it is constrained to $\Omega\in[0,360]^{\circ}$ to account for the intrinsic degeneracy existing in this type of data.}
\label{fig:orbitscov}
\end{figure}

\autoref{fig:orbits} shows 50 orbits randomly sampled from the posterior of the RV fit compared to the data points. The associated marginalized posteriors can be found in \autoref{fig:orbitspost} and the values of the estimated parameters are reported in \autoref{tab:RVresults}. The covariance between the inclination and the longitude of ascending node for the orientation of the orbital plane, and the eccentricity and semi-major axis for the shape of the orbits are shown in \autoref{fig:orbitscov}.

\begin{table}
  \centering
\begin{tabular}{ |c|ccc|l| } 
\hline
Parameters & \multicolumn{3}{c|}{HR 8799} & Priors\\ 
 & b & c & d &\\ 
\hline\hline
$a\,(\mathrm{au})$ & $64.6^{+3.7}_{-3.4}$ & $40.9^{+1.2}_{-1.8}$  & $24.4^{+0.7}_{-1.0}$ & Inverse ; $1/a$ with $a\in[10^{-3},10^{7}]$ \\
$e$ & $0.1^{+0.08}_{-0.04}$ & $<0.05$ & $0.23^{+0.4}_{-0.03}$ & Uniform ; $e\in[0,1]$\\
$\omega$ & ${135^{\circ}}^{+21}_{-30}$ & ${56}^{+249}_{-30}$ & $>{322^{\circ}}$ & Uniform; $\omega\in[0,360]^{\circ}$ \\
$\tau$ & $0.42^{+0.06}_{-0.04}$ & $0.35^{+0.07}_{-0.19}$ & $0.59^{+0.03}_{-0.02}$ & Uniform; $\tau\in[0,1]$\\
 \hline
$i$ && ${19.2^{\circ}}^{\,+4.6}_{\,-4.5}$ & & geometric ; $sin(i)$ with $i\in[0,180]^{\circ}$ \\
$\Omega$ && ${88^{\circ}}^{\,+18}_{\,-17}$ & & Uniform; $\Omega\in[0,360]^{\circ}$ if using RV, \\
  & & & & $\Omega\in[0,180]^{\circ}$ otherwise\\
 \hline
Parallax && $24.2^{+0.08}_{-0.1}\,\mathrm{mas}$& & Gaussian; $24.22\pm0.09\,\textrm{mas}$ \\
$M_{\mathrm{tot}}$ && $1.43^{0.33}_{-0.2}M_\odot$& & Gaussian; $1.52\pm0.15 M_\odot$\\
$\mathrm{RV}_\star$ && $-10.5^{\,+0.3}_{\,-0.2}\,\mathrm{km/s}$& & Gaussian; $-12.6\pm1.4\,\textrm{km/s}$\\
 \hline
\end{tabular}
\caption{Estimated orbital parameters from joint coplanar fit of HR 8799b, c, and d using planetary RVs. The error bars and upper/lower limits represent the $68\%$ confidence interval. The last five parameters are shared by the two planets. The parallax, stellar mass, and stellar RV priors are respectively defined from \citet{Gaia2018}, \citet{Baines2012}, and \citet{Gontcharov2006}}
\label{tab:RVresults}
\end{table}

\section{Atmosphere characterization}
\label{sec:atmcharac}

\subsection{Atmospheric model grid}
\label{sec:atmcharacgrid}
\begin{figure}
  \centering
  \includegraphics[trim={0cm 0cm 0cm 0cm},clip,width=0.4\linewidth]{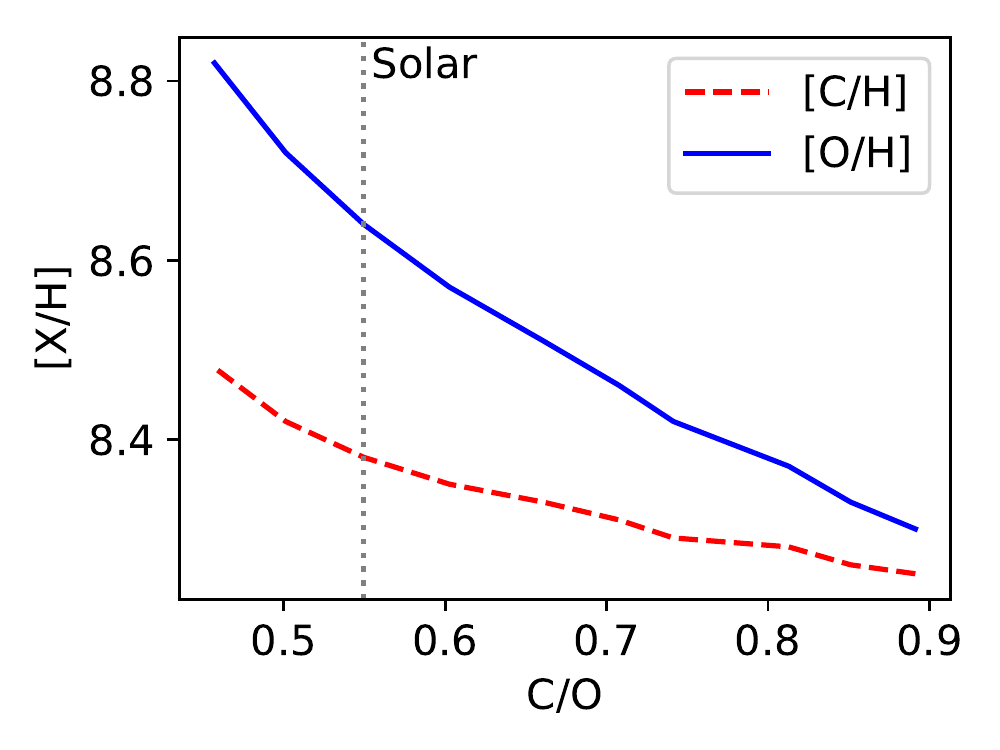}
  \caption{Log abundance by number of carbon and oxygen relative to hydrogen for the atmospheric model grid used in this work. }
  \label{fig:CtoO}
\end{figure}

We now measure the carbon to oxygen ratios (C/O) of HR 8799b, c, and d using a grid of PHOENIX atmospheric models initially described in \cite{Barman2011}. The models use the updated methane line list from \cite{Yurchenko2014} and the optical opacities from \cite{Karkoschka2010}, as in \cite{Barman2015}. Disequilibrium chemistry is modeled by a coefficient of eddy diffusion; $K_{\mathrm{zz}}=10^8\,\mathrm{cm}^2\mathrm{s}^{-1}$. Clouds are included using the parameterized cloud model described in \cite{Barman2011}. The lower cloud boundary is determined assuming chemical equilibrium and the upper cloud boundary is modeled by an exponential decay starting at a specified pressure \citep{Barman2015}. The specified pressure is a free parameter, set to $4\times10^6\,\mathrm{dynes/cm}^2$ in this work, that determines the thickness of the clouds, allowing for the models to have clouds that range from low opacity or cloud-free (COND) to high opacity (DUSTY, \citealt{Allard2012}). The particle size distribution is log-normal with a mean size chosen to be 5$\mu$m. 
Due to the computational cost of generating the models, the oxygen and carbon abundances are not varied individually but using only the C/O ratio. The C/O traces the ratio of accreted solids to gas in the planet atmosphere assuming it formed between the CO\textsubscript{2} and CO snowlines (see Figure 10 in \citealt{Barman2015}; \citealt{Oberg2011}).
In the original grid, temperature varies from 800 K to 1200 K in steps of 100 K, surface gravity $\mathrm{log}(g/[1\,\mathrm{cm.s}^{-2}])$ varies from 3 to 4.5 in step of 0.5, and C/O varies from 0.45 to 0.90 in steps of 0.05.
C/O is defined as a ratio of number of molecules. The wavelength sampling of the models is $\delta \lambda \approx 5\times10^{-6} \mu\mathrm{m}$ ($\lambda/\delta \lambda \sim 426,000$) compared to OSIRIS' pixels that have $\delta \lambda \approx 2.5\times10^{-4} \mu\mathrm{m}$ ($\lambda/\delta \lambda \sim 8,500$). The [C/H] and [O/H] are given as a function of C/O in \autoref{fig:CtoO}.

\subsection{Forward modelling with OSIRIS}
\label{sec:atmcharacOSIRIS}

We infer the four parameters $[T_{\mathrm{eff}}, \mathrm{log}(g), \mathrm{C/O}, \mathrm{RV}]$ according to \Secref{sec:inference} by linearly interpolating the models on a finer grid. 
The best fit models are compared to the combined extracted spectra in \autoref{fig:spectra}. These spectra are not used in the definition of the likelihood, but only used as visualization tool. They feature well-resolved H\textsubscript{2}O and CO spectral signatures. By defining the likelihood from the spectral cubes directly, hence avoiding the spectral extraction step, and jointly modeling the starlight and the planet, we prevent the potential biases caused by over-subtraction of the companion signal that are very common in high-contrast imaging \citep{Marois2014,Pueyo2016}. Moreover, an interesting property of this framework is the marginalization over the starlight subtraction itself.

The different likelihoods for each exposures can be combined assuming statistical independence. In this case, there is no difference between posterior and likelihood because we are not introducing any special prior. We evaluate the consistency of the C/O estimates by marginalizing the posterior for each exposure and the result is shown in \autoref{fig:CtoO_meas}. 
The ability to test the modeling of the noise by fitting every exposure individually is one of the advantages of using a grid of models compared to the more computationally costly full retrieval.
After marginalizing over the planetary RV, the $[T_{\mathrm{eff}}, \mathrm{log}(g), \mathrm{C/O}]$ posteriors are shown in \autoref{fig:allpost} and the corner plots can be found in \Appref{app:corners}.

The best fit parameters corresponding to \autoref{fig:spectra} are summarized in \autoref{tab:atmcharacbestfit}. The combined C/O estimates for HR 8799b, c, and d are $0.572^{+0.008}_{-0.007}$, $0.562^{+0.003}_{-0.007}$, and $0.55^{+0.02}_{-0.03}$ respectively. We recognize that these formal errors are highly underestimated and do not include the broader level of uncertainty surrounding their estimation due to imperfect data calibrations and the definition of the model grid. 
First, inferred temperature resulting from this fit is meeting the upper end of the grid at 1,200 K, which means that any covariance between temperature and C/O would bias the values quoted previously. Simply extending the grid to include higher temperatures is not necessarily a solution as these values are thought to be unlikely in view of evolutionary models and the use of inaccurate temperatures can also bias the inference. As discussed in \citet{Konopacky2013} and \citet{Barman2015}, these issues are known and resolving them is outside the scope of this work. 
Second, another well documented issue is the coarseness of the grid of atmospheric models \citep{Czekala2015}. The derived formal uncertainties are almost an order of magnitude smaller that the step-size in each dimension ($\Delta T_{\mathrm{eff}} = 100\,\mathrm{K}$, $\Delta \mathrm{log}(g) = 0.5$, $\Delta C/O = 0.05 $). The local minima in the C/O posteriors of HR 8799b and other kinks in \autoref{fig:allpost} are not real and match the sampling of grid.
In the following section, we fit the same grid of models to published photometry and low resolution spectra as an independent way to evaluate the consistency of the posteriors. While acknowledging these issues, we can conclude that the three planets appear to have similar C/O without large deviations from the stellar value.

\begin{rotatepage}
\begin{sidewaysfigure}
\includegraphics[width=\linewidth]{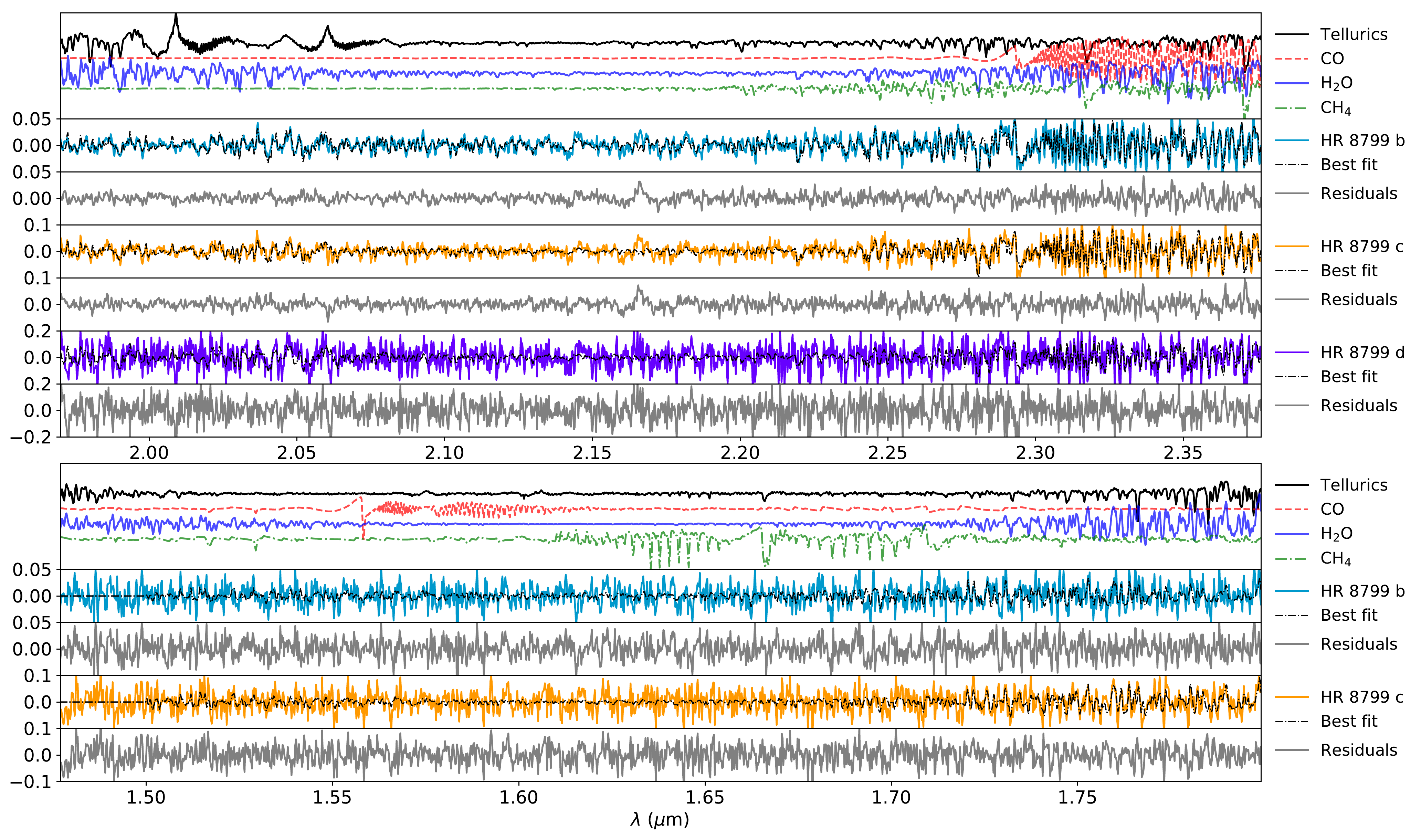}
\caption{Extracted spectra of HR 8799b, c, and d. These spectra have not been divided by a transmission spectrum of the atmosphere and therefore include telluric lines. The top panel shows the K-band combined spectra and the bottom panel is the same for H band. From top to bottom, each panel includes the molecular signatures used for molecule detection as well as a high pass filtered spectrum of the tellurics, the spectrum of each planet and its best fit model (see \autoref{tab:atmcharacbestfit}), followed by the residuals of the fit. The best fit model is however not derived from these extracted spectra, but using a forward model of each individual spectral cube instead.}\label{fig:spectra}
\end{sidewaysfigure}
\end{rotatepage}

\begin{figure}[bth]
\fig{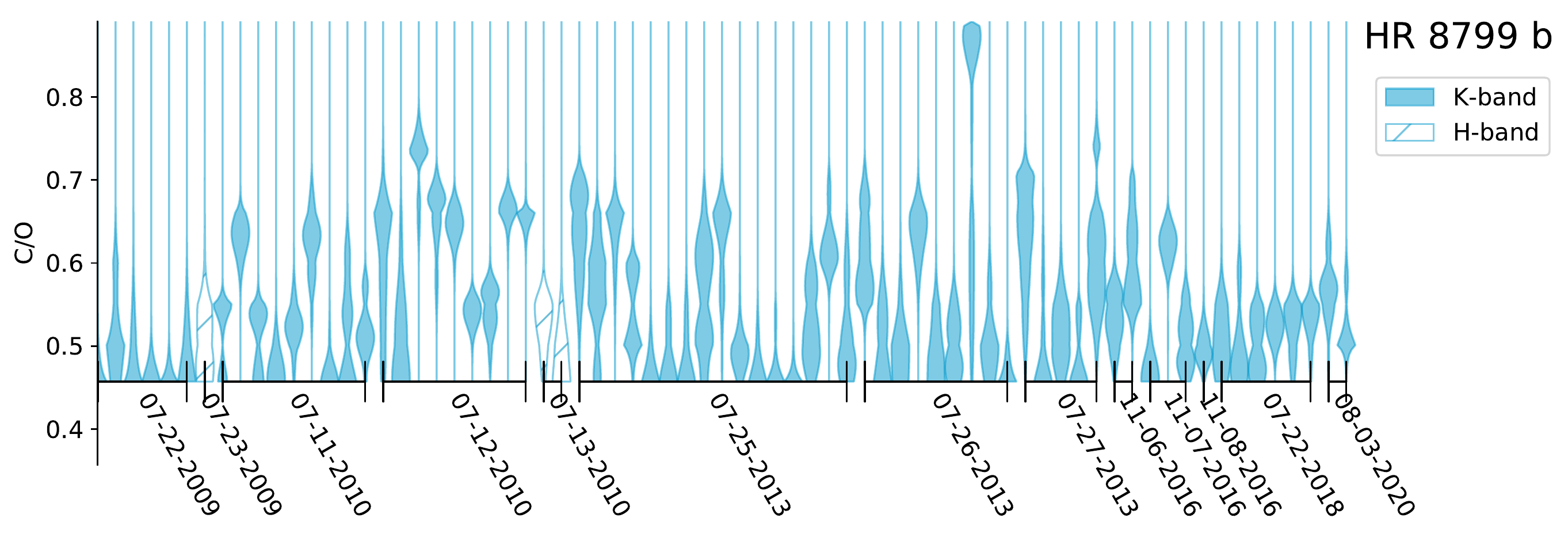}{1\textwidth}{}
\fig{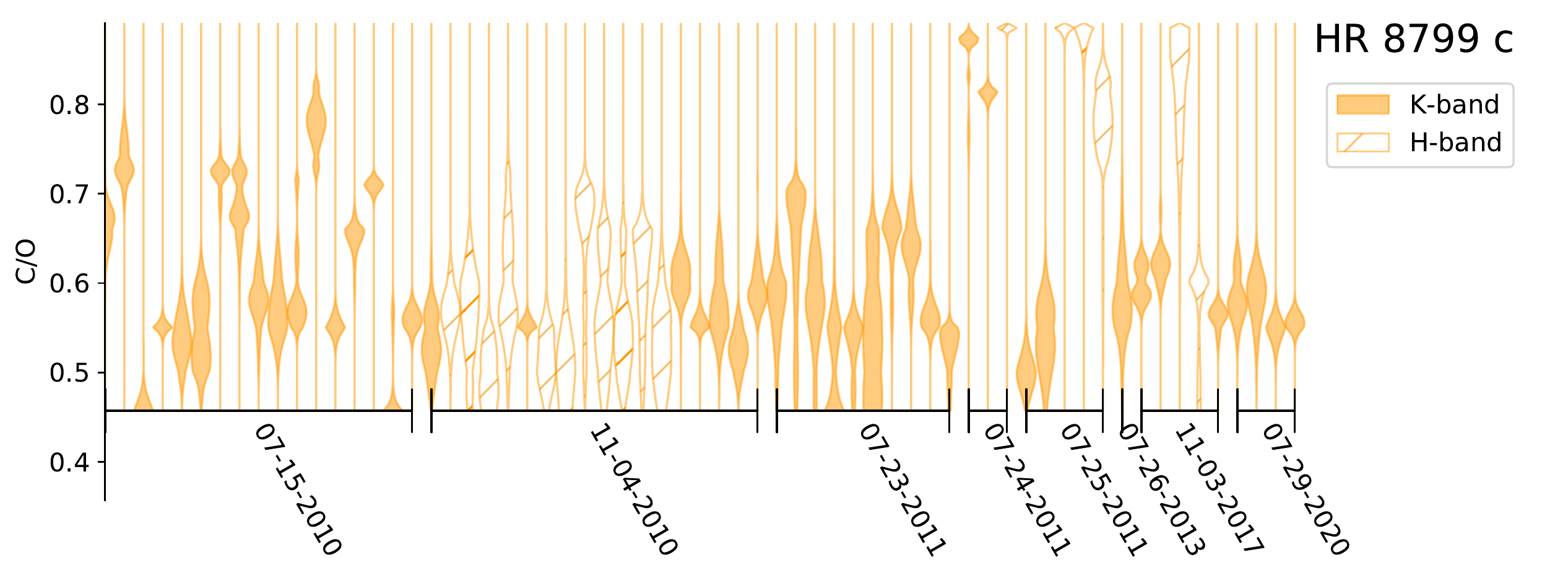}{1\textwidth}{}
\fig{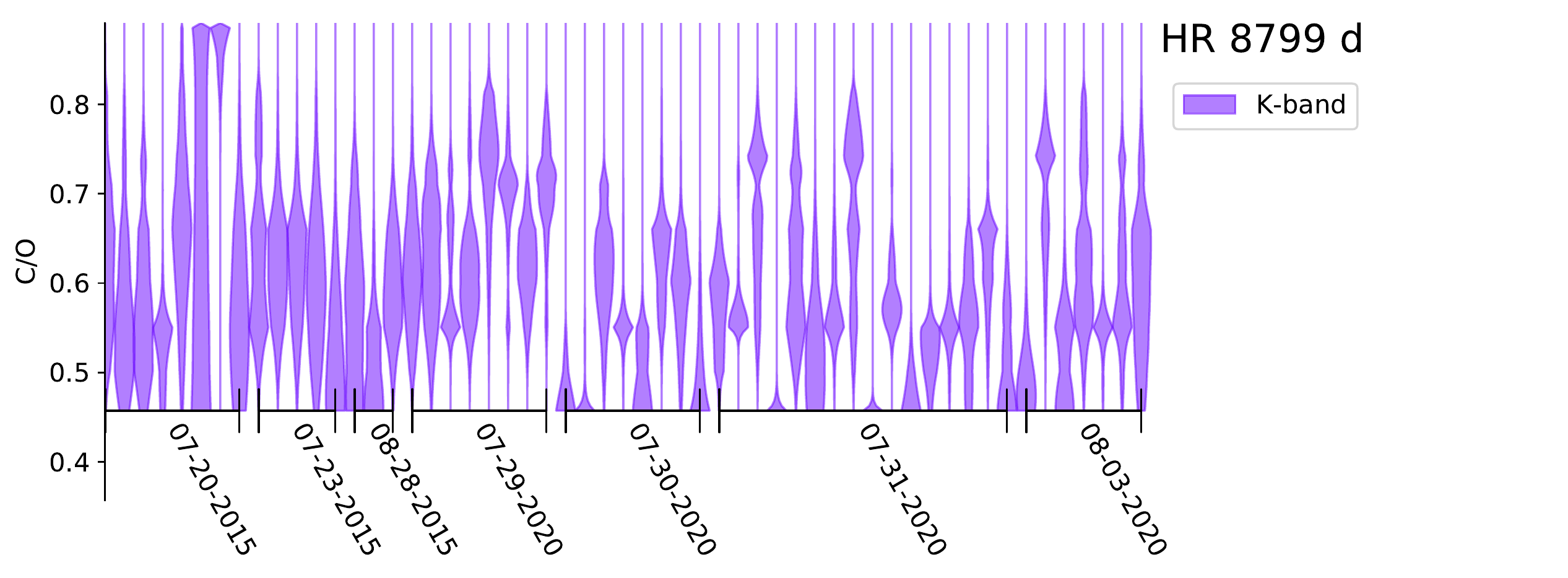}{1\textwidth}{}
\caption{Marginalized C/O posteriors for each exposure of HR 8799b, c, and d as a function of the date of observation. The width of the hatched (H band) or shaded (K band) area is proportional to the probability density of the posterior after normalization of its peak value. The limits of the y-axis correspond to the edge of the atmospheric model grid along the C/O dimension.}
\label{fig:CtoO_meas}
\end{figure}

\begin{figure}
  \centering
  \includegraphics[width=1\linewidth]{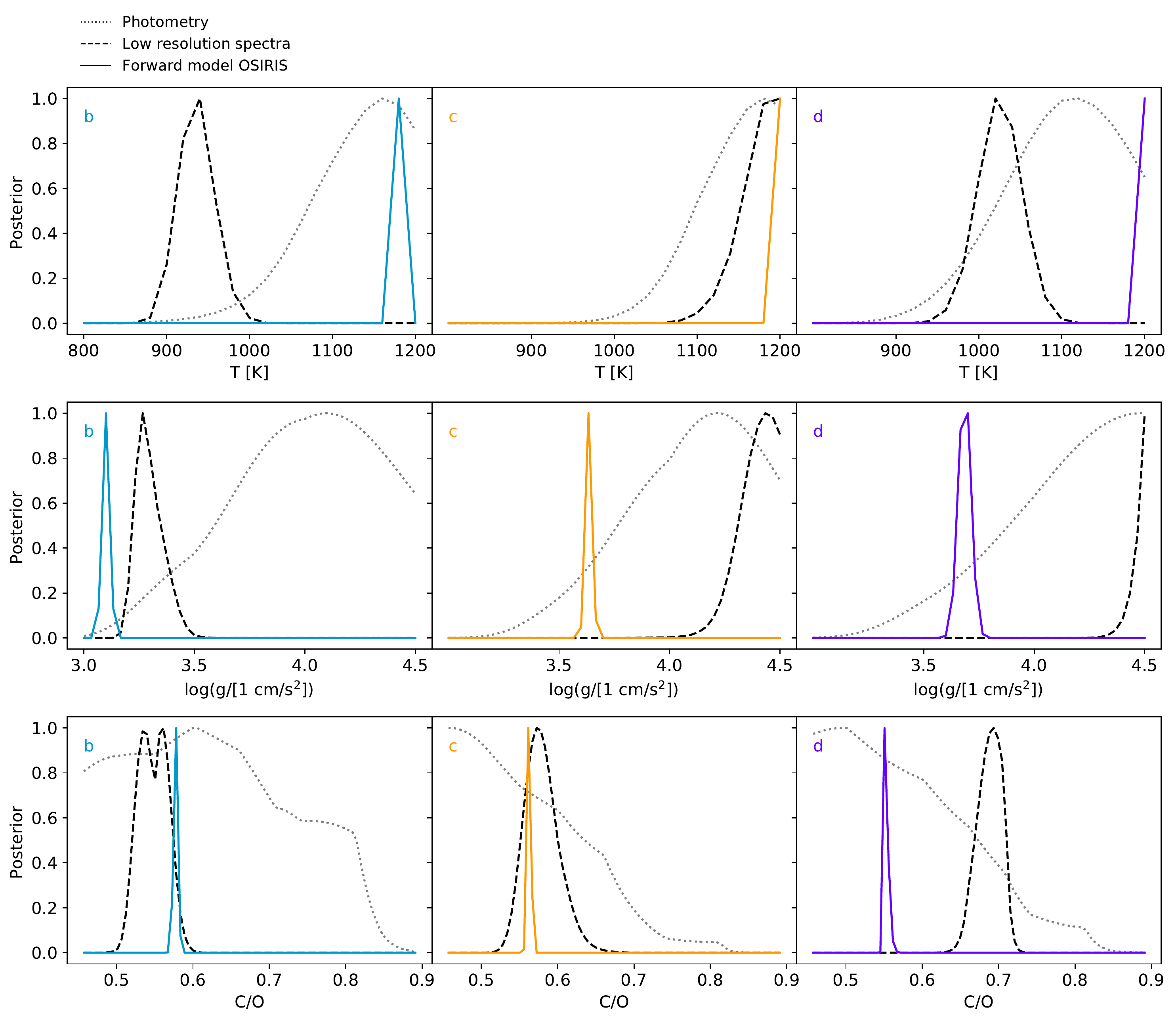}
  \caption{Marginalized posteriors for temperature, surface gravity, and C/O of HR 8799b, c, and d. Each row corresponds to a different parameter and each column represents to a different planet. Each panels contains three posteriors corresponding to fits to the photometry (solid line), low resolution spectra (black and grey dashed lines), and the forward model of OSIRIS spectra (colored lines).}
  \label{fig:allpost}
\end{figure}

\begin{table}
  \centering
\begin{tabular}{ |c|ccc| } 
 \hline
  Planet & $T_{\mathrm{eff}}$ (K) & $\mathrm{log}(g/[1\,\mathrm{cm.s}^{-2}]) $ & C/O \\ 
 \hline\hline
b & $1180^{+14}_{-14}$ & $3.10^{+0.03}_{-0.03}$ & $0.578^{+0.004}_{-0.005}$  \\
c & $1200^{*}_{-14}$ & $3.63^{+0.03}_{-0.02}$ & $0.562^{+0.004}_{-0.004}$  \\
d & $1200^{*}_{-14}$ & $3.7^{+0.03}_{-0.05}$ & $0.551^{+0.005}_{-0.004}$  \\
 \hline
\end{tabular}
\caption{Best fit parameters based on the forward model of the OSIRIS data only. The star $*$ signifies that the best fit value corresponds to the edge of the grid and should therefore be seen as an upper, or lower, limit instead.}
\label{tab:atmcharacbestfit}
\end{table}

\subsection{Fitting the spectral energy distribution}
\label{sec:atmcharacSED}

We can compare the atmospheric model fits from OSIRIS to lower resolution published data.
\autoref{fig:allpost} contains the marginalized posteriors for temperature, surface gravity, and C/O for the three planets as a function of the dataset used: photometry, low resolution spectra, and moderate resolution with OSIRIS. The best fit models are compared to the data in \autoref{fig:lowresspectra}. The different fits are described below.

First, we fit the photometric data points from the literature and summarized in \autoref{tab:photometry}. The C/O model grid used in this work is limited to H and K bands, which is why we cannot include the existing $3-5\,\mu\mathrm{m}$ photometry in this analysis.
The conversion between absolute magnitudes and flux is done using each instrument filter\footnote{\url{http://svo2.cab.inta-csic.es/theory/fps/}} and a reference Vega spectrum\footnote{\url{https://www.stsci.edu/hst/instrumentation/reference-data-for-calibration-and-tools/astronomical-catalogs/calspec}} \citep{Bohlin2014}. Variations of the atmosphere water content and transmission is assumed negligible and was not accounted for. 
We generated a grid of photometric points from the high resolution atmospheric models, which was then linearly interpolated on a finer grid. The fluxes were converted to magnitudes and then compared to the data with a $\chi^2$. In the absence of priors, the posterior is equal to the likelihood ($\propto \exp{-\chi^2/2}$).

Then, for the low resolution spectra, we fit the H ($R=46$), K1 ($R=66$), and K2-band ($R=79$) GPI spectra from \citet{Greenbaum2018} for HR 8799c and d, and the H-band spectrum ($R=60$) from \citet{Barman2011} combined with a smoothed ($R=200$) OSIRIS spectrum from \citet{Barman2015}. The GPI linespread function is oversampled so we are only using a fourth of the data points to limit possible issues related to correlation. We first convolved the model to match the sampling and resolution of the low resolution spectra. The broadened model grid was then linearly interpolated and a $\chi^2$ computed over a more finely sampled grid. We use a similar approach in the case of the photometric data points. The atmospheric models are only defined over H+K band so we cannot include existing observations at shorter and longer wavelength.

The results of this analysis of the photometry and low resolution spectra compared to the higher resolution fits are discussed in \autoref{sec:combiningdata}.

\begin{table}
\centering
\begin{tabular}{l|l|ll|ll|ll|l}
\hline
\multicolumn{1}{l}{Filter} & \multicolumn{1}{c}{\multirow{2}{*}{$\lambda$}} & \multicolumn{2}{l}{HR 8799b}    & \multicolumn{2}{l}{HR 8799c}    & \multicolumn{2}{l}{HR 8799d}    & \multicolumn{1}{l}{\multirow{2}{*}{Reference}} \\
\cline{3-8}
\multicolumn{1}{l}{Photometry} &  \multicolumn{1}{c}{}  & Mag.  & \multicolumn{1}{l}{Err.} & Mag.  & \multicolumn{1}{l}{Err.} & Mag.  & \multicolumn{1}{l}{Err.} &   \multicolumn{1}{c}{}  \\ 
\hline
H2         & 1.593 & 15.1  & 0.14 & 14.11 & 0.12 & 14.04 & 0.17 & \citet{Zurlo2016}    \\
H3         & 1.667 & 14.8  & 0.1  & 13.8  & 0.1  & 13.87 & 0.16 & \citet{Zurlo2016}    \\
K1         & 2.11  & 14.17 & 0.06 & 13.21 & 0.05 & 13.22 & 0.07 & \citet{Zurlo2016}    \\
K2         & 2.251 & 13.99 & 0.09 & 12.88 & 0.07 & 12.86 & 0.1  & \citet{Zurlo2016}    \\
H          & 1.633 & 14.87 & 0.17 & 13.93 & 0.17 & 13.86 & 0.22 & \citet{Marois2008science}\tablenotemark{a}   \\
CH4S       & 1.592 & 15.18 & 0.17 & 14.25 & 0.19 & 14.03 & 0.3  & \citet{Marois2008science}   \\
CH4L       & 1.681 & 14.89 & 0.18 & 13.9  & 0.19 & 14.57 & 0.23 & \citet{Marois2008science}   \\
Ks         & 2.146 & 14.05 & 0.08 & 13.13 & 0.08 & 13.11 & 0.12 & \citet{Marois2008science}   \\
H          & 1.633 & 15.08 & 0.13 & 14.18 & 0.14 & 14.23 & 0.2  & \citet{Skemer2012}\tablenotemark{a}   \\
\hline
\end{tabular}
\begin{tabular}{l|l|c|c|c|c|l}
\hline
\multicolumn{1}{l}{Spectra} & \multicolumn{1}{|c|}{$\lambda$} & \multicolumn{1}{|c|}{$R=\lambda/\delta \lambda $} & b & c & d  & \multicolumn{1}{l}{Reference} \\
\hline
H         & 1.48-1.82 & 46 &    & $\times$ & $\times$ & GPI; \citet{Greenbaum2018}    \\
K1         & 1.86-2.22 & 66 &    & $\times$ & $\times$ & GPI; \citet{Greenbaum2018}    \\
K2         & 2.08-2.43 & 79 &   & $\times$ & $\times$ & GPI; \citet{Greenbaum2018}    \\
Hbb         & 1.45-1.83 & 200 & $\times$  & & & OSIRIS; \citet{Barman2011}    \\
Kbb         & 1.94-2.42 & 200 & $\times$  & & & OSIRIS; \citet{Barman2011}    \\
\hline
\end{tabular}
\caption{Published H and K band absolute magnitudes and low resolution spectra for HR 8799b, c, and d.}
\label{tab:photometry}
\tablenotetext{a}{The filters of these photometry point is broader than the definition of the model spectra. They are therefore not included in the fits but shown in \autoref{fig:lowresspectra}.}
\end{table}

\begin{rotatepage}
\begin{sidewaysfigure}
\includegraphics[width=\linewidth]{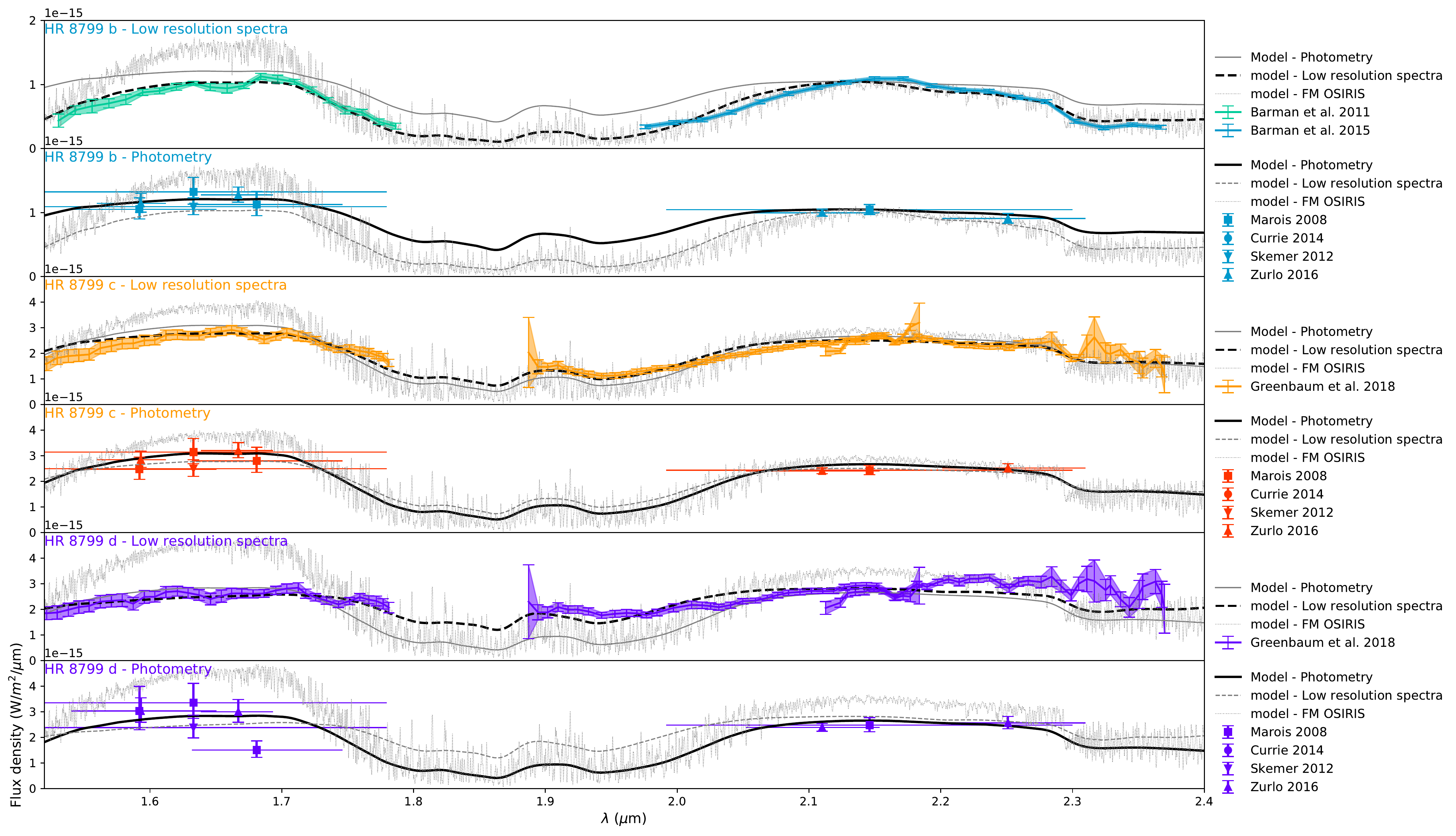}
\caption{Published photometry and low resolution spectra for HR 8799b, c, and d. Each panel contains three best fit models respectively from a fit to the photometry (solid), the low resolution spectra (dashed), and the OSIRIS data presented in this work (dotted). The best fit models for the photometry and the low resolution spectra were convolved to a resolution of $R=100$ in this vizualization. In each panel, the best fit model associated to the data being plotted is drawn with a thick black line as opposed to grey. The data used in this plot are listed in \autoref{tab:photometry}.}\label{fig:lowresspectra}
\end{sidewaysfigure}
\end{rotatepage}

\section{Discussion}
\label{sec:discussion}

\subsection{Molecule detection}

We used a method similar to cross correlation analysis to look for H\textsubscript{2}O, CO, and CH\textsubscript{4} in the atmosphere of HR 8799b, c, and d.
H\textsubscript{2}O and CO are detected in all three planets and these detection are at a higher S/N compared to past analysis of the same data for HR 8799b and c \citep{Konopacky2013,Barman2015,PetitditdelaRoche2018}. As expected, only H\textsubscript{2}O is detected in H-band spectra. Significant efforts were made to limit the effects of potential systematics such as imperfect telluric correction (see \Secref{sec:ccf}).

Detecting and measuring the abundance of methane as a function of planet mass, age, and temperature is important to constrain non-equilibrium chemistry resulting from atmospheric mixing. The mixing efficiency is parametrized by the coefficient of eddy diffusion $K_\textrm{zz}$. For HR 8799 like planets, an efficient mixing will result in the dissociation of methane in the hotter layer of the atmosphere decreasing its overall abundance. 
A detection of methane in HR 8799b was presented in \citet{Barman2015} based on cross correlation analysis and visible lines matching those of methane. A weak absorption band consistent with methane was also identified in a low resolution K-band OSIRIS spectrum in \citet{Barman2011}. Using only a subset of this dataset, \citet{PetitditdelaRoche2018} performed an independent cross correlation analysis that did not produce a similar detection of methane. They proposed that an imperfect telluric correction could explain the detection in the original work. To verify this hypothesis, we performed a new analysis of the results presented in \citet{Barman2015} that differed only by the telluric correction. Instead of computing the telluric transmission from the standard star observations, it is computed from the science spectral cubes themselves using the speckles from HR 8799A to account for telluric variability during the night. As shown in \Appref{app:hr8799bmethane}, the methane signal in the CCF not only remains but it is significantly stronger when compared to \citet{Barman2015}. In the present work and using a slightly different kind of analysis, we were also unable to produce a strong CCF detection of CH\textsubscript{4} while including more observations. 
While the strength of the methane signal in the different CCF analyses is puzzling, we estimate that the lack of strong CCF detection of methane in this work is not inconsistent with the derived methane abundance from \citet{Barman2015} within the uncertainty. 
The abundance of methane derived in \citet{Barman2015} (mole fraction between $10^{-5.06}$ and $10^{-5.85}$) suggested an eddy diffusion coefficient of $\mathrm{log}(K_{\mathrm{zz}})\approx7$. If the methane abundance was overestimated, a larger value of $K_{\mathrm{zz}}$ would be expected. A complete re-analysis of the individual molecular abundances of H\textsubscript{2}O, CO, and CH\textsubscript{4}, as performed in \citet{Barman2015}, is not within the scope of this work.  The newly commissioned Keck Planet Imager and Characterizer (KPIC; $R\approx37,500$; \citealt{Mawet2017SPIE}) instrument is a promising avenue to resolve this issue thanks to its order of magnitude higher spectral resolution. 

\subsection{Radial velocity and orbits}

In \citet{Ruffio2019RV}, we measured the RV of the outer two planets of the system. Here, we measured the RV of HR 8799 d; the next closest planet to the star. Similar to \citet{wilcomb2020}, we showed that imperfect telluric correction can result in significant RV systematics, which we corrected by included principal components of the residuals in the modeling of the data. The measurements from \citet{Ruffio2019RV} for HR 8799b and c were also updated with the new model.

The inferred RVs for the three planets remain consistent with the astrometric measurements and assumptions of coplanarity and stability of the system \citep{Wang2018}. 
We used the new RV to revisit the orbit analysis from \citet{Ruffio2019RV} while extending it to HR 8799b, c, and d instead of HR 8799b and c only. The three planets are assumed to be coplanar to speed up convergence. 
The past analysis already showed how the RV measurements of b and c ruled out the family of orbits with the longitude of ascending node ($\Omega \sim 280^{\circ}$). The new RV measurement of HR 8799d confirms this result, but it is otherwise providing marginal additional constraints on the other orbital parameters.
Better constraining the eccentricity is important as a non-zero eccentricity while in resonance could point to the fact the planets migrated while in resonance \citep{Wang2018}. Also differences in the eccentricity distribution of planets and brown-dwarf could highlight different formation processes \citep{Bowler2020}. Planetary RVs are expected to be comparatively more powerful in cases for which the visual orbits are not as well constrained as HR 8799. The RV from \autoref{tab:RVepochs} can be used to better constrain the mutual inclination of the HR 8799 planets, but this analysis is left for future work.

Measuring the RV of several components of a multi-planet system is interesting because it removes the need for a precise RV measurement of the host-star. 
The latter is often challenging as most stars orbited by directly imaged planets are young, sometimes variable, rapidly rotating, and early type stars. \citet{Snellen2014} and \citet{wilcomb2020} are two examples in which the relative RV measurement of the planet is limited by our knowledge of the absolute RV of the star. 
In the near future, higher resolution spectrographs such as KPIC should provide more precise planetary RVs. These measurements could be combined with extremely precise astrometric measurements such as the ones acquired by GRAVITY for HR 8799e \citep{Gravity2019}. Altogether, these would improve our knowledge of the orbits of the system.
These results are also important to demonstrate the feasibility of measuring RVs of planetary mass companions at high contrast. Such RV measurements have been proposed as a way to detect moons in the future \citep{Vanderburg2018}.

\subsection{Combining data with different spectral resolutions}
\label{sec:combiningdata}

Combining low and high-resolution spectra is a hallmark of the study of exoplanet atmospheres \citep{Brogi2019,Gravity2020,Molliere2020,WangJi2020}, but developing models that can consistently fit both has proven to be challenging. 
For example, \citet{WangJi2020} highlighted the difficulty in combining spectra by presenting separate and combined analyses of different instruments: GPI (H, $R=46$; K, $R=66-79$), CHARIS (JHK, $R=20$) and an OSIRIS spectrum binned down to $R=1,000$ to characterize HR 8799c. 
The prerequisite to combine spectra with different resolutions is to ensure that the posteriors from fitting them individually are consistent which each other. 
When jointly fitting the different resolutions, \citet{WangJi2020} used a strong prior to fix the planet temperature and radius and to yield a planet mass for HR 8799c consistent with the dynamical stability of the planetary system. Fixing a subset of parameters such as the temperature or surface gravity to better match the spectral energy distribution or cooling tracks of the planet is common practice. \citet{Zhang2020} also highlights the difficulty to retrieve surface gravity for brown-dwarfs. However, this does not resolve the underlying inconsistencies of the model and the data, which therefore remains subject to systematics that are not well characterized.

The highlights of this analysis compare to past work are the uniform analysis of the three planets, and the higher spectral resolution from OSIRIS. The native resolution of OSIRIS ($R=4,000$) resolves the molecular lines of H\textsubscript{2}O and CO, the dominant C and O carriers, which can unambiguously constrain C/O. For example, CO lines can be visually resolved in the $2.29\,\mu\mathrm{m}$ CO band-head for the three planets in the high S/N spectra shown in \autoref{fig:spectra}; this is a first for HR 8799 d. 

The downside of self-consistent models is the computational cost of generating the grid, which limits the number of parameters that can be explored. 
A recurring issue in atmospheric characterization of directly imaged low mass companions is the radius or under-luminosity problem \citep[e.g.,][]{Marois2008science,Bowler2010}. 
It is not uncommon for higher effective temperatures to be favored in the models, which lead to unrealistically small planet radii due to the low luminosity of the objects. These inferred parameters are inconsistent with evolutionary models. Evidence shows that the discrepancy can be improved with better modeling of the clouds and disequilibrium chemistry \citep{Currie2011,Skemer2011,Barman2011,Barman2011b}. \citet{WangJi2020} also showed that extending the wavelength coverage (e.g $3-5\,\mu\mathrm{m}$) helps to better constrain these parameters. In this work, the inferred temperatures in \autoref{tab:atmcharacbestfit} from OSIRIS are higher than the evolutionary model predictions, and therefore hitting the edge of our grid at $1,200\,\mathrm{K}$. 
Unfortunately, this systematic uncertainty cannot be self-consistently included in our estimates without additional parameters in the model (a.k.a. omitted-variable bias), which is likely a reason why formal error bars are unrealistically small. While free atmospheric retrievals are prone to over fitting, self-consistent models with few varied parameters are under fitting.
Going beyond the grid approach is paramount to better model the atmosphere and include more parameters that would help resolve this issue; e.g. clouds and disequilibrium chemistry.

Another potential problem plaguing inference of atmospheric parameters is the data calibration and modeling. The high contrast between the planet and the star can result in residual starlight in the planet spectrum. To only cite a few examples, photometric calibration can be challenging, speckle noise is common place in lower resolution spectra, and the telluric lines can add systematics to the data at higher resolution.
In \Appref{app:noiseanalysis}, we identify a moderate level of correlation between neighboring pixels in high-pass filtered OSIRIS spectra, which could point toward improper calibrations of the wavelength solution and the resolution of the instrument or correlated noise from the interpolation in the spectral cube reconstruction from the detector images. The need for principal components in the forward model, the residual stellar Brackett-$\gamma$ in \autoref{fig:spectra}, or the outliers in \autoref{fig:RV_bcd} and \autoref{fig:CtoO_meas} are all signs of imperfect modeling of the data. 
Hierachical Bayesian models are a great way to address these issues \citep[\S4.2 in][]{Czekala2017}, but the limited number of free parameters in the atmospheric grid is more likely the dominant factor to explain inconsistencies between the low and moderate resolution spectra in this analysis (see \autoref{fig:allpost}).

\subsection{Carbon to oxygen number ratio}

In this work, we estimated the C/O ratio of HR 8799b, c, and d using the framework described in \Secref{sec:model}. Out of the four planets in the system, HR 8799d is the only planet that did not have a reliable C/O measurement in the literature.

A feature of the modeling approach used is that it does not rely on a prior starlight suppression, spectral extraction, and telluric correction. The likelihood is directly defined from a forward model of the spectral cubes as produced by the OSIRIS data reduction pipeline.
The C/O estimates ($[C/O]_\textrm{b}=0.578^{+0.008}_{-0.005}$, $[C/O]_\textrm{c}=0.562^{+0.007}_{-0.004}$, and $[C/O]_\textrm{d}=0.55^{+0.011}_{-0.014}$) are consistent with past work from our group ($[C/O]_\textrm{b}=0.66^{+0.04}_{-0.08}$, $[C/O]_\textrm{c}=0.562^{+0.14}_{-0.11}$; \citealt{Barman2015}; \citealt{Konopacky2013}). 
As discussed previously, the uncertainties are underestimated but they really emphasize the constraining power of these spectra.
We note that C/O is the most consistent of the three parameters in \autoref{fig:allpost} with the only discrepancy being for the GPI spectrum of HR 8799 d.

Stellar C/Os were also found by \citet{Lavie2017} for HR 8799c ($[C/O]_c = 0.55^{0.1}_{-0.12}$), \citet{WangJi2020} ($[C/O]_c = 0.58^{0.06}_{-0.06}$ with weak priors), and by \citet{Molliere2020} for HR 8799 e ($[C/O]_e = 0.6^{+0.07}_{-0.08}$). As discussed in \Secref{sec:combiningdata}, such atmospheric retrievals are powerful, because they can marginalize over a wider range of parameters including clouds, which is a limitation in the grid based methodology used here. However, the computational cost of going to high spectral resolution can become prohibitive.
We note the discrepancy between the OSIRIS inferred C/O of HR 8799b and the results from atmospheric retrieval of low resolution spectra, which are close to unity: $[C/O]_b = 0.92\pm0.01$ \citep{Lavie2017} and $[C/O]_b = 0.97\pm0.01$ \citep{Lee2013}. Such high values of the C/O are ruled out by \citet{Barman2015} and the high S/N CO band head measurement from OSIRIS.

Despite the possible systematics, our analysis suggests that HR 8799 d has a similar C/O to HR 8799 b and c. Combined with literature values, it can conclude that all four known planets orbiting HR 8799 have a stellar C/O ratio. We discuss the implication on planet formation in the \Secref{sec:formation}.

\subsection{Implication for planet formation models}
\label{sec:formation}

Based on our work with OSIRIS for HR 8799b, c, and d, and the recent analysis of HR 8799e by \citet{Molliere2020}, all four planets in HR 8799 have C/O ratios consistent with the host star ($C/O = 0.54^{+0.12}_{-0.09}$; \citep{WangJi2020}). Though systematic uncertainties remain in the models which might alter these values somewhat, all analysis and extracted spectra point to spectral morphologies that do not vary in any significant way from solar values. We note that \citet{Molliere2020} performed a retrieval analysis in which they also derived an enhanced bulk metallicity relative for HR 8799e ([Fe/H] $\sim$ 0.5). We have not fit for metallicity, but solar metallicity templates appear to be a fine match to all of our spectra. The CO lines, for instance, do not appear to be significantly deeper than a solar metallicity, stellar C/O ratio requires. However, since $H$ and $K$ band spectra are primarily sensitive to the abundance of carbon and oxygen, future observations of shorter wavelengths, such as J band, where additional species such as potassium and iron (via FeH) may be present could shed light on the bulk metallicity of the planets.

The measurement of a consistent C/O ratio for all four planets is an important milestone and result. For all practical purposes, these values do not sufficiently deviate from stellar C/O to be decisive with respect to specific formation scenarios. A stellar ratio is consistent both with formation through gravitational instability, or with core/pebble accretion scenarios that include planetesimal enrichment. Indeed, it is clear that when taking into account the impact of a few of the many parameters that might impact this ratio in a planetary atmosphere, including pebble drift across ice lines and planet migration, almost any value of C/O (and C/H and O/H) is plausible (e.g., \citealt{Booth2017,Cridland2017,Madhu2017,Mordasini2016}). Our goal is to begin to compile a statistical ensemble of C/O ratio measurements for the directly imaged planets in order to begin to constrain which core/pebble scenarios may be feasible, starting with the four HR 8799 planets. If it is the case that gas giants that form via core/pebble accretion should notionally have an elevated C/O ratio in their atmospheres after runaway gas accretion, a clear possibility here is that these very large gas giant planets like the HR 8799 planets (see also $\kappa$ Andromedae b, \citealt{wilcomb2020}) must also undergo extremely significant planetesimal enrichment in order to grow so large, leading to a stellar C/O. 
. 

\citet{Molliere2020} describe a formation scenario in which all of the HR 8799 planets form beyond the CO ice line as a way to achieve the observed C/O ratios. This is predicated on an understanding of where the ice lines would have been around HR 8799. Since HR 8799 is a hotter A-type star, values based on the Sun are likely inappropriate. However, the values we assumed for the snow lines based on \citet{Oberg2011} predictions for the A0 star AB Auriga (Figure 2), which places the CO frost line $>$300 AU, is also likely incorrect for HR 8799. The best way to estimate these locations it to use measurements of real ice lines from similar, very young stars. For instance, \citet{Qi2015} directly measure the CO snow line at ~90 AU around the A2 star HD 163296. HR 8799 is cooler than this star, so the CO frost line is likely interior to this. However, \citet{Qi2015} also note that the CO freezes out at a higher temperature in this disk than in TW Hydra, perhaps due to compositional differences, and suggest that actually measuring snow lines is the best way to get a true statistical handle on their location. Based on the available information, a safe assumption is that CO snow line in the planet-forming disk around HR 8799 was probably in the range of $\sim$50-100 AU. 

If the snow line is closer to 50 AU, it is certainly plausible that all four planets started their initial formation beyond that range. Indeed, the simulation shown in \citet{Johansen17} for the formation of HR 8799 based on pebble accretion has the cores forming between $\sim50 - 160$ AU. However, in this simulation the bulk to the runaway accretion of the envelope for both HR 8799e and HR 8799d occurs interior to 50 AU as the planets all migrate inward to their current location. The fact that the current planet locations span more than one frost line complicates the overall picture of the formation history of the HR 8799 system.

Altogether, the results presented here and also in \citet{wilcomb2020} have not yet offered direct evidence of core/pebble accretion formation for directly imaged planets, which would have come in the form of an elevated C/O ratio. While we still seek to measure this ratio for all directly imaged planets currently discovered, if all values end up being consistent with stellar, additional formation diagnostics will be needed to shed more light on the formation of wide gas giants. Bulk metallicity and nitrogen abundance, as suggested by \citet{Molliere2020} based on \citet{Oberg2019} and \citet{Bosman2019} as well as \citet{Turrini2021}, are two possible diagnostics to pursue next.

\section{Summary and conclusion}
\label{sec:conclusion}

We performed a comprehensive uniform reanalysis of the HR 8799 data acquired by Keck/OSIRIS, a $R\approx4,000$ infrared integral field spectrograph, at H ($1.97-2.38\mu\mathrm{m}$) and K band ($1.47-1.80\mu\mathrm{m}$). 
The highlights of the this work are the detection and deep observations of HR 8799d, the second closest known planet in the system, as well as the new H band data for HR 8799c.

Using our recently developed framework \citep{Ruffio2019thesis,Ruffio2019RV,wilcomb2020}, we first looked for signatures of water (H\textsubscript{2}O), carbon monoxide (CO), and methane (CH\textsubscript{4}) in the atmosphere of HR 8799b, c, and d. H\textsubscript{2}O and CO were unambiguously detected in the three cases at high S/N. Due to conflicting results, we remain unable to unambiguously conclude on the detection of methane in the atmosphere of HR 8799b. We note that a lower abundance of methane would suggest a more efficient vertical mixing with an eddy diffusion coefficient $\mathrm{log}(K_{\mathrm{zz}})\ge7$.

Then, we measured the radial velocity of HR 8799d and revisited the estimates of HR 8799b and c from \citet{Ruffio2019RV} using an improved data model to better correct for systematics. The RV of the three planets are consistent with predictions from astrometric monitoring of the system as well as coplanar and stable orbits \citep{Wang2018}. We also showed how the planetary RVs allow better constraints on the orientation of the orbital plane of the system and the eccentricity of HR 8799b, c ,and d.

Finally, we fit a grid of atmosphere model with varying effective temperature, surface gravity, and C/O. Unfortunately, such measurements are still dominated by modeling uncertainties that are difficult to estimate, but they rule out significant deviations from stellar C/O for the three planets. 
A stellar C/O does not point to a specific formation scenario as both gravitational instability or core/pebble accretion are consistent with such value. However, if core accretion is the mechanism by which these planets formed, then a significant planetesimal enrichment is likely.

\vspace{5mm}
\acknowledgments
{\bf Acknowledgements:}
We would like to thank Karin {{\"O}berg} for her helpful discussions about planet formation.
J.-B. R. acknowledges support from the David and Ellen Lee Prize Postdoctoral Fellowship.
The research was supported by grants from NSF, including AST-1411868 (J.-B.R., B.M.) and 1614492 (T.S.B.).
Material presented in this work is supported by the National Aeronautics and Space Administration under Grants/Contracts/Agreements No. NNX17AB63G (Q.M.K., T.S.B., and K.K.W.) issued through the Astrophysics Division of the Science Mission Directorate and NNX15AD95G (J.-B.R., R.J.D.R.). Any opinions, findings, and conclusions or recommendations expressed in this work are those of the author(s) and do not necessarily reflect the views of the National Aeronautics and Space Administration.
This work benefited from NASA’s Nexus for Exoplanet System Science (NExSS) research coordination network sponsored by NASA’s Science Mission Directorate.
This work made use of the sky background models made available by the Gemini observatory \footnote{\url{http://www.gemini.edu/sciops/telescopes-and-sites/observing-condition-constraints/ir-background-spectra}}. 
This research has made use of the SVO Filter Profile Service (http://svo2.cab.inta-csic.es/theory/fps/) supported from the Spanish MINECO through grant AYA2017-84089.
The W. M. Keck Observatory is operated as a scientific partnership among the California Institute of Technology, the University of California, and NASA. The Keck Observatory was made possible by the generous financial support of the W. M. Keck Foundation. We also wish to recognize the very important cultural role and reverence that the summit of Mauna Kea has always had within the indigenous Hawaiian community. We are most fortunate to have the opportunity to conduct observations from this mountain.

\vspace{5mm}
\facilities{KeckI (OSIRIS)}

\software{astropy\footnote{\url{http://www.astropy.org}} \citep{Astropy2013}, Matplotlib\footnote{\url{https://matplotlib.org}} \citep{Hunter2007},
orbitize!\footnote{\url{https://github.com/sblunt/orbitize}} \citep{blunt2019},
ptemcee\footnote{\url{https://github.com/willvousden/ptemcee}} \citep{Vousden2016,ForemanMackey2013}, \texttt{whereistheplanet}\footnote{\url{https://www.whereistheplanet.com/}} \citep{whereistheplanet}
}

\clearpage
\appendix

\section{Observations Summary Table}
\label{app:apposirisobs}

\begin{figure}
  \centering
  \includegraphics[width=1\linewidth]{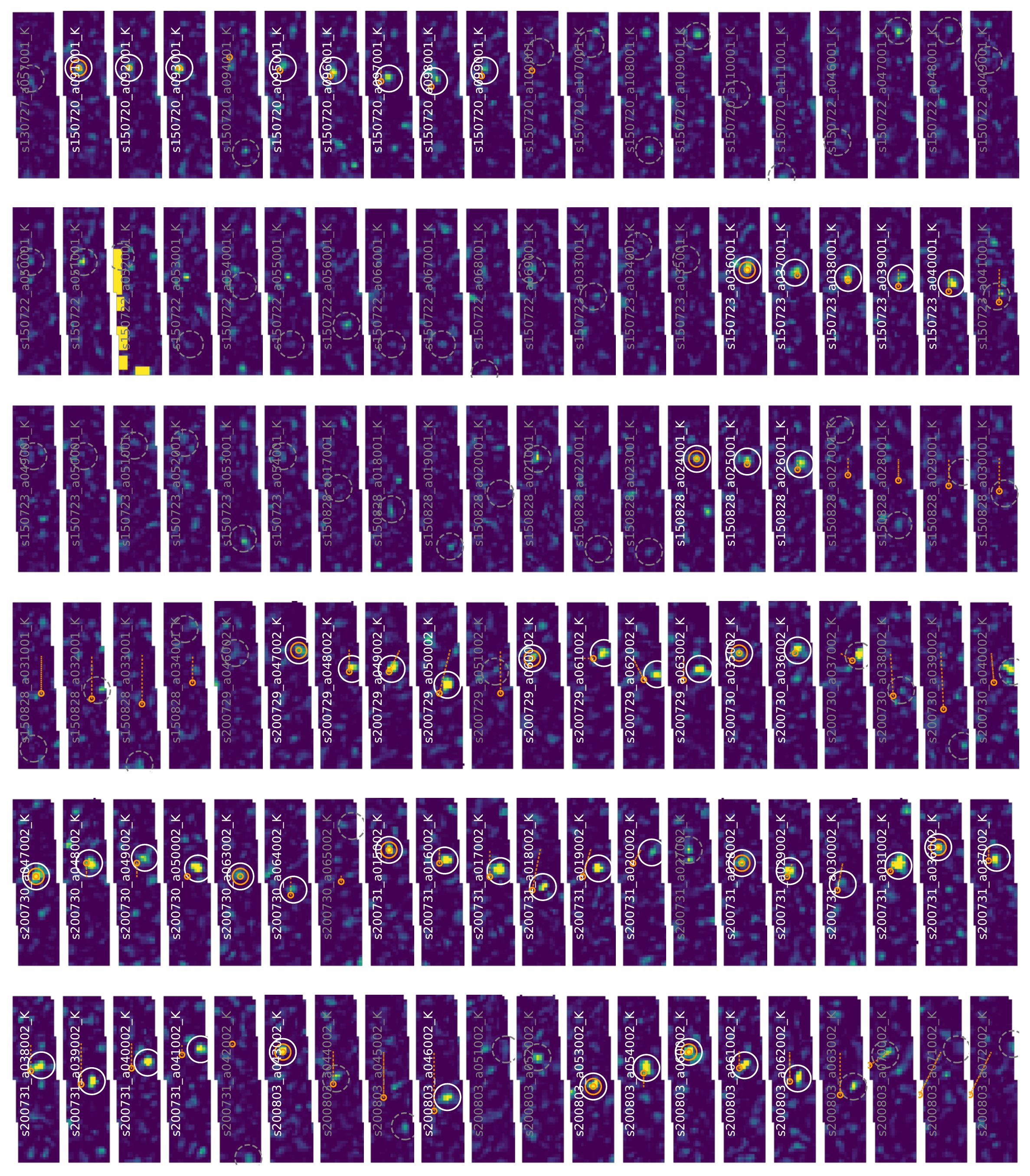}
  \caption{HR 8799 d S/N maps for each $10\,\mathrm{min}$ exposures. The data model only includes the companion and the starlight and 10 principal components. Detections are circled in white. The grey dark circles highlight the brightest pixel in the image when there are no detections. The orange marker indicates the expected location of the planet based on the first detection in the sequence and the offsets that are saved in the fits file headers by the instrument. The offsets have a precision of only a few pixels, but they can still be used as a visual guide.}
  \label{fig:mozaickl10}
\end{figure}

The observing nights summary can be found in \autoref{tab:obssummary} for the science and in \autoref{tab:standardstarsummary1}-\ref{tab:standardstarsummary2} for the telluric standards.
\autoref{fig:mozaickl10} show S/N maps of HR 8799 d at the expected RV of the star for each exposure. These detection maps are computed according to \autoref{sec:plmoldetec}.

\begin{table}[ht]
  \centering
\begin{tabular}{ |c|ccccccc|c| } 
 \hline
  Planet & date & band & Cubes & time & Skies & RV  & RV  & Notes \\ 
  &  &  &  & (h) &  & mean  & residual &  \\ 
  &  &  &  &  &  & offset  & error &  \\ 
  &  &  &  &  &  &  (km/s) & (km/s) &  \\ 
 \hline\hline
\multirow{15}{*}{b} 
 & 2009-07-22 & Kbb & 12 & 2.8 & 1 & 3.1 & 4 & Not used for RV \\ 
 & 2009-07-23 & Hbb & 6 & 1.5 & 1 & 3.3 & 4 & Not used for RV  \\ 
 & 2009-07-30 & Hbb & 8 & 2.0 & 3 & 4.9 & 4 & No detection.  \\ 
 &  &  &  &  &  &  &  & Cooling issue.  \\ 
 & 2009-09-03 & Hbb & 12 & 2.8 & 1 & -5.3 & 4 &  No detection. \\ 
 & 2010-07-11 & Kbb & 9 & 2.2 & 2 & 6.5 & 2.2 &   \\ 
 & 2010-07-12 & Kbb & 9 & 2.2 & 1 & 5.3 & 2.2 &   \\ 
 & 2010-07-13 & Hbb & 2 & 0.5 & 1 & 2.9 & 1.1 &   \\ 
 & 2013-07-25 & Kbb & 16 & 2.7 & 2 & 2.7 & 1.6 &   \\ 
 & 2013-07-26 & Kbb & 9 & 1.5 & 3 & 3.4 & 1.6 &   \\ 
 & 2013-07-27 & Kbb & 7 & 1.2 & 2 & 3.1 & 1.6 &   \\ 
 & 2016-11-06 & Kbb & 2 & 0.3 & 2 & 6.9 & 0.6 &   \\ 
 & 2016-11-07 & Kbb & 3 & 0.5 & 2 & 6.9 & 0.6 &   \\ 
 & 2016-11-08 & Kbb & 1 & 0.2 & 1 & 6.9 & 0.6 &   \\ 
 & 2018-07-22 & Kbb & 6 & 0.5 & 1 & -0.8 & 1.0 & 35mas platescale \\
 & 2020-08-03 & Kbb & 2 & 0.3 &  3 & 27.9 & 2 &  \\ 
\cline{2-6}
\hline
\hline
 \multirow{13}{*}{c}
 & 2010-07-15 & Kbb & 17 & 2.8 & 3 & 5.7 & 2.2 &   \\ 
 & 2010-10-28 & Hbb & 5 & 0.8 & 1 & 2.0 & 1.1 &  No detection. \\ 
 & 2010-11-04 & Hbb & 12 & 2.0 & 2 & 2.9 & 1.1 &   \\ 
 &  & Kbb & 8 & 1.3 & 2 & 5.7 & 2.2 &   \\ 
 & 2011-07-23 & Kbb & 12 & 2.0 & 3 & 3.4 & 2.9 &   \\ 
 & 2011-07-24 & Hbb & 12 & 2.0 & 3 & 0.8 & 1.5 &   \\ 
 &  & Kbb & 3 & 0.5 & 1 & 3.4 & 2.9 &   \\ 
 & 2011-07-25 & Hbb & 14 & 2.3 & 3 & 0.8 & 1.5 &   \\ 
 &  & Kbb & 2 & 0.3 & 2 & 3.4 & 2.9 &   \\ 
 & 2013-07-26 & Kbb & 1 & 0.2 & 3 & 3.4 & 1.6 &   \\ 
 & 2017-11-03 & Hbb & 14 & 2.3 & 3 & -11.4 & 0.8 &   \\ 
 &  & Kbb & 3 & 0.5 & 2 & -13.4 & 1.0 &   \\ 
 & 2020-07-29 & Kbb & 4 & 0.7 &  3 & 27.1 & 2 &    \\ 
\cline{2-6}
\hline
\hline
 \multirow{9}{*}{d}
 & 2013-07-27 & Kbb & 1 & 0.2 &  2 & 3.1 & 1.6 & No detection.  \\ 
 & 2015-07-20 & Kbb & 15 & 2.5 & 4 & -2.7 & 1.6 &   \\ 
 & 2015-07-22 & Kbb & 15 & 2.5 & 4 & -2.3 & 1.6 &  No detection.  \\ 
 & 2015-07-23 & Kbb & 15 & 2.5 & 4 & -2.7 & 1.6 &   \\ 
 & 2015-08-28 & Kbb & 18 & 3.0 & 2 & -1.9 & 1.6 &   \\ 
 & 2020-07-29 & Kbb & 10 & 1.7 & 3 & 27.1 & 2 &   \\ 
 & 2020-07-30 & Kbb & 13 & 2.2 & 3 & 26.7 & 2 &   \\ 
 & 2020-07-31 & Kbb & 18 & 3.0 & 3 & 26.7 & 2 &   \\ 
 & 2020-08-03 & Kbb & 15 & 2.5 & 3 & 27.9 & 2 &   \\ 
\hline
\end{tabular}
\caption{Night-by-night summary of the HR 8799 observations with OSIRIS/Keck. The RV mean offset corresponds to the spatially averaged wavelength calibration offset calculated from the $\mathrm{OH}^-$ emission lines within a given night. The RV residual error is an estimated error of the spatially dependent wavelength offset relative to the spatial average \citep[see][]{Ruffio2019RV}.}
\label{tab:obssummary}
\end{table}

\begin{table}[ht]
  \centering
\begin{tabular}{ |cccccc| } 
 \hline
  date & star name & band & Cubes & Int. time & Mode\\ 
  &  &  &  & (s) &   \\ 
 \hline\hline
2009-07-22 & BD+14 4774 & Kbb & 3 & 30.0 & PSF  \\ 
  & HD 210501 & Kbb & 3 & 30.0 & flux only \\ 
2009-07-23 & BD+14 4774 & Hbb & 4 & 30.0 & PSF  \\ 
  & HD 210501 & Hbb & 6 & 30.0 & flux only \\ 
2009-07-30 & BD+14 4774 & Hbb & 1 & 30.0 & PSF  \\ 
  & BD+14 4774 & Hbb & 2 & 30.0 & flux only \\ 
2009-09-03 & BD+14 4774 & Hbb & 3 & 30.0 & PSF  \\ 
  & HD 7215 & Hbb & 3 & 3.0 & PSF \\ 
2010-07-11 & HD 210501 & Kbb & 5 & 10.0 & PSF  \\ 
  & HIP 1123 & Kbb & 3 & 2.0 & PSF \\ 
  & HR 8799 & Kbb & 2 & 2.0 & PSF \\ 
2010-07-12 & HD 210501 & Kbb & 3 & 30.0 & PSF  \\ 
  & HIP 1123 & Kbb & 3 & 2.0 & PSF \\ 
2010-07-13 & HIP 1123 & Hbb & 3 & 2.0 & PSF  \\ 
  & HR 8799 & Hbb & 2 & 2.0 & PSF \\ 
2010-07-15 & HD 210501 & Kbb & 2 & 30.0 & PSF  \\ 
  & HIP 1123 & Kbb & 3 & 2.0 & PSF \\ 
2010-10-28 & HD 210501 & Hbb & 3 & 30.0 & PSF  \\ 
2010-11-04 & HD 210501 & Hbb & 4 & 30.0 & PSF  \\ 
  & HIP 1123 & Hbb & 3 & 2.0 & PSF \\ 
  & HIP 116886 & Hbb & 3 & 30.0 & PSF \\ 
  & HD 210501 & Kbb & 6 & 30.0 & PSF \\ 
  & HIP 1123 & Kbb & 3 & 2.0 & PSF \\ 
2011-07-23 & HD 210501 & Kbb & 3 & 30.0 & PSF  \\ 
  & HIP 1123 & Kbb & 3 & 2.0 & PSF \\ 
2011-07-24 & HD 210501 & Hbb & 2 & 15.0 & PSF  \\ 
  & HIP 1123 & Hbb & 2 & 2.0 & PSF \\ 
  & HIP 1123 & Kbb & 2 & 2.0 & PSF \\ 
2011-07-25 & HD 210501 & Hbb & 3 & 5.0 & PSF  \\ 
  & HIP 1123 & Hbb & 3 & 2.0 & PSF \\ 
  & HD 210501 & Kbb & 3 & 5.0 & PSF \\ 
  & HIP 1123 & Kbb & 3 & 2.0 & PSF \\ 
2013-07-25 & HD 210501 & Kbb & 5 & 10.0 & PSF  \\ 
  & HD 210501 & Kbb & 1 & 10.0 & flux only \\ 
  & HR 8799 & Kbb & 3 & 2.0 & flux only \\ 
2013-07-26 & HD 210501 & Kbb & 4 & 10.0 & PSF  \\ 
  & HR 8799 & Kbb & 4 & 10.0 & flux only \\ 
2013-07-27 & HD 210501 & Kbb & 4 & 10.0 & PSF  \\ 
  & HR 8799 & Kbb & 3 & 10.0 & PSF \\ 
  & HR 8799 & Kbb & 3 & 10.0 & flux only \\ 
\hline
\end{tabular}
\caption{ To be continued in \autoref{tab:standardstarsummary2}. }
\label{tab:standardstarsummary1}
\end{table}

\begin{table}[ht]
  \centering
\begin{tabular}{ |cccccc| } 
 \hline
  date & star name & band & Cubes & Int. time & Mode\\ 
  &  &  &  & (s) &   \\ 
 \hline\hline
2015-07-20 & HD 210501 & Kbb & 6 & 5.0 & PSF  \\ 
2015-07-22 & HD 210501 & Kbb & 9 & 5.0 & PSF  \\ 
2015-07-23 & HD 210501 & Kbb & 6 & 5.0 & PSF  \\ 
2015-08-28 & HD 210501 & Kbb & 4 & 5.0 & PSF  \\ 
  & HR 8799 & Kbb & 3 & 2.0 & flux only \\ 
2016-11-06 & HD 210501 & Kbb & 4 & 2.0 & PSF  \\ 
2016-11-07 & HD 210501 & Kbb & 3 & 2.0 & PSF  \\ 
  & HD 210501 & Kbb & 1 & 2.0 & flux only \\ 
2016-11-08 & HD 210501 & Kbb & 3 & 2.0 & PSF  \\ 
2017-11-03 & HD 210501 & Hbb & 1 & 2.0 & PSF  \\ 
  & HD 210501 & Hbb & 6 & 2.0 & flux only \\ 
  & HD 210501 & Kbb & 4 & 2.0 & PSF \\ 
2018-07-22 & HD 210501 & Kbb & 3 & 2.0 & PSF  \\ 
2020-07-29 & HD 210501 & Kbb & 5 & 2.0 & PSF  \\ 
  & HR 8799 & Kbb & 8 & 2.0 & PSF \\ 
2020-07-30 & HR 8799 & Kbb & 9 & 2.0 & PSF  \\ 
2020-07-31 & HR 8799 & Kbb & 15 & 2.0 & PSF  \\ 
2020-08-03 & HR 8799 & Kbb & 18 & 2.0 & PSF  \\ 
\hline
\end{tabular}
\caption{End of \autoref{tab:standardstarsummary1}. List of standard star observations used in this work. The quoted integration time is the one of a single exposure. The last column indicates of the data was used to calibrate the PSF and the transmission (PSF) or solely the transmission (flux only). }
\label{tab:standardstarsummary2}
\end{table}

\clearpage

\section{Noise Analysis}
\label{app:noiseanalysis}

In this appendix, we explore the noise budget for OSIRIS and our forward model at the location of HR 8799 d. We use a single exposure from July 31, 2020 (\textit{s200731\_a016002.fits}) with a good detection of HR 8799 d. We recognize the approximate nature of this analysis, and only aim at identifying the dominant source of noise.

\begin{figure}[bth]
\fig{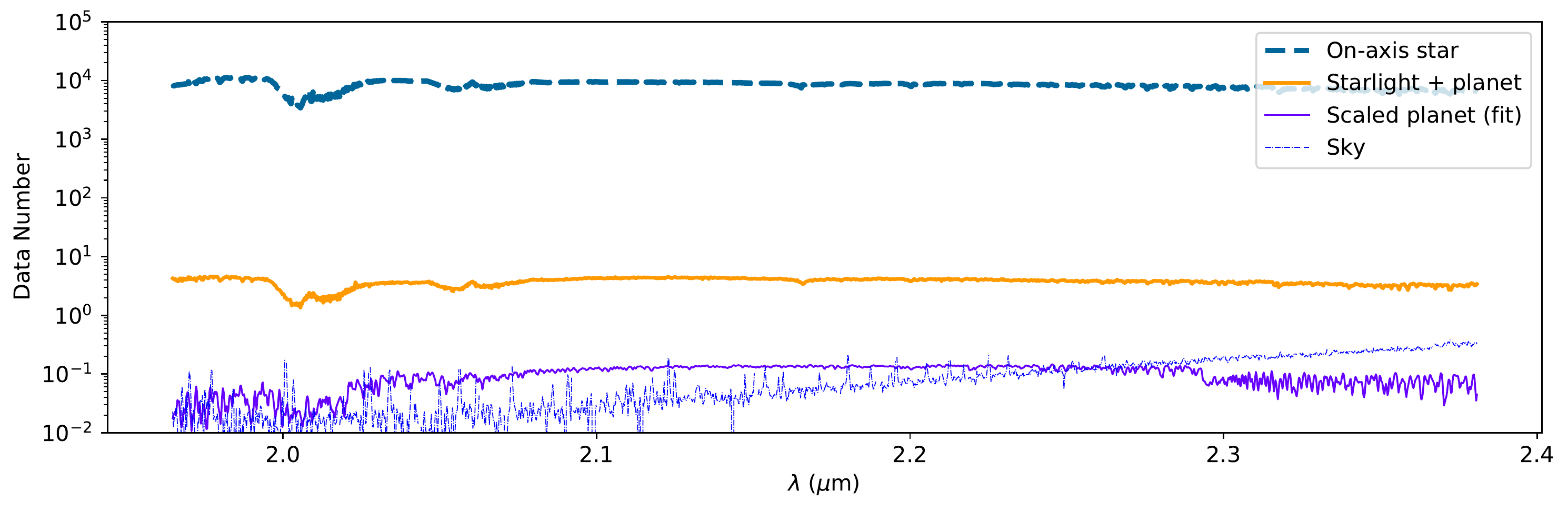}{1\textwidth}{(a) Flux levels}
\fig{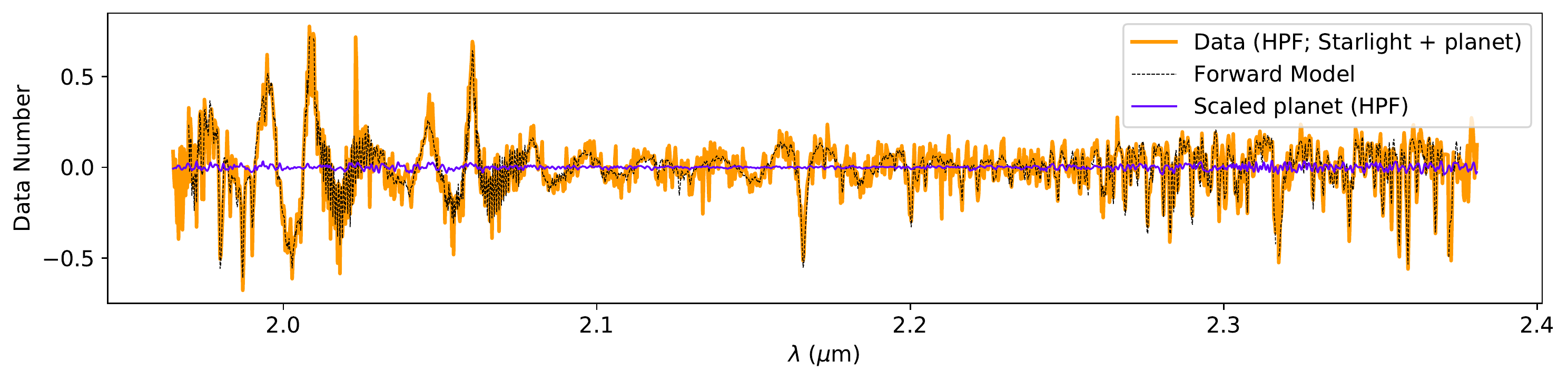}{1\textwidth}{(b) Forward Modeling}
\fig{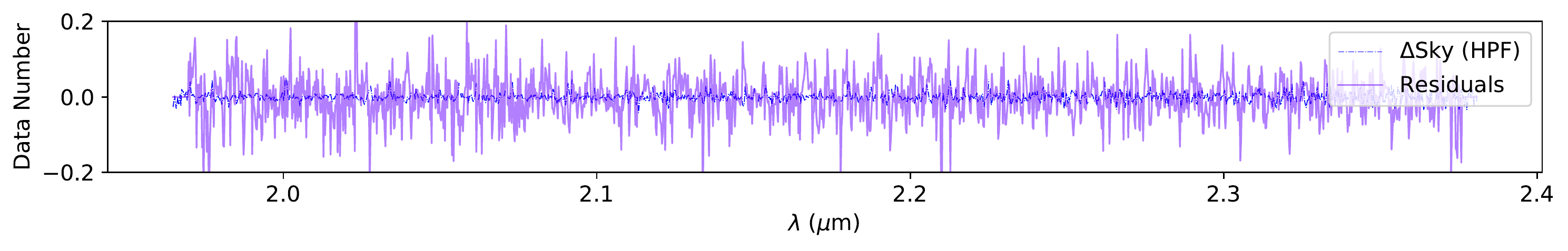}{1\textwidth}{(c) Forward Modeling - Residuals}
\caption{Starlight suppression and noise sources for a spaxel colocated with HR 8799 d. 
(a) The top panel shows the intensity of the on-axis star, the starlight at the location of the planet (i.e., data), the estimated amplitude of the planet signal, and the sky and instrument background. 
(b) Comparison of the high-pass filtered (HPF) data to the forward model; the planet-only component of the model is also shown.
(c) Residuals of the forward model compared to the pixel-to-pixel noise contributions of the dark and the sky background. The right sub-panel is a zoomed-in portion of the left panel around the CO band-head.
All the values are in data number per spaxel in a 10-minute OSIRIS spectral data cube as generated by the OSIRIS data reduction pipeline \citep{Krabbe2004,Lyke2017}.}
\label{fig:noiseanalysis}
\end{figure}

\begin{figure}
  \centering
  \includegraphics[width=0.5\linewidth]{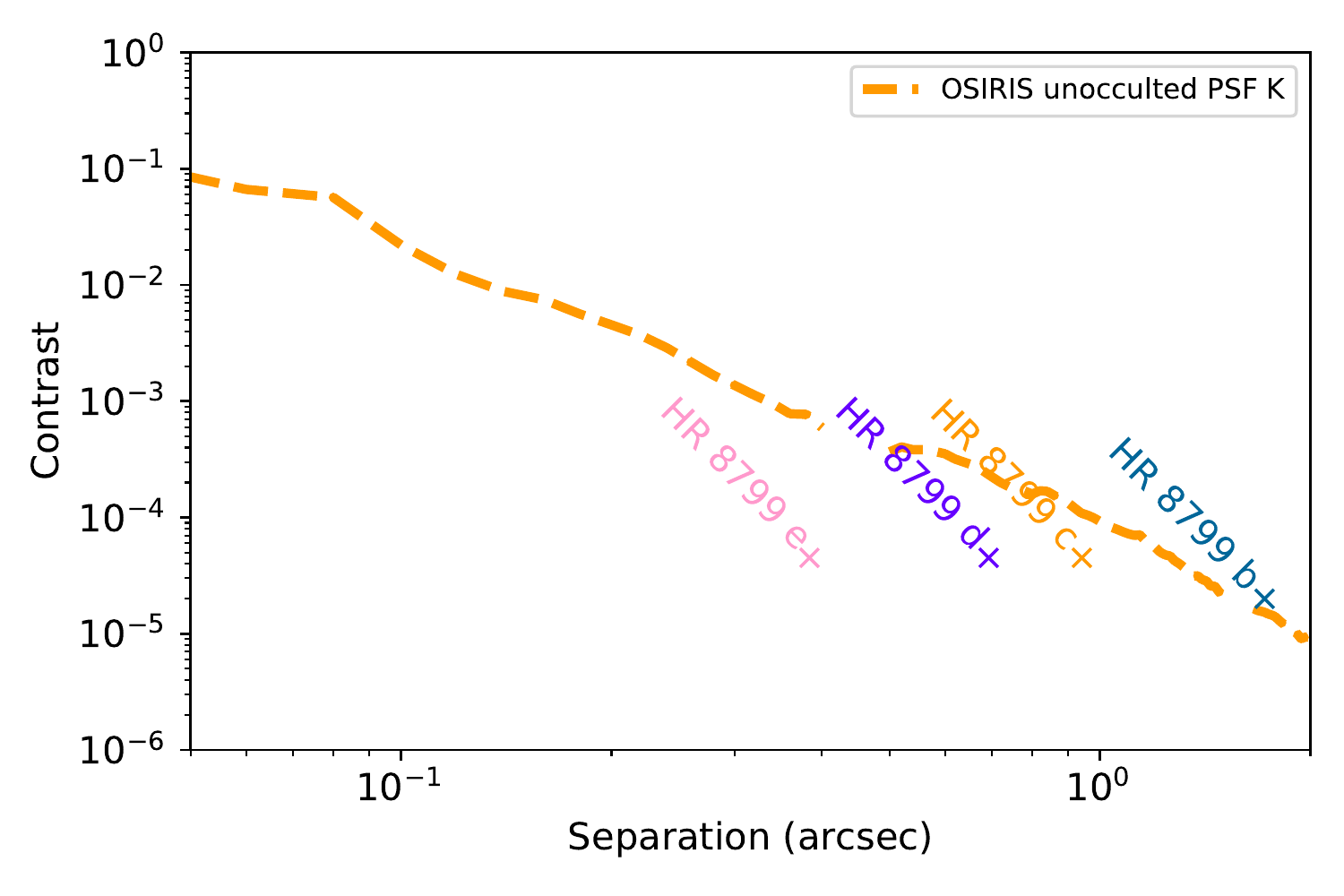}
  \caption{OSIRIS point spread function profile compared to contrast of the HR 8799 planets.}
  \label{fig:osirispsf}
\end{figure}

\begin{figure}
  \centering
  \includegraphics[width=1\linewidth]{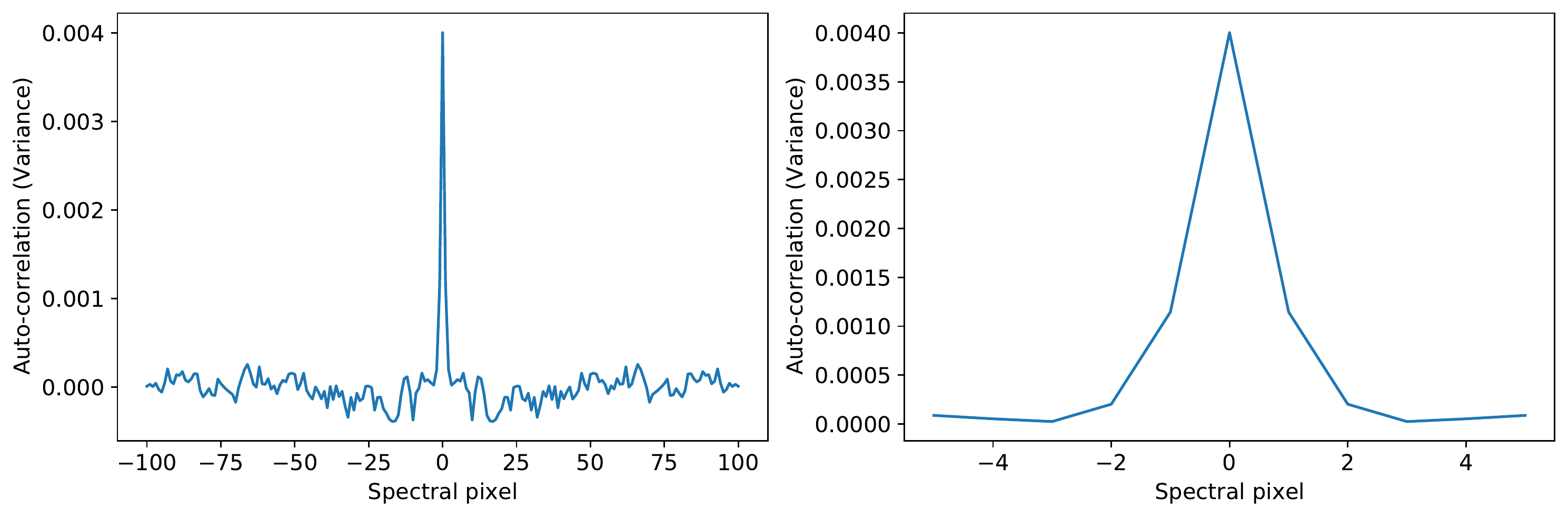}
  \caption{Autocorrelation of the residuals shown in \autoref{fig:noiseanalysis} (d). The right panel is a close-up of the central peak.}
  \label{fig:resautocorr}
\end{figure}

\autoref{fig:noiseanalysis} illustrates the starlight suppression and noise sources for the spaxel at the location of the planet. In this particular example, the forward model is able to detect a planet that is $\sim8$ times fainter than the starlight at the same location in only 10 minutes. We provide the OSIRIS PSF profile from the same night in \autoref{fig:osirispsf}. 
The bottom panel of \autoref{fig:noiseanalysis} shows that the forward model is a good fit of the data with a standard deviation of the residuals $\sigma_{\mathrm{res}}=0.06\,\mathrm{DN}$, and that the background noise ($\sigma_{\mathrm{bg}}=0.01\,\mathrm{DN}$) is negligible in this case. The background noise is estimated by taking the high-pass filtered difference of two sky cubes.
The spectral bin S/N for the planet continuum component of the signal is $\sim2$. This S/N is calculated by dividing the median value of the scaled planet model by the standard deviation of the residuals. The similar S/N for the starlight at the location of the planet is $\sim60$. The autocorrelation of the residuals plotted in \autoref{fig:resautocorr} suggests that there is a significant correlation between neighboring pixels, which could be explained by the inteprolations in the OSIRIS data reduction pipeline. The level of the residuals is consistent with being photon noise limited.

\section{Corner plots}
\label{app:corners}

\begin{figure}[bth]
\gridline{\fig{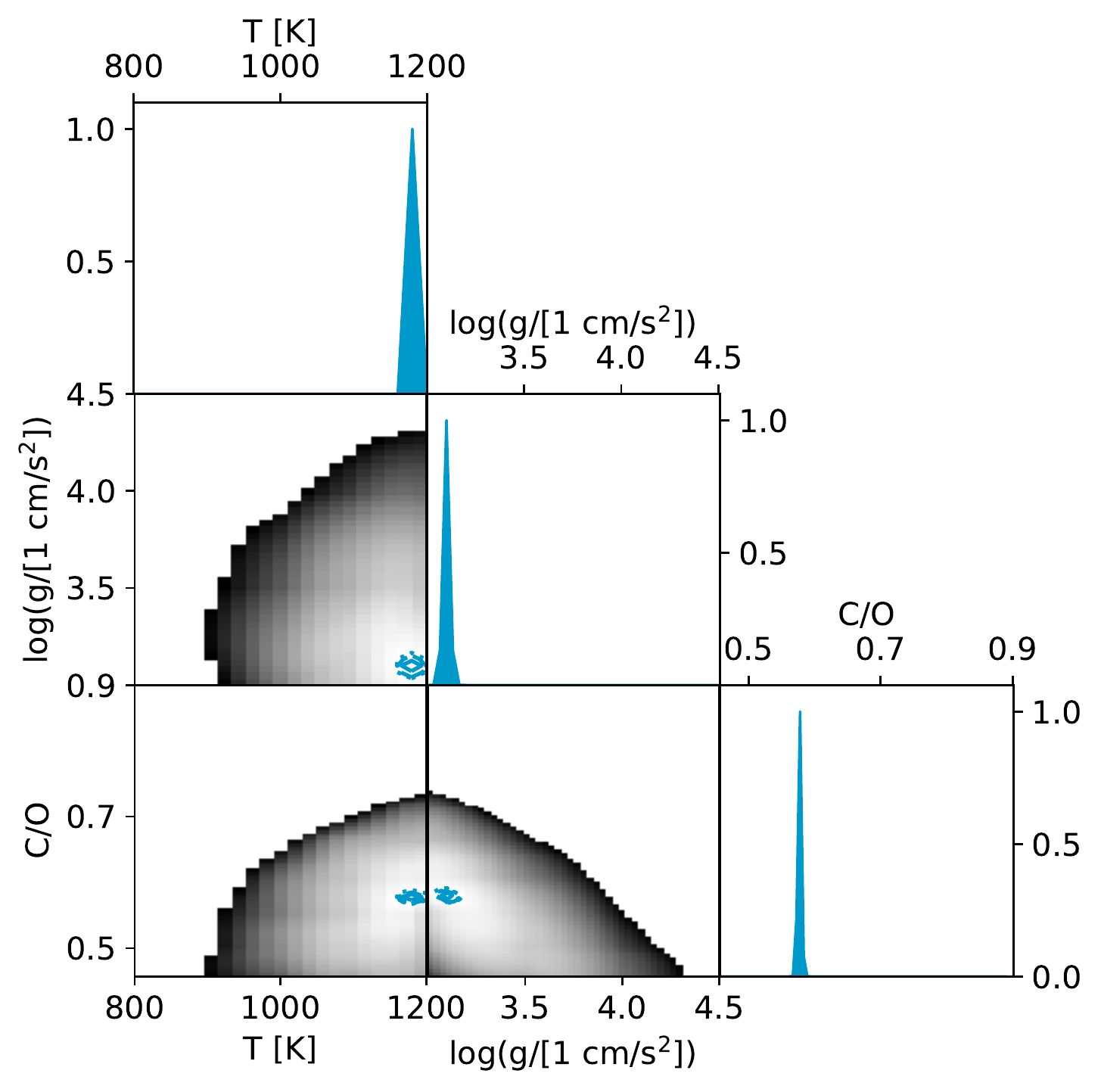}{0.49\textwidth}{(a) HR 8799 b}
		\fig{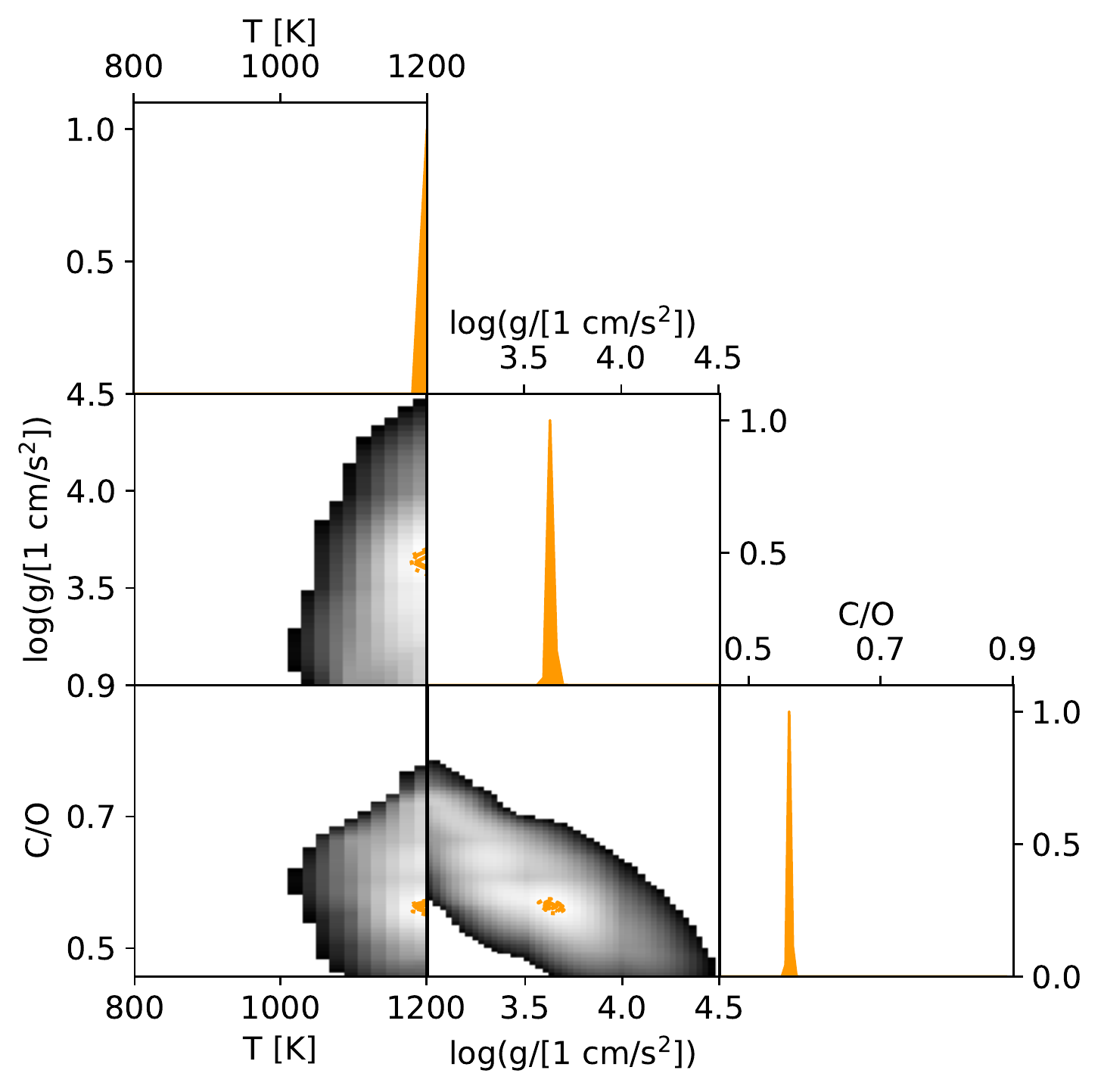}{0.49\textwidth}{(b) HR 8799 c}
          }
\gridline{\fig{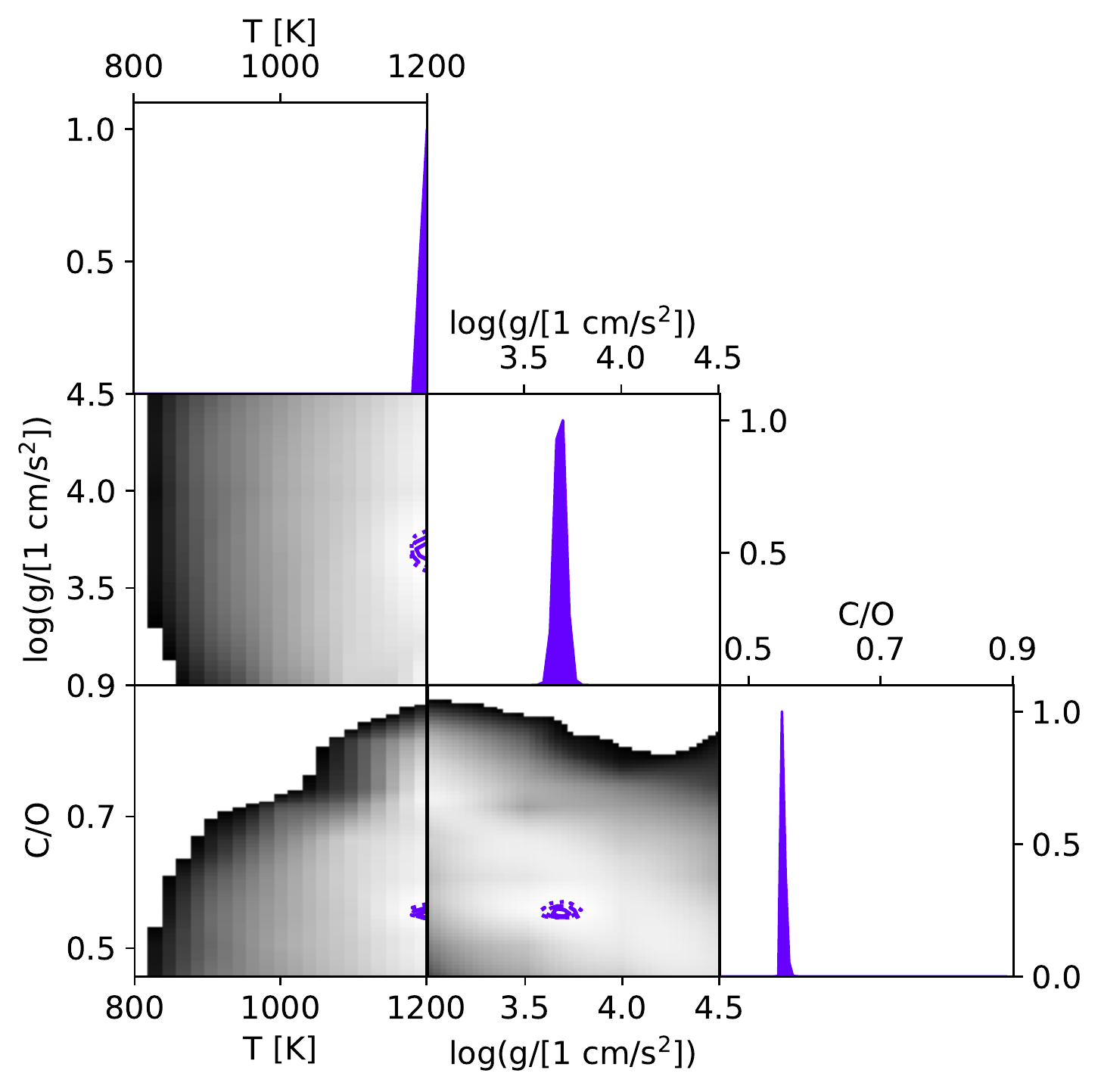}{0.49\textwidth}{(2)  HR 8799 d}
          }
\caption{Marginalized posteriors for $T_{\mathrm{eff}}$, $\mathrm{log}(g)$, and C/O for HR 8799 b, c, and d. The 2D maps represent the logarithm of the posterior probability density and include the contours are the $68\%$, $68\%$, and $99.7\%$ contours.}
\label{fig:hr8799_corners}
\end{figure}

\autoref{fig:hr8799_corners} include the corner plots for HR 8799 b, c, and d resulting from \Secref{sec:atmcharac} and marginalized over the RV of the planet.

\section{HR 8799 methane detection}
\label{app:hr8799bmethane}

In light of the analysis of a subset of our OSIRIS data on HR 8799b (18 of the 49 frames) by \citet{PetitditdelaRoche2018}, we performed a minimal reanalysis of all of our data for this source following the prescription from \citet{Barman2011}, \citet{Konopacky2013}, and \citet{Barman2015}. The primary change is implementation of a simultaneous telluric extraction and absorption as was performed in \citet{Konopacky2013}. We originally did not do this in \citet{Barman2015} because the speckles are significantly fainter at the location of HR 8799b than HR 8799c and we were concerned about adding more noise to our spectrum. We also note that while the CCF was used to help us assess whether a molecule was detected in \citet{Barman2015}, it was not used for any subsequent analysis. The calculation of methane abundance was based entirely on modeling the extracted spectra. 

Our original, data pipeline-processed datacubes for HR 8799b were first multiplied by the original telluric calibration spectrum used to remove the telluric features. Then, following \citet{Konopacky2013}, we extracted a spectrum at every location in the datacube, avoiding only those pixels surrounding HR 8799b. We median combined these spectra to obtain a spectrum of HR 8799A for each cube. We then interpolated over the Br$\gamma$ absorption feature and divided the spectrum by a blackbody with the stellar temperature to get our telluric calibration spectrum. We then divided each cube by its extracted telluric spectrum.

The datacubes were then cleaned for speckles using an identical procedure to \citet{Barman2015}, with our polynomial fitting procedure (note this type of analysis was not used elsewhere in this work). Upon removing the speckles, individual spectra were extracted from each datacube with a 3x3 spaxel box. All spectra were then median combined to obtain our final spectrum for analysis. To compute the CCF, we subtracted a smoothed version of the spectrum from our final spectrum to remove the continuum, as in \citet{Konopacky2013}.

\begin{figure}
  \centering
  \includegraphics[width=0.5\linewidth]{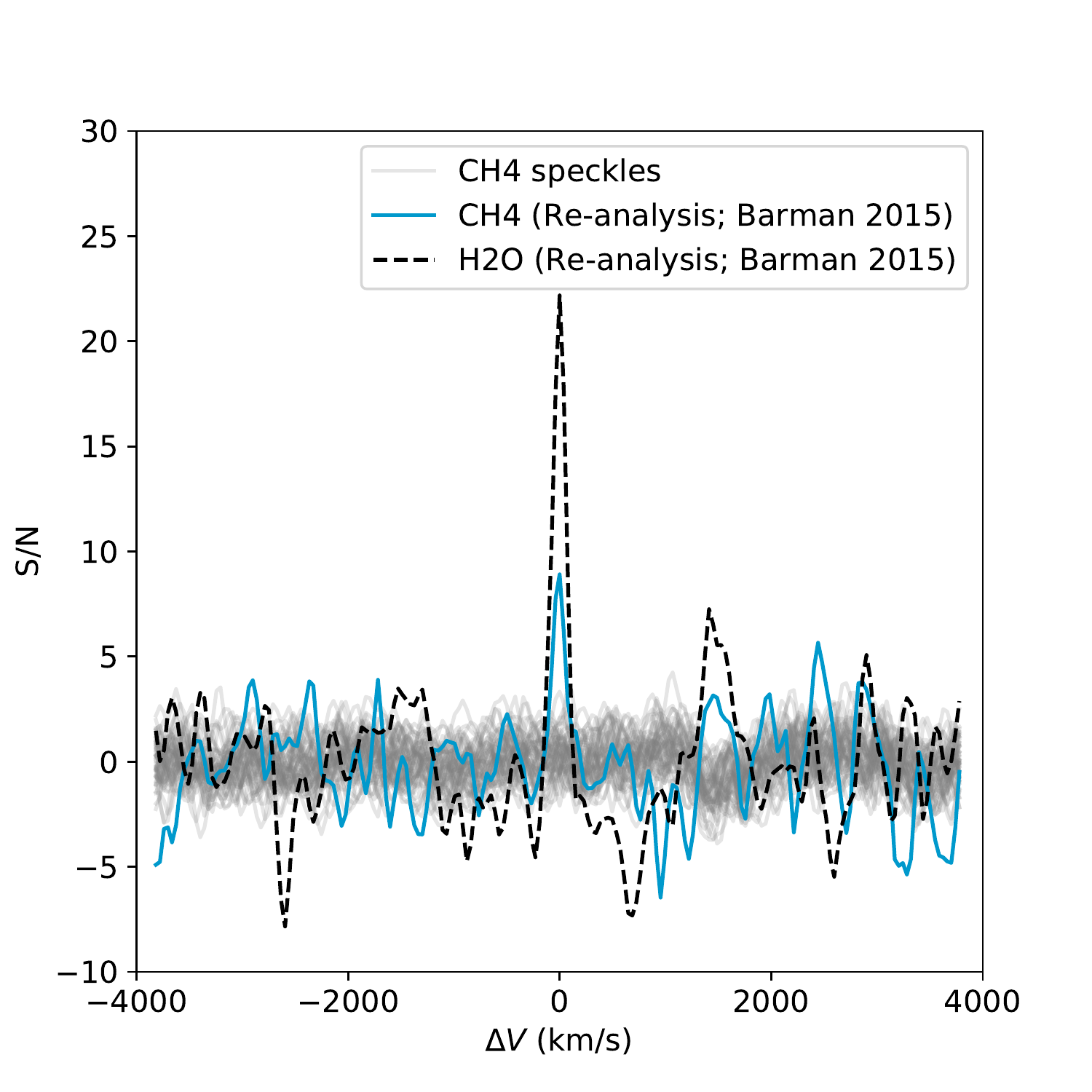}
  \caption{Revisited detection of CH\textsubscript{4} and H\textsubscript{2}O in the atmosphere of HR 8799b with OSIRIS in K band from \citet{Barman2015}. The CH\textsubscript{4} cross correlation function (CCF) of the planet is compared to those of the speckles to estimate the S/N. We do not identify any systematics from the speckle field that could explain the methane signal of HR 8799b.}
  \label{fig:HR8799bmethane}
\end{figure}

We then performed a cross-correlation of our continuum-removed spectrum with a pure methane template at the appropriate temperature and surface gravity. The results are shown in \autoref{fig:HR8799bmethane}, where the resulting detection of methane remains and is stronger than the original detection. To check whether residual tellurics were still playing a role in this detection, we generated 30 different "noise" spectra by randomly choosing different locations in each speckle-processed cube to extract a spectrum from using a 3x3 spaxel box, avoiding the location of the planet. All randomly selected spectra were then combined to generate a final "noise" spectrum. We cross-correlated all 30 spectra with the same methane template, and found no peak in the CCF at zero like we see for HR 8799b. These functions are shown as thin gray lines in \autoref{fig:HR8799bmethane}.

Because our original CCF methodology returned a peak where our newer forward modeling approach did not return a statistically significant detection, we decided to verify the "detection" threshold in each method by using model spectra with known amounts of methane in them. We generated fake planets using these models and a Gaussian flux distribution with a FWHM and peak flux based on the real HR 8799b in the cubes. These were placed in each of our 49 reduced cubes at locations more than 7 spaxels away from the real planet. We tried this with three different models - one with a methane abundance based on our fits to the extracted spectra from \citet{Barman2015}, one with twice that abundance of methane, and one with four times that abundance. We then ran the data cubes through each of our analysis pipelines, one following the method of \citet{Barman2015}, and one following the methods presented in this paper. The peak of the methane CCF in both methods scaled with the relative abundance of methane, with the highest peaks for the 4x abundance of methane. When performing an analysis of the simulated data using the methodology presented in \autoref{sec:ccf}, we detected peaks for the two higher methane models with SNR equal to 8.8 and 12.8. For the model that is similar to the HR 8799b model from \citet{Barman2015}, the peak in the center of the CCF has an SNR of 6.4, slightly higher than the value of 5.2 shown in Figure \autoref{fig:CCFsummary}, which we do not claim as a detection. The difference in these two values is small, and suggests that a small amount of methane in the range of abundances given in \citet{Barman2015} is consistent with the methane CCF in this paper. The mole fraction of methane from fitting the extracted spectrum, including the continuum, was found to be 10$^{-5.06}$ - 10$^{-5.85}$. Given that in our test we used a value in the middle of this range, it is certainly feasible that the abundance of methane is low enough in the spectrum to still be not give a statistically significant detection in our new method, as in Figure \autoref{fig:CCFsummary}. Thus, we conclude that our result is consistent with the results of \citet{Barman2015} and does not preclude the presence of weak methane. Note that the lack of detection of methane in H band was also the case in \citet{Barman2015} and was not inconsistent with our methane mole fraction, as the features are not expected to be stronger in H than in K band.

\clearpage

\newpage
\bibliographystyle{aasjournal}
\bibliography{bibliography}

\begin{thebibliography}{}
\expandafter\ifx\csname natexlab\endcsname\relax\def\natexlab#1{#1}\fi
\providecommand{\url}[1]{\href{#1}{#1}}

\bibitem[{{Allard} {et~al.}(2012){Allard}, {Homeier}, \&
  {Freytag}}]{Allard2012}
{Allard}, F., {Homeier}, D., \& {Freytag}, B. 2012, Philosophical Transactions
  of the Royal Society of London Series A, 370, 2765

\bibitem[{{Astropy Collaboration} {et~al.}(2013){Astropy Collaboration},
  {Robitaille}, {Tollerud}, {Greenfield}, {Droettboom}, {Bray}, {Aldcroft},
  {Davis}, {Ginsburg}, {Price-Whelan}, {Kerzendorf}, {Conley}, {Crighton},
  {Barbary}, {Muna}, {Ferguson}, {Grollier}, {Parikh}, {Nair}, {Unther},
  {Deil}, {Woillez}, {Conseil}, {Kramer}, {Turner}, {Singer}, {Fox}, {Weaver},
  {Zabalza}, {Edwards}, {Azalee Bostroem}, {Burke}, {Casey}, {Crawford},
  {Dencheva}, {Ely}, {Jenness}, {Labrie}, {Lim}, {Pierfederici}, {Pontzen},
  {Ptak}, {Refsdal}, {Servillat}, \& {Streicher}}]{Astropy2013}
{Astropy Collaboration}, {Robitaille}, T.~P., {Tollerud}, E.~J., {et~al.} 2013,
  \aap, 558, A33

\bibitem[{{Baines} {et~al.}(2012){Baines}, {White}, {Huber}, {Jones},
  {Boyajian}, {McAlister}, {ten Brummelaar}, {Turner}, {Sturmann}, {Sturmann},
  {Goldfinger}, {Farrington}, {Riedel}, {Ireland }, {von Braun}, \&
  {Ridgway}}]{Baines2012}
{Baines}, E.~K., {White}, R.~J., {Huber}, D., {et~al.} 2012, \apj, 761, 57

\bibitem[{{Baranne} {et~al.}(1996){Baranne}, {Queloz}, {Mayor}, {Adrianzyk},
  {Knispel}, {Kohler}, {Lacroix}, {Meunier}, {Rimbaud}, \& {Vin}}]{Baranne1996}
{Baranne}, A., {Queloz}, D., {Mayor}, M., {et~al.} 1996, \aaps, 119, 373

\bibitem[{{Barman} {et~al.}(2015){Barman}, {Konopacky}, {Macintosh}, \&
  {Marois}}]{Barman2015}
{Barman}, T.~S., {Konopacky}, Q.~M., {Macintosh}, B., \& {Marois}, C. 2015,
  \apj, 804, 61

\bibitem[{{Barman} {et~al.}(2011{\natexlab{a}}){Barman}, {Macintosh},
  {Konopacky}, \& {Marois}}]{Barman2011}
{Barman}, T.~S., {Macintosh}, B., {Konopacky}, Q.~M., \& {Marois}, C.
  2011{\natexlab{a}}, \apj, 733, 65

\bibitem[{{Barman} {et~al.}(2011{\natexlab{b}}){Barman}, {Macintosh},
  {Konopacky}, \& {Marois}}]{Barman2011b}
---. 2011{\natexlab{b}}, \apjl, 735, L39

\bibitem[{{Bell} {et~al.}(2015){Bell}, {Mamajek}, \& {Naylor}}]{Bell2015}
{Bell}, C.~P.~M., {Mamajek}, E.~E., \& {Naylor}, T. 2015, \mnras, 454, 593

\bibitem[{{Blunt} {et~al.}(2019){Blunt}, {Ngo}, {Wang}, {Angelo}, {Cody}, {De
  Rosa}, {Malena}, {Rubenzahl}, \& Logan}]{blunt2019}
{Blunt}, S., {Ngo}, H., {Wang}, J., {et~al.} 2019, {sblunt/orbitize: API Update
  to Prepare for RV+Astrometry Fits}, , , doi:10.5281/zenodo.3337378.
\newblock \url{https://doi.org/10.5281/zenodo.3337378}

\bibitem[{{Blunt} {et~al.}(2020){Blunt}, {Wang}, {Angelo}, {Ngo}, {Cody}, {De
  Rosa}, {Graham}, {Hirsch}, {Nagpal}, {Nielsen}, {Pearce}, {Rice}, \&
  {Tejada}}]{Blunt2020}
{Blunt}, S., {Wang}, J.~J., {Angelo}, I., {et~al.} 2020, \aj, 159, 89

\bibitem[{{Boehle} {et~al.}(2016){Boehle}, {Larkin}, {Adkins}, {Aliado},
  {Fitzgerald}, {Johnson}, {Lyke}, {Magnone}, {Sohn}, {Wang}, \&
  {Weiss}}]{Boehle2016}
{Boehle}, A., {Larkin}, J.~E., {Adkins}, S.~M., {et~al.} 2016, in Society of
  Photo-Optical Instrumentation Engineers (SPIE) Conference Series, Vol. 9908,
  Ground-based and Airborne Instrumentation for Astronomy VI, 99082Q

\bibitem[{{Bohlin} {et~al.}(2014){Bohlin}, {Gordon}, \&
  {Tremblay}}]{Bohlin2014}
{Bohlin}, R.~C., {Gordon}, K.~D., \& {Tremblay}, P.~E. 2014, \pasp, 126, 711

\bibitem[{{Boley}(2009)}]{Boley2009}
{Boley}, A.~C. 2009, \apjl, 695, L53

\bibitem[{{Bonnefoy} {et~al.}(2016){Bonnefoy}, {Zurlo}, {Baudino}, {Lucas},
  {Mesa}, {Maire}, {Vigan}, {Galicher}, {Homeier}, {Marocco}, {Gratton},
  {Chauvin}, {Allard}, {Desidera}, {Kasper}, {Moutou}, {Lagrange}, {Antichi},
  {Baruffolo}, {Baudrand }, {Beuzit}, {Boccaletti}, {Cantalloube}, {Carbillet},
  {Charton}, {Claudi}, {Costille}, {Dohlen}, {Dominik}, {Fantinel},
  {Feautrier}, {Feldt}, {Fusco}, {Gigan}, {Girard}, {Gluck}, {Gry}, {Henning},
  {Janson}, {Langlois}, {Madec}, {Magnard}, {Maurel}, {Mawet}, {Meyer},
  {Milli}, {Moeller-Nilsson}, {Mouillet}, {Pavlov}, {Perret}, {Pujet}, {Quanz},
  {Rochat}, {Rousset}, {Roux}, {Salasnich}, {Salter}, {Sauvage}, {Schmid},
  {Sevin}, {Soenke}, {Stadler}, {Turatto}, {Udry}, {Vakili}, {Wahhaj}, \&
  {Wildi}}]{Bonnefoy2016}
{Bonnefoy}, M., {Zurlo}, A., {Baudino}, J.~L., {et~al.} 2016, \aap, 587, A58

\bibitem[{{Booth} {et~al.}(2016){Booth}, {Jord{\'a}n}, {Casassus}, {Hales},
  {Dent}, {Faramaz}, {Matr{\`a}}, {Barkats}, {Brahm}, \& {Cuadra}}]{Booth2016}
{Booth}, M., {Jord{\'a}n}, A., {Casassus}, S., {et~al.} 2016, \mnras, 460, L10

\bibitem[{{Booth} {et~al.}(2017){Booth}, {Clarke}, {Madhusudhan}, \&
  {Ilee}}]{Booth2017}
{Booth}, R.~A., {Clarke}, C.~J., {Madhusudhan}, N., \& {Ilee}, J.~D. 2017,
  \mnras, 469, 3994

\bibitem[{{Bosman} {et~al.}(2019){Bosman}, {Cridland}, \&
  {Miguel}}]{Bosman2019}
{Bosman}, A.~D., {Cridland}, A.~J., \& {Miguel}, Y. 2019, \aap, 632, L11

\bibitem[{{Boss}(1997)}]{Boss1997}
{Boss}, A.~P. 1997, Science, 276, 1836

\bibitem[{{Bowler} {et~al.}(2020){Bowler}, {Blunt}, \& {Nielsen}}]{Bowler2020}
{Bowler}, B.~P., {Blunt}, S.~C., \& {Nielsen}, E.~L. 2020, \aj, 159, 63

\bibitem[{{Bowler} {et~al.}(2010){Bowler}, {Liu}, {Dupuy}, \&
  {Cushing}}]{Bowler2010}
{Bowler}, B.~P., {Liu}, M.~C., {Dupuy}, T.~J., \& {Cushing}, M.~C. 2010, \apj,
  723, 850

\bibitem[{{Brogi} {et~al.}(2014){Brogi}, {de Kok}, {Birkby}, {Schwarz}, \&
  {Snellen}}]{Brogi2014}
{Brogi}, M., {de Kok}, R.~J., {Birkby}, J.~L., {Schwarz}, H., \& {Snellen},
  I.~A.~G. 2014, \aap, 565, A124

\bibitem[{{Brogi} \& {Line}(2019)}]{Brogi2019}
{Brogi}, M., \& {Line}, M.~R. 2019, \aj, 157, 114

\bibitem[{{Bryan} {et~al.}(2018){Bryan}, {Benneke}, {Knutson}, {Batygin}, \&
  {Bowler}}]{Bryan2018}
{Bryan}, M.~L., {Benneke}, B., {Knutson}, H.~A., {Batygin}, K., \& {Bowler},
  B.~P. 2018, Nature Astronomy, 2, 138

\bibitem[{{Bryan} {et~al.}(2020){Bryan}, {Chiang}, {Bowler}, {Morley},
  {Millholland}, {Blunt}, {Ashok}, {Nielsen}, {Ngo}, {Mawet}, \&
  {Knutson}}]{Bryan2020}
{Bryan}, M.~L., {Chiang}, E., {Bowler}, B.~P., {et~al.} 2020, \aj, 159, 181

\bibitem[{{Cale} {et~al.}(2019){Cale}, {Plavchan}, {LeBrun}, {Gagn{\'e}},
  {Gao}, {Tanner}, {Beichman}, {Xuesong Wang}, {Gaidos}, {Teske}, {Ciardi},
  {Vasisht}, {Kane}, \& {von Braun}}]{Cale2019}
{Cale}, B., {Plavchan}, P., {LeBrun}, D., {et~al.} 2019, \aj, 158, 170

\bibitem[{{Contro} {et~al.}(2016){Contro}, {Horner}, {Wittenmyer}, {Marshall},
  \& {Hinse}}]{Contro2016}
{Contro}, B., {Horner}, J., {Wittenmyer}, R.~A., {Marshall}, J.~P., \& {Hinse},
  T.~C. 2016, \mnras, 463, 191

\bibitem[{{Cridland} {et~al.}(2017){Cridland}, {Pudritz}, {Birnstiel},
  {Cleeves}, \& {Bergin}}]{Cridland2017}
{Cridland}, A.~J., {Pudritz}, R.~E., {Birnstiel}, T., {Cleeves}, L.~I., \&
  {Bergin}, E.~A. 2017, \mnras, 469, 3910

\bibitem[{{Crossfield}(2014)}]{Crossfield2014}
{Crossfield}, I. J.~M. 2014, \aap, 566, A130

\bibitem[{{Crossfield} {et~al.}(2014){Crossfield}, {Biller}, {Schlieder},
  {Deacon}, {Bonnefoy}, {Homeier}, {Allard}, {Buenzli}, {Henning}, {Brandner},
  {Goldman}, \& {Kopytova}}]{Crossfield2014Nature}
{Crossfield}, I.~J.~M., {Biller}, B., {Schlieder}, J.~E., {et~al.} 2014, \nat,
  505, 654

\bibitem[{{Currie} {et~al.}(2011){Currie}, {Burrows}, {Itoh}, {Matsumura},
  {Fukagawa}, {Apai}, {Madhusudhan}, {Hinz}, {Rodigas}, {Kasper}, {Pyo}, \&
  {Ogino}}]{Currie2011}
{Currie}, T., {Burrows}, A., {Itoh}, Y., {et~al.} 2011, \apj, 729, 128

\bibitem[{{Currie} {et~al.}(2014){Currie}, {Burrows}, {Girard}, {Cloutier},
  {Fukagawa}, {Sorahana}, {Kuchner}, {Kenyon}, {Madhusudhan}, {Itoh},
  {Jayawardhana}, {Matsumura}, \& {Pyo}}]{Currie2014}
{Currie}, T., {Burrows}, A., {Girard}, J.~H., {et~al.} 2014, \apj, 795, 133

\bibitem[{{Czekala} {et~al.}(2015){Czekala}, {Andrews}, {Mandel}, {Hogg}, \&
  {Green}}]{Czekala2015}
{Czekala}, I., {Andrews}, S.~M., {Mandel}, K.~S., {Hogg}, D.~W., \& {Green},
  G.~M. 2015, \apj, 812, 128

\bibitem[{{Czekala} {et~al.}(2017){Czekala}, {Mandel}, {Andrews}, {Dittmann},
  {Ghosh}, {Montet}, \& {Newton}}]{Czekala2017}
{Czekala}, I., {Mandel}, K.~S., {Andrews}, S.~M., {et~al.} 2017, \apj, 840, 49

\bibitem[{{Fabrycky} \& {Murray-Clay}(2010)}]{Fabrycky2010}
{Fabrycky}, D.~C., \& {Murray-Clay}, R.~A. 2010, \apj, 710, 1408

\bibitem[{{Foreman-Mackey} {et~al.}(2013){Foreman-Mackey}, {Hogg}, {Lang}, \&
  {Goodman}}]{ForemanMackey2013}
{Foreman-Mackey}, D., {Hogg}, D.~W., {Lang}, D., \& {Goodman}, J. 2013, \pasp,
  125, 306

\bibitem[{{Gaia Collaboration}(2018)}]{Gaia2018}
{Gaia Collaboration}. 2018, VizieR Online Data Catalog, I/345

\bibitem[{{Galicher} {et~al.}(2011){Galicher}, {Marois}, {Macintosh}, {Barman},
  \& {Konopacky}}]{Galicher2011}
{Galicher}, R., {Marois}, C., {Macintosh}, B., {Barman}, T., \& {Konopacky}, Q.
  2011, \apjl, 739, L41

\bibitem[{{Gao} {et~al.}(2016){Gao}, {Plavchan}, {Gagn{\'e}}, {Furlan},
  {Bottom}, {Anglada-Escud{\'e}}, {White}, {Davison}, {Beichman}, {Brinkworth},
  {Johnson}, {Ciardi}, {Wallace}, {Mennesson}, {von Braun}, {Vasisht}, {Prato},
  {Kane}, {Tanner}, {Crawford}, {Latham}, {Rougeot}, {Geneser}, \&
  {Catanzarite}}]{Gao2016}
{Gao}, P., {Plavchan}, P., {Gagn{\'e}}, J., {et~al.} 2016, \pasp, 128, 104501

\bibitem[{{Geiler} {et~al.}(2019){Geiler}, {Krivov}, {Booth}, \&
  {L{\"o}hne}}]{Geiler2019}
{Geiler}, F., {Krivov}, A.~V., {Booth}, M., \& {L{\"o}hne}, T. 2019, \mnras,
  483, 332

\bibitem[{{Gontcharov}(2006)}]{Gontcharov2006}
{Gontcharov}, G.~A. 2006, Astronomy Letters, 32, 759

\bibitem[{{Go{\'z}dziewski} \& {Migaszewski}(2018)}]{Gozdziewski2018}
{Go{\'z}dziewski}, K., \& {Migaszewski}, C. 2018, \apjs, 238, 6

\bibitem[{{Go{\'z}dziewski} \& {Migaszewski}(2020)}]{Gozdziewski2020}
---. 2020, \apjl, 902, L40

\bibitem[{{Gravity Collaboration} {et~al.}(2019){Gravity Collaboration},
  {Lacour}, {Nowak}, {Wang}, {Pfuhl}, {Eisenhauer}, {Abuter}, {Amorim},
  {Anugu}, {Benisty}, {Berger}, {Beust}, {Blind}, {Bonnefoy}, {Bonnet},
  {Bourget}, {Brandner}, {Buron}, {Collin}, {Charnay}, {Chapron}, {Cl{\'e}net},
  {Coud{\'e} Du Foresto}, {de Zeeuw}, {Deen}, {Dembet}, {Dexter}, {Duvert},
  {Eckart}, {F{\"o}rster Schreiber}, {F{\'e}dou}, {Garcia}, {Garcia Lopez},
  {Gao}, {Gendron}, {Genzel}, {Gillessen}, {Gordo}, {Greenbaum}, {Habibi},
  {Haubois}, {Hau{\ss}mann}, {Henning}, {Hippler}, {Horrobin}, {Hubert},
  {Jimenez Rosales}, {Jocou}, {Kendrew}, {Kervella}, {Kolb}, {Lagrange},
  {Lapeyr{\`e}re}, {Le Bouquin}, {L{\'e}na}, {Lippa}, {Lenzen}, {Maire},
  {Molli{\`e}re}, {Ott}, {Paumard}, {Perraut}, {Perrin}, {Pueyo}, {Rabien},
  {Ram{\'\i}rez}, {Rau}, {Rodr{\'\i}guez-Coira}, {Rousset}, {Sanchez-Bermudez},
  {Scheithauer}, {Schuhler}, {Straub}, {Straubmeier}, {Sturm}, {Tacconi},
  {Vincent}, {van Dishoeck}, {von Fellenberg}, {Wank}, {Waisberg}, {Widmann},
  {Wieprecht}, {Wiest}, {Wiezorrek}, {Woillez}, {Yazici}, {Ziegler}, \&
  {Zins}}]{Gravity2019}
{Gravity Collaboration}, {Lacour}, S., {Nowak}, M., {et~al.} 2019, \aap, 623,
  L11

\bibitem[{{Gravity Collaboration} {et~al.}(2020){Gravity Collaboration},
  {Nowak}, {Lacour}, {Molli{\`e}re}, {Wang}, {Charnay}, {van Dishoeck},
  {Abuter}, {Amorim}, {Berger}, {Beust}, {Bonnefoy}, {Bonnet}, {Brandner},
  {Buron}, {Cantalloube}, {Collin}, {Chapron}, {Cl{\'e}net}, {Coud{\'e} Du
  Foresto}, {de Zeeuw}, {Dembet}, {Dexter}, {Duvert}, {Eckart}, {Eisenhauer},
  {F{\"o}rster Schreiber}, {F{\'e}dou}, {Garcia Lopez}, {Gao}, {Gendron},
  {Genzel}, {Gillessen}, {Hau{\ss}mann}, {Henning}, {Hippler}, {Hubert},
  {Jocou}, {Kervella}, {Lagrange}, {Lapeyr{\`e}re}, {Le Bouquin}, {L{\'e}na},
  {Maire}, {Ott}, {Paumard}, {Paladini}, {Perraut}, {Perrin}, {Pueyo}, {Pfuhl},
  {Rabien}, {Rau}, {Rodr{\'\i}guez-Coira}, {Rousset}, {Scheithauer},
  {Shangguan}, {Straub}, {Straubmeier}, {Sturm}, {Tacconi}, {Vincent},
  {Widmann}, {Wieprecht}, {Wiezorrek}, {Woillez}, {Yazici}, \&
  {Ziegler}}]{Gravity2020}
{Gravity Collaboration}, {Nowak}, M., {Lacour}, S., {et~al.} 2020, \aap, 633,
  A110

\bibitem[{{Gray} \& {Kaye}(1999)}]{Gray1999}
{Gray}, R.~O., \& {Kaye}, A.~B. 1999, \aj, 118, 2993

\bibitem[{{Greenbaum} {et~al.}(2018){Greenbaum}, {Pueyo}, {Ruffio}, {Wang}, {De
  Rosa}, {Aguilar}, {Rameau}, {Barman}, {Marois}, {Marley}, {Konopacky},
  {Rajan}, {Macintosh}, {Ansdell}, {Arriaga}, {Bailey}, {Bulger}, {Burrows},
  {Chilcote}, {Cotten}, {Doyon}, {Duch{\^e}ne}, {Fitzgerald}, {Follette},
  {Gerard}, {Goodsell}, {Graham}, {Hibon}, {Hung}, {Ingraham}, {Kalas},
  {Larkin}, {Maire}, {Marchis}, {Metchev}, {Millar-Blanchaer}, {Nielsen},
  {Norton}, {Oppenheimer}, {Palmer}, {Patience}, {Perrin}, {Poyneer},
  {Rantakyr{\"o}}, {Savransky}, {Schneider}, {Sivaramakrishnan}, {Song},
  {Soummer}, {Thomas}, {Wallace}, {Ward-Duong}, {Wiktorowicz}, \&
  {Wolff}}]{Greenbaum2018}
{Greenbaum}, A.~Z., {Pueyo}, L., {Ruffio}, J.-B., {et~al.} 2018, \aj, 155, 226

\bibitem[{{Hinz} {et~al.}(2010){Hinz}, {Rodigas}, {Kenworthy}, {Sivanandam},
  {Heinze}, {Mamajek}, \& {Meyer}}]{Hinz2010}
{Hinz}, P.~M., {Rodigas}, T.~J., {Kenworthy}, M.~A., {et~al.} 2010, \apj, 716,
  417

\bibitem[{{Hoeijmakers} {et~al.}(2018){Hoeijmakers}, {Ehrenreich}, {Heng},
  {Kitzmann}, {Grimm}, {Allart}, {Deitrick}, {Wyttenbach}, {Oreshenko}, {Pino},
  {Rimmer}, {Molinari}, \& {Di Fabrizio}}]{Hoeijmakers2018}
{Hoeijmakers}, H.~J., {Ehrenreich}, D., {Heng}, K., {et~al.} 2018, \nat, 560,
  453

\bibitem[{{Hughes} {et~al.}(2011){Hughes}, {Wilner}, {Andrews}, {Williams},
  {Su}, {Murray-Clay}, \& {Qi}}]{Hughes2011}
{Hughes}, A.~M., {Wilner}, D.~J., {Andrews}, S.~M., {et~al.} 2011, \apj, 740,
  38

\bibitem[{Hunter(2007)}]{Hunter2007}
Hunter, J.~D. 2007, Computing In Science \& Engineering, 9, 90

\bibitem[{{Husser} {et~al.}(2013){Husser}, {Wende-von Berg}, {Dreizler},
  {Homeier}, {Reiners}, {Barman}, \& {Hauschildt}}]{Husser2013}
{Husser}, T.~O., {Wende-von Berg}, S., {Dreizler}, S., {et~al.} 2013, \aap,
  553, A6

\bibitem[{{Ingraham} {et~al.}(2014){Ingraham}, {Marley}, {Saumon}, {Marois},
  {Macintosh}, {Barman}, {Bauman}, {Burrows}, {Chilcote}, {De Rosa}, {Dillon},
  {Doyon}, {Dunn}, {Erikson}, {Fitzgerald}, {Gavel}, {Goodsell}, {Graham},
  {Hartung}, {Hibon}, {Kalas}, {Konopacky}, {Larkin}, {Maire}, {Marchis},
  {McBride}, {Millar-Blanchaer}, {Morzinski}, {Norton}, {Oppenheimer},
  {Palmer}, {Patience}, {Perrin}, {Poyneer}, {Pueyo}, {Rantakyr{\"o}},
  {Sadakuni}, {Saddlemyer}, {Savransky}, {Soummer}, {Sivaramakrishnan}, {Song},
  {Thomas}, {Wallace}, {Wiktorowicz}, \& {Wolff}}]{Ingraham2014}
{Ingraham}, P., {Marley}, M.~S., {Saumon}, D., {et~al.} 2014, \apj, 794, L15

\bibitem[{{Johansen} \& {Lambrechts}(2017)}]{Johansen17}
{Johansen}, A., \& {Lambrechts}, M. 2017, Annual Review of Earth and Planetary
  Sciences, 45, 359

\bibitem[{{Karkoschka} \& {Tomasko}(2010)}]{Karkoschka2010}
{Karkoschka}, E., \& {Tomasko}, M.~G. 2010, \icarus, 205, 674

\bibitem[{{Konopacky} {et~al.}(2013){Konopacky}, {Barman}, {Macintosh}, \&
  {Marois}}]{Konopacky2013}
{Konopacky}, Q.~M., {Barman}, T.~S., {Macintosh}, B.~A., \& {Marois}, C. 2013,
  Science, 339, 1398

\bibitem[{{Konopacky} {et~al.}(2016){Konopacky}, {Marois}, {Macintosh},
  {Galicher}, {Barman}, {Metchev}, \& {Zuckerman}}]{Konopacky2016}
{Konopacky}, Q.~M., {Marois}, C., {Macintosh}, B.~A., {et~al.} 2016, \aj, 152,
  28

\bibitem[{{Krabbe} {et~al.}(2004){Krabbe}, {Gasaway}, {Song}, {Iserlohe},
  {Weiss}, {Larkin}, {Barczys}, \& {Lafreniere}}]{Krabbe2004}
{Krabbe}, A., {Gasaway}, T., {Song}, I., {et~al.} 2004, in Society of
  Photo-Optical Instrumentation Engineers (SPIE) Conference Series, Vol. 5492,
  Ground-based Instrumentation for Astronomy, ed. A.~F.~M. {Moorwood} \&
  M.~{Iye}, 1403--1410

\bibitem[{{Kratter} {et~al.}(2010){Kratter}, {Murray-Clay}, \&
  {Youdin}}]{Kratter2010}
{Kratter}, K.~M., {Murray-Clay}, R.~A., \& {Youdin}, A.~N. 2010, \apj, 710,
  1375

\bibitem[{{Larkin} {et~al.}(2006){Larkin}, {Barczys}, {Krabbe}, {Adkins},
  {Aliado}, {Amico}, {Brims}, {Campbell}, {Canfield}, {Gasaway}, {Honey},
  {Iserlohe}, {Johnson}, {Kress}, {LaFreniere}, {Magnone}, {Magnone},
  {McElwain}, {Moon}, {Quirrenbach}, {Skulason}, {Song}, {Spencer}, {Weiss}, \&
  {Wright}}]{Larkin2006}
{Larkin}, J., {Barczys}, M., {Krabbe}, A., {et~al.} 2006, New Astronomy
  Reviews, 50, 362

\bibitem[{{Lavie} {et~al.}(2017){Lavie}, {Mendon{\c{c}}a}, {Mordasini},
  {Malik}, {Bonnefoy}, {Demory}, {Oreshenko}, {Grimm}, {Ehrenreich}, \&
  {Heng}}]{Lavie2017}
{Lavie}, B., {Mendon{\c{c}}a}, J.~M., {Mordasini}, C., {et~al.} 2017, \aj, 154,
  91

\bibitem[{{Lee} {et~al.}(2013){Lee}, {Heng}, \& {Irwin}}]{Lee2013}
{Lee}, J.-M., {Heng}, K., \& {Irwin}, P. G.~J. 2013, \apj, 778, 97

\bibitem[{{Lissauer}(1987)}]{Lissauer1987}
{Lissauer}, J.~J. 1987, \icarus, 69, 249

\bibitem[{{Lissauer} \& {Kary}(1991)}]{Lissauer1991}
{Lissauer}, J.~J., \& {Kary}, D.~M. 1991, \icarus, 94, 126

\bibitem[{{Lockhart} {et~al.}(2019){Lockhart}, {Do}, {Larkin}, {Boehle},
  {Campbell}, {Chappell}, {Chu}, {Ciurlo}, {Cosens}, {Fitzgerald}, {Ghez},
  {Lu}, {Lyke}, {Mieda}, {Rudy}, {Vayner}, {Walth}, \& {Wright}}]{Lockhart2019}
{Lockhart}, K.~E., {Do}, T., {Larkin}, J.~E., {et~al.} 2019, \aj, 157, 75

\bibitem[{{Lyke} {et~al.}(2017){Lyke}, {Do}, {Boehle}, {Campbell}, {Chappell},
  {Fitzgerald}, {Gasawy}, {Iserlohe}, {Krabbe}, {Larkin}, {Lockhard}, {Lu},
  {Mieda}, {McElwain}, {Perrin}, {Rudy}, {Sitarski}, {Vayner}, {Walth},
  {Weiss}, {Wizanski}, \& {Wright}}]{Lyke2017}
{Lyke}, J., {Do}, T., {Boehle}, A., {et~al.} 2017, {OSIRIS Toolbox:
  OH-Suppressing InfraRed Imaging Spectrograph pipeline}, , , ascl:1710.021

\bibitem[{{Madhusudhan} {et~al.}(2017{\natexlab{a}}){Madhusudhan}, {Bitsch},
  {Johansen}, \& {Eriksson}}]{Madhusudhan2017}
{Madhusudhan}, N., {Bitsch}, B., {Johansen}, A., \& {Eriksson}, L.
  2017{\natexlab{a}}, \mnras, 469, 4102

\bibitem[{{Madhusudhan} {et~al.}(2017{\natexlab{b}}){Madhusudhan}, {Bitsch},
  {Johansen}, \& {Eriksson}}]{Madhu2017}
---. 2017{\natexlab{b}}, \mnras, 469, 4102

\bibitem[{{Madhusudhan} {et~al.}(2011){Madhusudhan}, {Burrows}, \&
  {Currie}}]{Madhusudhan2011}
{Madhusudhan}, N., {Burrows}, A., \& {Currie}, T. 2011, \apj, 737, 34

\bibitem[{{Malo} {et~al.}(2013){Malo}, {Doyon}, {Lafreni{\`e}re}, {Artigau},
  {Gagn{\'e}}, {Baron}, \& {Riedel}}]{Malo2013}
{Malo}, L., {Doyon}, R., {Lafreni{\`e}re}, D., {et~al.} 2013, \apj, 762, 88

\bibitem[{{Marois} {et~al.}(2014){Marois}, {Correia}, {Galicher}, {Ingraham},
  {Macintosh}, {Currie}, \& {De Rosa}}]{Marois2014}
{Marois}, C., {Correia}, C., {Galicher}, R., {et~al.} 2014, in Society of
  Photo-Optical Instrumentation Engineers (SPIE) Conference Series, Vol. 9148,
  \procspie, 91480U

\bibitem[{{Marois} {et~al.}(2008){Marois}, {Macintosh}, {Barman}, {Zuckerman},
  {Song}, {Patience}, {Lafreni{\`e}re}, \& {Doyon}}]{Marois2008science}
{Marois}, C., {Macintosh}, B., {Barman}, T., {et~al.} 2008, Science, 322, 1348

\bibitem[{{Marois} {et~al.}(2010){Marois}, {Zuckerman}, {Konopacky},
  {Macintosh}, \& {Barman}}]{Marois2010}
{Marois}, C., {Zuckerman}, B., {Konopacky}, Q.~M., {Macintosh}, B., \&
  {Barman}, T. 2010, \nat, 468, 1080

\bibitem[{{Matthews} {et~al.}(2014){Matthews}, {Kennedy}, {Sibthorpe}, {Booth},
  {Wyatt}, {Broekhoven-Fiene}, {Macintosh}, \& {Marois}}]{Matthews2014}
{Matthews}, B., {Kennedy}, G., {Sibthorpe}, B., {et~al.} 2014, \apj, 780, 97

\bibitem[{{Mawet} {et~al.}(2017){Mawet}, {Delorme}, {Jovanovic}, {Wallace},
  {Bartos}, {Wizinowich}, {Fitzgerald}, {Lilley}, {Ruane}, {Wang}, {Klimovich},
  \& {Xin}}]{Mawet2017SPIE}
{Mawet}, D., {Delorme}, J.~R., {Jovanovic}, N., {et~al.} 2017, in Society of
  Photo-Optical Instrumentation Engineers (SPIE) Conference Series, Vol. 10400,
  1040029

\bibitem[{{Mieda} {et~al.}(2014){Mieda}, {Wright}, {Larkin}, {Graham},
  {Adkins}, {Lyke}, {Campbell}, {Maire}, {Do}, \& {Gordon}}]{Mieda2014}
{Mieda}, E., {Wright}, S.~A., {Larkin}, J.~E., {et~al.} 2014, \pasp, 126, 250

\bibitem[{{Molli{\`e}re} {et~al.}(2020){Molli{\`e}re}, {Stolker}, {Lacour},
  {Otten}, {Shangguan}, {Charnay}, {Molyarova}, {Nowak}, {Henning}, {Marleau},
  {Semenov}, {van Dishoeck}, {Eisenhauer}, {Garcia}, {Garcia Lopez}, {Girard},
  {Greenbaum}, {Hinkley}, {Kervella}, {Kreidberg}, {Maire}, {Nasedkin},
  {Pueyo}, {Snellen}, {Vigan}, {Wang}, {de Zeeuw}, \& {Zurlo}}]{Molliere2020}
{Molli{\`e}re}, P., {Stolker}, T., {Lacour}, S., {et~al.} 2020, arXiv e-prints,
  arXiv:2006.09394

\bibitem[{{Mordasini} {et~al.}(2016){Mordasini}, {van Boekel}, {Molli{\`e}re},
  {Henning}, \& {Benneke}}]{Mordasini2016}
{Mordasini}, C., {van Boekel}, R., {Molli{\`e}re}, P., {Henning}, T., \&
  {Benneke}, B. 2016, \apj, 832, 41

\bibitem[{{Nielsen} {et~al.}(2019){Nielsen}, {De Rosa}, {Macintosh}, {Wang},
  {Ruffio}, {Chiang}, {Marley}, {Saumon}, {Savransky}, {Ammons}, {Bailey},
  {Barman}, {Blain}, {Bulger}, {Burrows}, {Chilcote}, {Cotten}, {Czekala},
  {Doyon}, {Duch{\^e}ne}, {Esposito}, {Fabrycky}, {Fitzgerald}, {Follette},
  {Fortney}, {Gerard}, {Goodsell}, {Graham}, {Greenbaum}, {Hibon}, {Hinkley},
  {Hirsch}, {Hom}, {Hung}, {Dawson}, {Ingraham}, {Kalas}, {Konopacky},
  {Larkin}, {Lee}, {Lin}, {Maire}, {Marchis}, {Marois}, {Metchev},
  {Millar-Blanchaer}, {Morzinski}, {Oppenheimer}, {Palmer}, {Patience},
  {Perrin}, {Poyneer}, {Pueyo}, {Rafikov}, {Rajan}, {Rameau}, {Rantakyr{\"o}},
  {Ren}, {Schneider}, {Sivaramakrishnan}, {Song}, {Soummer}, {Tallis},
  {Thomas}, {Ward-Duong}, \& {Wolff}}]{Nielsen2019}
{Nielsen}, E.~L., {De Rosa}, R.~J., {Macintosh}, B., {et~al.} 2019, \aj, 158,
  13

\bibitem[{{{\"O}berg} {et~al.}(2011){{\"O}berg}, {Murray-Clay}, \&
  {Bergin}}]{Oberg2011}
{{\"O}berg}, K.~I., {Murray-Clay}, R., \& {Bergin}, E.~A. 2011, \apjl, 743, L16

\bibitem[{{{\"O}berg} \& {Wordsworth}(2019)}]{Oberg2019}
{{\"O}berg}, K.~I., \& {Wordsworth}, R. 2019, \aj, 158, 194

\bibitem[{{O'Neil} {et~al.}(2019){O'Neil}, {Martinez}, {Hees}, {Ghez}, {Do},
  {Witzel}, {Konopacky}, {Becklin}, {Chu}, {Lu}, {Matthews}, \&
  {Sakai}}]{ONeil2019}
{O'Neil}, K.~K., {Martinez}, G.~D., {Hees}, A., {et~al.} 2019, \aj, 158, 4

\bibitem[{{Petit dit de la Roche} {et~al.}(2018){Petit dit de la Roche},
  {Hoeijmakers}, \& {Snellen}}]{PetitditdelaRoche2018}
{Petit dit de la Roche}, D.~J.~M., {Hoeijmakers}, H.~J., \& {Snellen}, I.~A.~G.
  2018, \aap, 616, A146

\bibitem[{{Pueyo}(2016)}]{Pueyo2016}
{Pueyo}, L. 2016, \apj, 824, 117

\bibitem[{{Pueyo} {et~al.}(2015){Pueyo}, {Soummer}, {Hoffmann}, {Oppenheimer},
  {Graham}, {Zimmerman}, {Zhai}, {Wallace}, {Vescelus}, {Veicht}, {Vasisht},
  {Truong}, {Sivaramakrishnan}, {Shao}, {Roberts}, {Roberts}, {Rice}, {Parry},
  {Nilsson}, {Lockhart}, {Ligon}, {King}, {Hinkley}, {Hillenbrand}, {Hale},
  {Dekany}, {Crepp}, {Cady}, {Burruss}, {Brenner}, {Beichman}, \&
  {Baranec}}]{Pueyo2015}
{Pueyo}, L., {Soummer}, R., {Hoffmann}, J., {et~al.} 2015, \apj, 803, 31

\bibitem[{{Qi} {et~al.}(2015){Qi}, {{\"O}berg}, {Andrews}, {Wilner}, {Bergin},
  {Hughes}, {Hogherheijde}, \& {D'Alessio}}]{Qi2015}
{Qi}, C., {{\"O}berg}, K.~I., {Andrews}, S.~M., {et~al.} 2015, \apj, 813, 128

\bibitem[{{Read} {et~al.}(2018){Read}, {Wyatt}, {Marino}, \&
  {Kennedy}}]{Read2018}
{Read}, M.~J., {Wyatt}, M.~C., {Marino}, S., \& {Kennedy}, G.~M. 2018, \mnras,
  475, 4953

\bibitem[{{Reidemeister} {et~al.}(2009){Reidemeister}, {Krivov}, {Schmidt},
  {Fiedler}, {M{\"u}ller}, {L{\"o}hne}, \& {Neuh{\"a}user}}]{Reidemeister2009}
{Reidemeister}, M., {Krivov}, A.~V., {Schmidt}, T.~O.~B., {et~al.} 2009, \aap,
  503, 247

\bibitem[{{Ruffio}(2019)}]{Ruffio2019thesis}
{Ruffio}, J.-B. 2019, PhD thesis, Stanford University.
\newblock \url{http://purl.stanford.edu/yq071tj0740}

\bibitem[{{Ruffio} {et~al.}(2019){Ruffio}, {Macintosh}, {Konopacky}, {Barman},
  {De Rosa}, {Wang}, {Wilcomb}, {Czekala}, \& {Marois}}]{Ruffio2019RV}
{Ruffio}, J.-B., {Macintosh}, B., {Konopacky}, Q.~M., {et~al.} 2019, \aj, 158,
  200

\bibitem[{{Schwarz} {et~al.}(2016){Schwarz}, {Ginski}, {de Kok}, {Snellen},
  {Brogi}, \& {Birkby}}]{Schwarz2016}
{Schwarz}, H., {Ginski}, C., {de Kok}, R.~J., {et~al.} 2016, \aap, 593, A74

\bibitem[{{Skemer} {et~al.}(2011){Skemer}, {Close}, {Sz{\H{u}}cs}, {Apai},
  {Pascucci}, \& {Biller}}]{Skemer2011}
{Skemer}, A.~J., {Close}, L.~M., {Sz{\H{u}}cs}, L., {et~al.} 2011, \apj, 732,
  107

\bibitem[{{Skemer} {et~al.}(2012){Skemer}, {Hinz}, {Esposito}, {Burrows},
  {Leisenring}, {Skrutskie}, {Desidera}, {Mesa}, {Arcidiacono}, {Mannucci},
  {Rodigas}, {Close}, {McCarthy}, {Kulesa}, {Agapito}, {Apai}, {Argomedo},
  {Bailey}, {Boutsia}, {Briguglio}, {Brusa}, {Busoni}, {Claudi}, {Eisner},
  {Fini}, {Follette}, {Garnavich}, {Gratton}, {Guerra}, {Hill}, {Hoffmann},
  {Jones}, {Krejny}, {Males}, {Masciadri}, {Meyer}, {Miller}, {Morzinski},
  {Nelson}, {Pinna}, {Puglisi}, {Quanz}, {Quiros-Pacheco}, {Riccardi},
  {Stefanini}, {Vaitheeswaran}, {Wilson}, \& {Xompero}}]{Skemer2012}
{Skemer}, A.~J., {Hinz}, P.~M., {Esposito}, S., {et~al.} 2012, \apj, 753, 14

\bibitem[{{Skemer} {et~al.}(2014){Skemer}, {Marley}, {Hinz}, {Morzinski},
  {Skrutskie}, {Leisenring}, {Close}, {Saumon}, {Bailey}, {Briguglio},
  {Defrere}, {Esposito}, {Follette}, {Hill}, {Males}, {Puglisi}, {Rodigas}, \&
  {Xompero}}]{Skemer2014}
{Skemer}, A.~J., {Marley}, M.~S., {Hinz}, P.~M., {et~al.} 2014, \apj, 792, 17

\bibitem[{{Snellen} {et~al.}(2014){Snellen}, {Brandl}, {de Kok}, {Brogi},
  {Birkby}, \& {Schwarz}}]{Snellen2014}
{Snellen}, I. A.~G., {Brandl}, B.~R., {de Kok}, R.~J., {et~al.} 2014, \nat,
  509, 63

\bibitem[{{Soummer} {et~al.}(2011){Soummer}, {Hagan}, {Pueyo}, {Thormann},
  {Rajan}, \& {Marois}}]{Soummer2011}
{Soummer}, R., {Hagan}, J.~B., {Pueyo}, L., {et~al.} 2011, \apj, 741, 55

\bibitem[{{Su} {et~al.}(2009){Su}, {Rieke}, {Stapelfeldt}, {Malhotra},
  {Bryden}, {Smith}, {Misselt}, {Moro-Martin}, \& {Williams}}]{Su2009}
{Su}, K.~Y.~L., {Rieke}, G.~H., {Stapelfeldt}, K.~R., {et~al.} 2009, \apj, 705,
  314

\bibitem[{{Turrini} {et~al.}(2021){Turrini}, {Schisano}, {Fonte}, {Molinari},
  {Politi}, {Fedele}, {Pani{\'c}}, {Kama}, {Changeat}, \&
  {Tinetti}}]{Turrini2021}
{Turrini}, D., {Schisano}, E., {Fonte}, S., {et~al.} 2021, \apj, 909, 40

\bibitem[{{Vanderburg} {et~al.}(2018){Vanderburg}, {Rappaport}, \&
  {Mayo}}]{Vanderburg2018}
{Vanderburg}, A., {Rappaport}, S.~A., \& {Mayo}, A.~W. 2018, \aj, 156, 184

\bibitem[{{Vousden} {et~al.}(2016){Vousden}, {Farr}, \& {Mandel}}]{Vousden2016}
{Vousden}, W.~D., {Farr}, W.~M., \& {Mandel}, I. 2016, \mnras, 455, 1919

\bibitem[{{Wang} {et~al.}(2018{\natexlab{a}}){Wang}, {Mawet}, {Fortney},
  {Hood}, {Morley}, \& {Benneke}}]{WangJi2018}
{Wang}, J., {Mawet}, D., {Fortney}, J.~J., {et~al.} 2018{\natexlab{a}}, \aj,
  156, 272

\bibitem[{{Wang} {et~al.}(2020){Wang}, {Wang}, {Ma}, {Chilcote}, {Ertel},
  {Guyon}, {Ilyin}, {Jovanovic}, {Kalas}, {Lozi}, {Macintosh}, {Stone}, \&
  {Strassmeier}}]{WangJi2020}
{Wang}, J., {Wang}, J., {Ma}, B., {et~al.} 2020, arXiv e-prints,
  arXiv:2007.02810

\bibitem[{{Wang} {et~al.}(2021){Wang}, {Kulikauskas}, \&
  {Blunt}}]{whereistheplanet}
{Wang}, J.~J., {Kulikauskas}, M., \& {Blunt}, S. 2021, {whereistheplanet:
  Predicting positions of directly imaged companions}, , , ascl:2101.003

\bibitem[{{Wang} {et~al.}(2018{\natexlab{b}}){Wang}, {Graham}, {Dawson},
  {Fabrycky}, {De Rosa}, {Pueyo}, {Konopacky}, {Macintosh}, {Marois}, {Chiang},
  {Ammons}, {Arriaga}, {Bailey}, {Barman}, {Bulger}, {Chilcote}, {Cotten},
  {Doyon}, {Duch{\^e}ne}, {Esposito}, {Fitzgerald}, {Follette}, {Gerard},
  {Goodsell}, {Greenbaum}, {Hibon}, {Hung}, {Ingraham}, {Kalas}, {Larkin},
  {Maire}, {Marchis}, {Marley}, {Metchev}, {Millar-Blanchaer}, {Nielsen},
  {Oppenheimer}, {Palmer}, {Patience}, {Perrin}, {Poyneer}, {Rajan}, {Rameau},
  {Rantakyr{\"o}}, {Ruffio}, {Savransky}, {Schneider}, {Sivaramakrishnan},
  {Song}, {Soummer}, {Thomas}, {Wallace}, {Ward-Duong}, {Wiktorowicz}, \&
  {Wolff}}]{Wang2018}
{Wang}, J.~J., {Graham}, J.~R., {Dawson}, R., {et~al.} 2018{\natexlab{b}}, \aj,
  156, 192

\bibitem[{{Wertz} {et~al.}(2017){Wertz}, {Absil}, {G{\'o}mez Gonz{\'a}lez},
  {Milli}, {Girard}, {Mawet}, \& {Pueyo}}]{Wertz2017}
{Wertz}, O., {Absil}, O., {G{\'o}mez Gonz{\'a}lez}, C.~A., {et~al.} 2017, \aap,
  598, A83

\bibitem[{{Wilcomb} {et~al.}(2020){Wilcomb}, {Konopacky}, {Barman}, {Theissen},
  {Ruffio}, {Brock}, {Macintosh}, \& {Marois}}]{wilcomb2020}
{Wilcomb}, K.~K., {Konopacky}, Q.~M., {Barman}, T.~S., {et~al.} 2020, \aj, 160,
  207

\bibitem[{{Wilner} {et~al.}(2018){Wilner}, {MacGregor}, {Andrews}, {Hughes},
  {Matthews}, \& {Su}}]{Wilner2018}
{Wilner}, D.~J., {MacGregor}, M.~A., {Andrews}, S.~M., {et~al.} 2018, \apj,
  855, 56

\bibitem[{{Yurchenko} \& {Tennyson}(2014)}]{Yurchenko2014}
{Yurchenko}, S.~N., \& {Tennyson}, J. 2014, \mnras, 440, 1649

\bibitem[{{Zhang} {et~al.}(2020){Zhang}, {Liu}, {Hermes}, {Magnier}, {Marley},
  {Tremblay}, {Tucker}, {Do}, {Payne}, \& {Shappee}}]{Zhang2020}
{Zhang}, Z., {Liu}, M.~C., {Hermes}, J.~J., {et~al.} 2020, \apj, 891, 171

\bibitem[{{Zuckerman} {et~al.}(2011){Zuckerman}, {Rhee}, {Song}, \&
  {Bessell}}]{Zuckerman2011}
{Zuckerman}, B., {Rhee}, J.~H., {Song}, I., \& {Bessell}, M.~S. 2011, \apj,
  732, 61

\bibitem[{{Zurlo} {et~al.}(2016){Zurlo}, {Vigan}, {Galicher}, {Maire}, {Mesa},
  {Gratton}, {Chauvin}, {Kasper}, {Moutou}, {Bonnefoy}, {Desidera}, {Abe},
  {Apai}, {Baruffolo}, {Baudoz}, {Baudrand}, {Beuzit}, {Blancard},
  {Boccaletti}, {Cantalloube}, {Carle}, {Cascone}, {Charton}, {Claudi},
  {Costille}, {de Caprio}, {Dohlen}, {Dominik}, {Fantinel}, {Feautrier},
  {Feldt}, {Fusco}, {Gigan}, {Girard}, {Gisler}, {Gluck}, {Gry}, {Henning},
  {Hugot}, {Janson}, {Jaquet}, {Lagrange}, {Langlois}, {Llored}, {Madec},
  {Magnard}, {Martinez}, {Maurel}, {Mawet}, {Meyer}, {Milli},
  {Moeller-Nilsson}, {Mouillet}, {Orign{\'e}}, {Pavlov}, {Petit}, {Puget},
  {Quanz}, {Rabou}, {Ramos}, {Rousset}, {Roux}, {Salasnich}, {Salter},
  {Sauvage}, {Schmid}, {Soenke}, {Stadler}, {Suarez}, {Turatto}, {Udry},
  {Vakili}, {Wahhaj}, {Wildi}, \& {Antichi}}]{Zurlo2016}
{Zurlo}, A., {Vigan}, A., {Galicher}, R., {et~al.} 2016, \aap, 587, A57

\end{thebibliography}

\end{document}